\documentclass[12pt,preprint]{aastex}

\usepackage{color}
\usepackage{graphicx}
 
\newcommand{\Lya}{\mbox{Ly$\alpha$}}
\newcommand{\lya}{\mbox{Ly$\alpha$}}

\newcommand{\kms}{\mbox{km s$^{-1}$}}
\newcommand{\cmm}{\mbox{cm$^{-2}$}}

\newcommand{\flux}{$\mathrm{ergs}
        \mathrm{\ cm}^{-2}
        \mathrm{\ s}^{-1}
        \mathrm{\ Hz}^{-1}
        \mathrm{\ sr}^{-1}$}

\newcommand{\ob}{\mbox{$\Omega_b$}}
\newcommand{\obh}{\mbox{$\Omega_bh^{2}$}}
\newcommand{\om}{\mbox{$\Omega_m$}}
\newcommand{\ol}{\mbox{$\Omega_{\Lambda } $}}
\newcommand{\sig}{\mbox{$\sigma_8$}}
\newcommand{\gammahe}{\mbox{$X_{228}$}}
\newcommand{\gammah}{\mbox{$\gamma_{912}$}}

\newcommand{\nhi}{\mbox{N$_{\rm H I}$}}
\newcommand{\lnhi}{\mbox{log \nhi}}

\newcommand{\fbar}{\mbox{$\bar{F}$}}
\newcommand{\taueff}{\mbox{$\tau _{eff}$}}

\newcommand{\bsig}{\mbox{$b_{\sigma}$}}
\newcommand{\pl}{\mbox{$P_{-2}$}}
\newcommand{\pmed}{\mbox{$P_{-1.5}$}}  
\newcommand{\ps}{\mbox{$P_{-1}$}}

\newcommand{\lyaf} {\lya\ forest}

\newif\ifdraftmodep
\draftmodepfalse

\newcommand{\NOTE}[1]{\ifdraftmodep {\color {red} [{\it #1}]} \fi}

\newif\ifapjp
\apjpfalse


\begin{document}

\title{A Concordance Model of The Lyman-Alpha Forest at z = 1.95}

\author{
Tridivesh Jena,
Michael L. Norman,
David Tytler,
David Kirkman,
Nao Suzuki,
Angela Chapman,
Carl Melis,
Pascal Paschos,
Brian O'Shea,
Geoffrey So,
Dan Lubin, 
Wen-Ching Lin
\\
Center for Astrophysics and Space Sciences;\\
University of California, San Diego; \\
MS 0424; La Jolla; CA 92093-0424\\
Dieter Reimers, 
Eckardt Janknecht and
Cora Fechner
\\
Hamburger Sternwarte,Universitaet Hamburg,\\
Gojenbergsweg 112, D-21029, Hamburg, Germany
}

%\altaffiltext{1}{Physics Department, University of California at San Diego, 
%  San Diego, CA 92093}

\begin{abstract}
  
We present 40 fully hydrodynamical numerical simulations of the
intergalactic gas that gives rise to the Ly$\alpha$ forest. The 
simulation code, input and output files are available at 
http://www.cosmos.ucsd.edu/\~gso/index.html. For each
simulation we predict the observable properties of the H~I absorption
in QSO spectra. We then find the sets of cosmological and
astrophysical parameters that match the QSO spectra.  We present our results as
scaling relationships between input and output parameters. The input
parameters include the main cosmological parameters \ob , \om , \ol ,
$H_0$ and \sig ; and two astrophysical parameters \gammah\ and \gammahe . 
The parameter
\gammah\ controls the rate of ionization of H~I, He~I and He~II and is
equivalent to the intensity of the UV background.  The second
parameter \gammahe\ controls the rate of heating from the
photoionization of He~II and can be related to the shape of the UVB at
$\lambda < 228$ \AA .  We show how these input parameters; especially
\sig , \gammah\ and \gammahe ; effect the output parameters that we
measure in simulated spectra. These parameters are the mean flux \fbar , 
a measure of the most common \lya\ line width ($b$-value) \bsig , and the 
1D power spectrum of the flux on scales from 0.01 -- 0.1 s/km.  We 
compare the simulation output to
data from Kim et al. (2004) and Tytler et al. (2004) and we give a new
measurement of the flux power from HIRES and UVES spectra for the low
density IGM alone at $z=1.95$.

We find that simulations with a wide variety of \sig\ values, from at
least 0.8 -- 1.1, can fit the small scale flux power and $b$-values
when we adjust \gammahe\ to compensate for the \sig\ change.  We can
also use \gammah\ to adjust the H~I ionization rate to simultaneously
match the mean flux.  When we examine only the mean flux, $b$-values
and small scale flux power we can not readily break the strong degeneracy
between \sig\ and \gammahe .

We can break the degeneracy using large scale flux power or other data to
fix \sig .  When we pick a specific \sig\ value the simulations give
\gammahe\ and hence the IGM temperature that we need to match the
observed small scale flux power and $b$-values.  We can then also find
the \gammah\ required to match the mean flux for that combination of
\sig\ and \gammahe .  We derive scaling relations that give the output parameter
values expected for a variety of input parameters. We predict the line
width parameter \bsig\ with an error of 1.4\% and the mean amount of H~I
absorption to 2\%, equivalent to a 0.27\% error on \fbar\ ar $z=1.95$. 
These errors are four times smaller than those on the best current
measurement. We can readily
calculate the sets of input parameters that give outputs that match
the data. For \sig $= 0.9$, with \ob $=0.044$, \om $=0.27$, \ol $=0.73$,
$h = 0.71$ and $n=1.0$ we find \gammahe $ = 1.26$ and \gammah $ =
1.00$, equivalent to $\Gamma_{912} = 1.33 \times 10^{-12} $
ionizations per H~I atom per second.  If we run an optically thin
simulation with these input parameter values in a box size of 76.8 Mpc
comoving with cells of 18.75 kpc comoving we expect that the simulated
spectra will match \lyaf\ data at $z=1.95$.  The rates predicted by
\citet{madau99a} correspond to \gammah $=1$ and \gammahe $=1$. We are
in accord for \gammah\ while the larger \gammahe\ is reasonable to
correct for the opacity that is missing from the optically thin
simulations. To match data for smaller \sig , structure is more
extended and we need a smaller \gammahe\ corresponding to a cooler
IGM.  We also need a larger \gammah\ to stop the neutral fraction from
increasing at the lower temperatures.

\end{abstract}


\section{Introduction}

Hydrodynamic cosmological simulations \citep{cen94,zhang95a,hernquist96,
miraldaescude96,zhang97,zhang98,wadsley97,theuns98c}
 have revolutionized our understanding of Lyman alpha forest 
absorption lines seen in the spectra of high redshift QSOs.   Based on the 
impressive agreement between simulations and observations on a variety of HI 
absorption line statistics (see \cite{rauch98} for a review) it is now widely accepted 
that the HI \lya\ absorption is caused by mildly overdense, highly photoionized 
intergalactic gas that closely traces the dark matter distribution in CDM models of 
structure formation. According to these simulations, on scales of a Mpc or more, 
the dark matter and 
baryons trace out a network of sheets and filaments (Fig. 1) referred to as the cosmic 
web arising from the growth of primordial matter fluctuations \citep{bond96}. 

The intimate physical connection between absorption and dark matter density 
has stimulated many researchers \citep{croft98,croft02b,mcdonald00a,mcdonald03,mcdonald04a,
zaldarriaga01b,mandelbaum03,seljak03,seljak04a,viel04a} to 
explore the possibility of using observations of the \lya\ forest as a cosmological 
probe of the z=2-4 universe in much the same way as observations of CMB 
anisotropies have been used to probe the z=1300 universe \citep{spergel03a}.
The forest probes the primordial power spectrum on scales an order 
of magnitude smaller than the highest resolution CMB experiments. The function 
that assumes the role of the CMB angular power spectrum is the flux power 
spectrum. This is essentially the Fourier transform of transmitted flux spectrum 
averaged over many lines of sight (we will define it more precisely below). The 
current state-of-the-art is the work of \citep{seljak04a} who have measured the 
amplitude and shape of the matter power spectrum by combining SDSS and WMAP
results. They find consistency with the best fit WMAP 
LCDM model with $\sig = 0.90\pm 0.03$ and a primordial slope $n=0.98\pm 0.02$.

Somewhat decoupled from this effort, other researchers have used observations 
of the \lya\ forest and hydrodynamic simulations to probe the thermal 
and ionization state of the IGM at high redshift 
\citep{rauch97,schaye99,schaye00a,bryan00,theuns00b,tytler04b,bolton04b}.
Simulations and analytic studies have shown that the thermal state of 
the low density photoionized 
IGM is well approximated by the so-called equation of state of the 
IGM \citep{hui97} $T=T_0(\rho/<\rho>)^{\gamma-1}$, where $T_0 \sim 10^4$ K, and
$ 1 \leq \gamma \leq 1.6$. 
$T_0$ depends on the hardness of the UV background spectrum or other heating
mechanisms and $\gamma$ depends on 
the reionization history of the gas \citep{hui97}. Well after reionization, 
photoionization equilibrium holds in the low density gas responsible for 
the \lyaf\ 
and therefore its local ionization state is determined by the 
photoionization rate $\Gamma_{912}$ through the equilibrium condition
$n_{HI}\Gamma_{912}=n_e n_{HII}\alpha(T)$, where 
$\alpha(T)$ is the temperature-dependent 
recombination rate. Broadly speaking, the authors listed above have explored 
various means of relating the physical parameters $T_0, \gamma,$ and $\Gamma_{912}$ to 
observables. Available observables include the
mean flux \fbar ; which is related the 
effective \lya\
opacity \taueff ~via $\taueff =-ln(\fbar)$; the moments of the transmitted flux spectrum,
the line width (b-parameter) distribution function $f(b)$,
the flux (or opacity) distribution function, and the flux power spectrum. 

To cite just a few 
results, \cite{theuns00b} used the b-parameter distribution to measure the temperature 
of the IGM at z=3.25, finding $T_0 \geq 15,000$ K. 
\cite{schaye00a} used the lower 
cutoff in the $b-N(HI)$ scatter diagram to measure the temperature evolution of the 
IGM over the redshift range 2-4.5. They found evidence of late reheating at z$\sim$3, 
which they ascribed to late He II reionization by quasars. \cite{bryan00}
independently explored the same diagnostics and found that temperature 
estimates were sensitive to the assumed cosmology, in particular the amplitude 
of mass fluctuations on a few Mpc scales. Since the ionization state 
of the IGM depends on the gas temperature through the recombination rate,
this implies that estimates of the photoionization rate $\Gamma_{912}$ from the
effective opacity will also depend on cosmology. 
If the Bryan \& Machacek result is correct, it has
important implications that the determination of cosmological parameters
and the thermal/ionization state of the IGM are coupled problems and should be
treated as such. The two types of investigations described above
cannot be done independent of one another. 

This paper is a sequel to \citet[T04b]{tytler04b} where we presented a 
measurement of the mean flux at $z=1.9$ and we compared the output of
five simulations to both the mean flux and the $b$-value distribution.
In T04b we found one set of parameters that gave an excellent fit to the data
and we measured the H~I ionization rate.
Here we give a more thorough exploration of the
influence of cosmology and astrophysical parameters
on the observed properties of the \lyaf\ using a large grid of high resolution
Eulerian hydrodynamic cosmological simulations. 
Our grid of models is more extensive in its
parameter coverage than those cited above, and goes further in attempting to explore
box size and resolution effects. 
We derive scaling relationships between input and output 
parameters. The input parameters include cosmological parameters,
in particular the amplitude of the matter
power spectrum \sig\ and 
two astrophysical variables that control the intensity and shape of the UV 
background: \gammah, the HI photoionization rate and
\gammahe, an extra heating parameter implemented as the He II photoheating
rate.
We examine how these input parameters change the measured 
properties of simulated spectra, including the
mean flux, b-values and flux power.

We find that simulations with a wide variety of \sig\ values,
from at least 0.7 -- 1.1, can fit the small scale flux power and $b$-value
distribution when we adjust 
 \gammahe\ to compensate for the \sig\ change.
We can also fit the mean flux simultaneously by adjusting the H ionization      
rate \gammah\ to match the mean flux.
When we examine just the mean flux, b-values and small scale power 
we cannot break the strong degeneracy between \sig\ and \gammahe.

The outline of this paper is as follows. In \S \ref{sec:CosSim}
we describe our hydrodynamic simulations and grid of models. In 
\S \ref{sec:Z} we summarize relevant scales at our 
operating redshift z=1.95. In \S \ref{subsec:SpecX} and \ref{sec:Meas}
we describe how we made spectra from the simulations and the measurements
that we made on them.
In \S \ref{sec:ObSmpl} we discuss the observational data. 
In \S \ref{sec:ConvTst}
we examine numerical convergence to the quantities of interest.
In \S \ref{sec:Out} we introduce the main results of the paper, which are the 
correlations between input and output parameters, which we present as 
scaling relations in \S \ref{sec:Scal}.
We discuss the joint determination of cosmological and astrophysical 
parameters in \S \ref{sec:JointDeterm}
and present our conclusions in \S \ref{sec:Conclude}.

\section{Hydrodynamical Simulations \NOTE{section\_b}}
\label{sec:CosSim}

Cosmological simulations enable us to precisely determine the physical
state of the Ly$\alpha$ forest, including density, temperature,
ionization state and peculiar velocities of the gas responsible for
Ly$\alpha$ absorption. We have performed several simulations of the
Ly$\alpha$ forest using different cosmological models. All simulations
were performed using our cosmological hydrodynamics code Enzo. Enzo
incorporates a Lagrangean particle-mesh (PM) algorithm to follow the
collisionless dark matter and a higher-order accurate piecewise
parabolic method (PPM) to solve the equations of gas dynamics. In
addition to the usual ingredients of baryonic and dark matter, Enzo
also solves a coupled system of non-equilibrium ionization equations
with radiative cooling for a gas with primordial abundances. Our
chemical reaction network includes six species: HI, HII, HeI, HeII,
HeIII and $e^{-}$ \citep{abel97a, anninos97a}.  

The simulation starts
with the initial perturbations originating from inflation-inspired
adiabatic fluctuations. The Eisenstein-Hu transfer function
\citep{eisenstein99} is employed with the standard Harrison-Zel'dovich
power spectrum with slope $n_s = 1.0$. 
The simulation is evolved
starting at a redshift of 60, where the perturbations on the scale of
the box are small, to a redshift of 1.9. We examined only the output at $z=2.0$.

Another important component in the simulations is an
ultraviolet (UV) radiation background which ionizes the neutral
intergalactic medium. \citet{madau99a} have provided a UV radiation
field with a radiation transfer model in a clumpy universe based upon
the observed quasar luminosity function and radiative contribution
from the early stars in galaxies as well.  Enzo starts to import their
homogeneous UV background spectra at redshift 7 and increases the
intensity of the spectra at redshift 6 to generate photoionization and
photoheating rates in our simulations.

We vary the amplitude of the UVB using a
parameter that changes the amplitude but not the shape
of the UVB. We measure this amplitude with the parameter
\gammah\ that is the rate of ionization of H~I
in units of that predicted by \citet{madau99a}. 
At $z=1.9$ we have
\begin{equation}
\Gamma _{912} = 1.329 \times 10^{-12}\gamma _{912} \rm~s^{-1},           
\end{equation}
where $\gamma _{912} $ is a dimensionless number.  \citet{madau99a}
predicted $\gamma _{912}=1$ and when we adopt their spectrum shape
the ionization rate is proportional to the intensity of the UVB, $J_{HI}$
\citep[Eqn. 3]{hui02}. Since we do not change the shape of the UVB
the rates of photoionization of He~I and He~II per atom of He~I and He~II are
also proportional to \gammah .

We use a second parameter, \gammahe , to control the heat input.
This does not explicitly change the shape of the UVB, and it does not change 
the He II photoionization rate.
The rate of ionization of He II to He III is given by 
\gammah\ and the shape of the UVB alone; it is independent of \gammahe . 
We decouple the He II ionization rate from the heat input by He II
ionizations to help us correct for the heat missing in the optically thin
limit.
The rate of He II photo-heating, per He II atom, is given by the 
product \gammah \gammahe .
In \S 12 of T04b we did not make this point explicitly, and the text could be 
read to incorrectly imply that the He II photo-heating was independent of
\gammah . We have changed the notation to make this point more clear. The
parameter \gammahe\ was called 
$\gamma _{228}$ % WARNING the gamma228 here is special
in T04b.

The heating of the gas in the simulations is from the compression of gas, 
shocks and photoionization.
To first order the heating of the IGM depends on the shape of the
UVB and the extra heating, but not on the intensity of the UVB. This is because 
increased intensity leads to decreased
H~I, He~I and He~II leaving the radiative heating per baryon unchanged
(\citealt{miraldaescude94}; \citealt[Eqn. 3]{abel99};
\citealt[Eqn. 21]{valageas02}).
Since we did not explicitly change the shape of the UVB, the heating 
per baryon depends on \gammahe\ alone and not on \gammah .
In T04b we incorrectly stated in \S 12.2.2 that the 
shape of the UVB depended on the ratio of \gammah\ to \gammahe .

Larger values for \gammahe\ give hotter gas.
\citet{bryan00} suggested that late He~II
reionization resulted in increased heating of Helium in turn resulting
in an increase in the b parameters. This increased heating of Helium is
simulated via the \gammahe\ parameter. The extra
heating could be due to any other reason as well because the \gammahe\ 
parameter is just a proxy for any source of extra heating.  

We can interpret \gammahe\ as a particular type of change in the shape of 
the UVB at wavelengths $<228$ \AA . The change should leave the rate of 
photoionization constant, while increasing the heating.
The rate of photoionization is proportional to
$\nu ^{-2.7}$ while the heat input is proportional to the energy of the 
photoelectron, $\nu - \nu _{228}$. We could use these two constraints to
calculate the UVB spectrum implied by any \gammahe\ value.
The simulation that we obtain using some \gammahe\ will be the same as one with
\gammahe $=1$ and the UVB changed in this way.

Values of \gammahe $>1$ are intended to simulate heating that
might be missing because of missing opacity.
We can interpret \gammahe $<1$ as a change in the UVB shape, but
not a correction for opacity. 
Three of our simulations; I, H and J; do use \gammahe $< 1$ and they appear
to follow the same trends as models with \gammahe $>1$.

Another assumption made in our simulations is that the UV background
is spatially uniform. To be more precise, we use a spatially averaged
photoionization rate and thus smooth over any fluctuations in the
background field. The effect of these inhomogeneities have been
studied by \citet{haardt96,croft03}; and \citet{meiksin04a}. \citet{haardt96}
find that UV background fluctuations effects are very small at $z=1.95$; 
for example, they contribute less
than 1\% of the signal in the flux power at $z = 2.5$ at $k = 0.002$ s/km.

\subsection{List of Different Hydrodynamical Simulations \NOTE{subsection\_b}}
\label{sec:CosSimList}

In this paper we determine the set of ${\gamma_{912}}$ and \gammahe\
values that produce simulated spectra that match the data.
We typically leave other parameters fixed at
values given for \citet{spergel03a} using WMAP data; 
namely ${\Omega_b}$ = 0.044, ${\Omega_m}$ = 0.27,
${\Omega_{\Lambda}}$ = 0.73 and {h} = 0.71. 
However, we do explore a wide range of \sig\ values.

In Table \ref{table:SimTable.tbl} we
list the input parameters that describe the simulations that we have run 
over the last 5 years for a variety of projects.
The input parameters include the usual cosmological parameters
and the two astrophysical parameters
${\gamma_{912}}$ and \gammahe .
We do not list the initial slope of the power spectrum; that we keep at $n=1.0$.
The box size $L$ is in comoving Mpc (not $h^{-1}$Mpc) and $N$ is the
number of cells in the simulation.
We also list the cell size in comoving kpc, $C=L/N$, because this has a 
significant effect on the b-values and the small scale flux power.

We sort the 40
simulations into groups according to the input 
parameter that 
is changing, shown in bold. We separate the groups with empty rows.
We name the simulations with letters, where
simulations A to E were used in T04b, and simulations with
the same letter mostly differ in only one input  parameter.
We list a simulation more than once if we use it to examine more than one
parameter.

Simulation A is our largest box. It is accompanied by A2, A3 and A4
that differ only in box size.
A5 is identical to A3, using our most common box and cell size, 
but A5 uses a random number seed that initializes the power spectrum 
different from all the other simulations.
W1 and K2, a second series, differ in box size but
have a larger ratio \gammahe\ and are hotter than the A series.
The series B2, B and A4 vary only in cell size, as do the 
K series that have larger \gammahe\ and are hotter.
We have 8 sets that differ in both box and cell size:
L4, U;
L5, Q1;
and 6 combinations using one of A, A2 or A3 with one of B or B2.

Simulations F, G, H, I, J and K1 are all in our most common 19.2 Mpc box, but
with a smaller 37.5~kpc cell size which is better suited for b-values and 
small scale power.
Four sets differ only in \gammahe :
I, H, J;
C, N, D, E;
K2, P5;
and the O series. We will find that models with \gammahe $\simeq 1.3$
and cell sizes $\leq 37.5$ kpc 
are the most like data for \sig $\simeq 0.9$.
Four sets differ only in $\sigma_8$:
the L series; 
M, K2;
U, Q1 and
P2, Q. 
Three sets differ only in $\gamma_{912}$:
O2, P1, P2; 
L3, P5, C;
and
Q, Q1. 
K2 joins the S series, differing in both \ob\ and $h$.
Set T, K2 differ in \om\ and \ol , as do the large cell size pair K3, V.

\subsection{Random Number Seeds \NOTE{subsection\_b}}
\label{sec:CosSimSeeds}

Nearly all the simulations that we ran began with the same random
number generator seed.
The constant seed means that the simulations have the same
overall initial conditions of density and velocities.
Two simulations with different cell or box sizes look
very similar on large scales if we overlay the entire boxes.
In larger boxes the overall structure becomes larger in Mpc,
it occupies the same proportions of the box length, and
it scales to smaller wavenumbers.
In simulations with smaller cells, we have more cells in a given
structure.

The mean density over all cells in each simulation is exactly the
mean density for that cosmological model. We do not vary this mean:
\begin{equation}
{1 \over N^3} \sum \rho = \rm mean ~density~ of~ model~ universe,
\end{equation}
where $N^3$ is the number of cells and $\rho $ is the comoving gas density in
each cell.
If we were to run many simulations with different seeds, they would
show no variation in the mean density.

There is a similar constraint on the variance of the initial
density in the cells,
\begin{equation}
{1 \over N^3}\sum (\rho -\bar {\rho })^2,  ~or~
{1 \over N^3}\sum \rho ^2 = \rm constant~for~ cosmology,~ box ~and~cell~size.
\end{equation}
This follows from Parseval's theorem, because the initial mass power that we 
put into each simulation is exactly that expected for the average over the
universe. This initial power does not change with the random seed.
As a simulation evolves, the mean
matter density remains a constant, but the total power, or variance of the
density, increases.
The initial mass power in the simulations does not explain why we will later
see the small scale flux power decrease as box size increases. Larger boxes 
contain more mass power on all scales. They contain
mass power on scales that did not fit in the smaller boxes,
they contain more mass power on scales that just fit into the smaller boxes,
and they contain no less power on small scales.

The simulations are too small to individually
contain the full variety of density in the universe.                  
Moreover, both the mean density and the total power at included wavenumbers
in each simulation, with any random seed, are identical to the cosmic means.
Hence, if we made many simulations with different seeds and averaged
their results, we would still not sample the full range of density and power.
Since none of the
boxes contain the lowest and highest densities found in the universe,
the errors in the parameters that we measure from the simulated spectra
may be larger than implied by the comparison of different simulations.
We will see less variation in many 
statistics than similarly sized portions of the universe and much 
less for our smallest boxes.  
Larger simulations or many small ones that explicitly sample the range of
densities and power expected in the universe would give improved accuracy.

We will return to the topic of random seeds at the end of our
discussion of the artificial spectra in 
\S \ref{subsec:SpecX}, in
\S \ref{subsec:MeasP} when we discuss measuring the flux power, and in
\S \ref{subsec:MeasErr} where we estimate the errors in parameters measured
from the simulated spectra.
\section{Redshift and Measures of Length \NOTE{section\_b1}}
\label{sec:Z}

The results that we present from simulations are effectively for $z=1.95$.
The simulations were evolved to $z=2.00$. However, when we make 
artificial spectra we further evolve the H~I gas density per comoving
Mpc$^3$ along the sight line
using $\rho (z) = \rho(2) ((1+z)/3)^{3}$.
We do not change the particle positions, density fluctuations, velocities,
temperatures, or the ionization rate per atom.
This is the scaling we expected if these other parameters are all held constant,
the space expands and the $H(z) \propto (1+z)^{3/2}$. 
The effective redshift of our spectra is near 1.95 because all spectra
extend from $z=2.00$ to 1.90.
We made this choice to help match real spectra in a different project,
although real spectra typically span a much larger range of redshift.

The natural unit of simulations is Mpc but for observations it is \kms .
A change in redshift from 1.95 to 1.951
corresponds to a change in the observed wavelength of \lya\ of 1.21567~\AA\ or 
a velocity difference of 101.607 \kms .
In a simulation of a universe with  
H$\rm _o=71$~\kms~Mpc$^{-1}$, 
\ol $= 0.73$, and \om $= 0.27$, 
this is an interval of 1.525 Mpc comoving, and an increase in lookback
time of 1.686Myr.
We then have 66.62~\kms\ per comoving Mpc, and 
$H(z) = 2.768H_o = 196.5$~\kms per proper Mpc.

Most of our simulations are in cubic boxes with side length 
$L=19.2$ Mpc comoving, or $L_v=LH(z)/(1+z)=1279$~\kms . 
If the boxes were 1D, they would contain power from modes
with wavelengths $\lambda _n = L_v/n$, $n=1,2,3,... $ and wavenumbers
$k_n = 2\pi n/L_v$~s/km, where $k_1 = 0.00491$~s/km. 
The power is less than it should be because only these periodic
modes are used, both at the start of the simulations and as they evolve. 
The error decreases with larger $n$.
Since the simulations are in 3D cubes, the precise wavenumbers that are 
periodic depend on direction relative to the box sides, and an 
artificial spectrum experiences a wide variety of
wavenumbers with $k > 2 \pi n/L_v$
\citep{tormen96a}, but still with a relative lack of power at small $n$.

Most of our simulations have $256^3$ cells, each of size 75~kpc 
comoving, or 5.0~\kms . Their Nyquist frequency is $k_{128}=0.628$~s/km.
We use several simulations with twice this resolution: cells of size 37.5~kpc 
comoving, or 2.50~\kms . 

Simulations A4, B and B2 use 9.6 Mpc boxes. Although they are typically
linear, with $\delta \rho /\rho << 1$ at $z=3$, they can become 
non-linear by $z=2$, and give unreliable results because they lack
long scale mode-mode coupling. In section 5.2 \citet[C02b]{croft02b}
state that the 
non-linear scale is $k \simeq 0.02$ s/km at z=2.72.
We return to this issue when we compare measurements for different box
sizes in \S \ref{subsec:Box}.

\section{Extraction of Simulated Spectra \NOTE{section\_c}} 
\label{subsec:SpecX}

Spectra are extracted from each box as described in
\citet{zhang97}. To summarize, the spectrum generator starts at the
point with the lowest neutral hydrogen density inside the box,
shooting photons along random lines of sight through the
box calculating the transmitted flux of a QSO
at redshift z as $e^{- \tau_{\nu}}$, with the optical depth
$\tau_{\nu}$ given by
\begin{equation}
   \tau_{\nu} (t) \equiv \int_{t}^{t_{0}}   n_{HI}(t) \sigma_{\nu} c dt
\label{eqn:tau.eqn}
\end{equation}
where c is the speed of light, $n_{HI}$ is the number density of the
HI absorbers, $\sigma_{\nu}$ is the absorption cross-section, t is the
corresponding cosmic time at redshift z and $t_{0}$ is the cosmic time
today. Integration is performed along the line of sight from the QSO
to the observer. This can be written in a form more suitable for
computation (Zhang et.al. 1997) as
${ \tau_\nu(z) = \frac{c^2 \sigma_o}{\sqrt{\pi} \nu_o} \int_{z}^{z_o} 
   \frac{n_{HI}(\acute{z})}{b} \frac{a^2}{\dot{a}} 
   \,exp \left\{ - \left[ (1+\acute{z})\frac{\nu}{\nu_o} - 1 + \frac{v}{c} 
   \right]^{2} \frac{c^2}{b^2} \right\} d\!\acute{z}        }$
where $z$ is the redshift, $\sigma_o$ is the resonant Ly-${\alpha}$
cross-section, $\nu_o$ is the Ly$\alpha$ rest frequency, $v$ is the
peculiar velocity along the line-of-sight and $\nu$ is the redshifted
frequency. The effect of Doppler broadening on the absorption cross
section is given by the parameter b and is equal to $\sqrt{2kT/m_p}$,
where k is the Boltzmann constant, T is the gas temperature and $m_p$
is the mass of a proton. This equation, parameterized to order ${v/c}$,
also needs the scale factor $a$ to be specified, which is given by the
Friedman equation,
${  \dot{a} = H_o \sqrt{1 + \Omega_m (\frac{1}{a}-1) + \Omega_{\Lambda} (a^2-1)} }$

For each redshift, we made 25 simulated spectra.
Averaging over multiple spectra is necessary to reduce the effects of
cosmic variance. All spectra are of length 
$\Delta$z = 0.1, and they travel around our 19.2 Mpc simulation box about 
8 times, changing direction each time they hit a box edge. 

Since the rays leave the starting point 
in random directions, we expect the spectra to be relatively independent on
scales $<L$.
If we chop the rays into segments of length $L$ and arrange them
on a uniform grid over one face of the box they would be separated by 
1.3 Mpc.
However, if we made many times more spectra they would become duplicative 
and add little information. We discuss the errors in measured quantities 
from the 25 spectra in \S \ref{subsec:MeasErr}.

Most of our simulations used the same initial random number seeds, and 
hence their overall power distributions are very similar in units of $kL$,
specially for large modes.
We also send simulated spectra down exactly the same directions.
However, the spectra do differ in other ways. The location of the cell
with the lowest density will change slightly with differing evolution 
coming from different input parameters.
All spectra cover a distance in Mpc equivalent to a redshift path of 0.1. 
The corresponding distance
in the boxes will depend on the \ol\ and $h$ because the simulation
grid is measured in Mpc, not km/s.
Also, in smaller boxes the spectra must pass through the box more times to 
accumulate the redshift path.

Other than these differences, the spectra differ primarily in the 
the initial density field that is given by the cosmology, and in the
evolution, but less so in the ways in which the spectra sample the box.
When we compare simulation to simulation, random variations are 
suppressed, because the initial conditions and the sight lines
are similar. The suppression is a major factor for the large scale power.
However, when we compare to simulations using different seeds
or to data, we will see much larger differences,
because of the different random fluctuations in the 
small boxes \citep{barkana04a}.

\section{Measurement of the Simulated Spectra \NOTE{subsection\_c}}
\label{sec:Meas}

In this section we describe how we measure the mean flux, b-values, and
flux power in the simulated spectra.
We work with the ''raw" simulated spectra, with native resolution,
exceedingly high S/N, no continuum level error. We do not rebin these 
spectra to HIRES sized bins for any of the measurements presented here.

The mean flux is simplest. We work with flux, and also with the
effective optical depth, \taueff $ = -ln $ \fbar , because the former
is easier to comprehend while the latter gives better scalings.
The other parameters require more discussion.

\subsection{Measurement of the b-values \NOTE{subsection\_c}}
\label{subsec:MeasB}

We fit Voigt profiles to the \lya\ lines in each simulated spectrum using 
the code described in \citet{zhang97}.
As in T04b we represent the b-value distribution with a single
parameter, \bsig , that is proportional to the position of the peak of the
distribution. We measure this parameter by fitting
\begin{equation}
dn/db = B_{HR}
{b_{\sigma}^4 \over b^5}             
exp \left( -{b_{\sigma}^4 \over b^4} \right),
\end{equation}
from \citet{hui99c}
to a distribution of lines \citep[Eqn. 3]{kim01}
where $dn/db$ is the number of lines per \kms\ and
we can use $b _{\sigma}$ to describe the velocity of the peak of the function,
since $b_{peak} = \sqrt(2) b_{\sigma} 5^{-1/4} = 0.9457 b_{\sigma }$.
We bin the b-values in bins each 2 km/s wide, where $b_n$ extends
from $2n$ to $2(n+1)$ km/s. 
The binning is significant for small samples when we work with the peak.
In T04b we found for the first time, as excellent agreement between the
entire distribution given by data, our simulation B and the fitting formula.
We will confine our discussion to \bsig\ rather than the entire b-value 
distributions.

\subsection{Calculation of the Power Spectrum \NOTE{subsection\_c}}
\label{subsec:Pk}

The systematic errors involved in calculating the flux power spectrum     
from HIRES QSO spectra has been investigated in detail before           
C02b and we make full use of those results. They conclude
that noise and unremoved metal lines in the spectra does affect the
power on scales $k < 0.15$ s/km.  We also use
their result to be confident that our scaling of the observed spectra
to a mean redshift, as described in \S \ref{subsubsec:ObSmplNewP}, has 
not introduced any systematic biases.           

Given an absorption spectrum, which is the transmitted flux as a
function of wavelength, one can take it's Fourier transform to get the
flux power spectrum. However, this quantity would depend on the mean
flux of the spectrum which is a strong function of redshift, and uncertain. 
Therefore
it is better to first define a flux overdensity 
\begin{equation}
\delta_f \equiv \frac{f-\bar{f}}{\bar{f}}, 
\end{equation}
where $f$ is the flux in units of the continuum level (unity in the 
simulated spectra) as a function of wavelength, and
$\bar{f}$ is the mean flux of the spectrum in a given redshift 
interval and is calculated by averaging
all data points in the spectrum. 
In terms of \citet[K04e]{kim04a,kim04e}, $f = F1$ and $\delta_f = F2$.

In this work, we chose to make
$\delta_f$ a function of velocity. One can do this by first
transforming wavelength, $\lambda$, to redshift via the relation:
\begin{equation}
\lambda = \lambda_o (1+z)
\end{equation}
where z is the redshift and ${\lambda_o}$ is the rest frame Ly$\alpha$
wavelength of 1215.67 Angstrom. For small spectrum lengths we can then
approximate the velocity separation $\Delta v_\parallel$ as:
\begin{equation}
\Delta v_\parallel \equiv c\frac{\mid z_{j} - z_{i} \mid}{1+\bar{z}}
\label{eqn:Vp_def.eqn}
\end{equation}
where $\mid z_{j} - z_{i} \mid$ $\ll$ $\bar{z}$, $\bar{z}$ =
$\frac{z_1 + z_n}{2}$ and the spectra is given at redshifts ${z_1,
  z_2, ... , z_n}$. One can now define the flux power, $P_{f}(k)$, as
the complex conjugate square of the Fourier transform of $\delta_f$; k
is defined as 2$\pi$/($\Delta v_{\parallel}$).

In this work, we perform the Fourier transform using a publicly
available code: FFTW ("The Fastest Fourier Transform of The West")
\citep{FFTW98}. FFTW computes the unnormalized discrete Fourier
transform, so to get the physical transform $\delta_f(k)$, we need to
multiply the output of FFTW by ${\frac{\Delta v}{N}}$, where N is the
number of data points in the spectra and $\Delta v$ is the velocity
separation between the two ends of the spectrum.  We will be comparing
the power spectrum of spectra of different lengths, so we include a
factor of ${(1/L)^{1/2}}$, where L is the length of the spectrum.
Finally therefore, the power spectrum $P_{f}(k)$ = ${\delta_f(k)^*
  \delta_f(k)}$ = ${\frac{\Delta v}{N^2}}$ $\times$ (FFTW-Output), where
N is the number of data points in the spectra.

The overall shape of the flux power spectrum is understood. 
Though the initial mass power has larger
amplitudes at smaller scales, at later redshifts modes of shorter
wavelengths have their amplitudes reduced relative to those of long
wavelengths. This is due to Jeans mass effects where pressure opposes
growth of perturbations at small scales and damping effects where free
streaming of collisionless dark matter particles erases perturbations
on small scales.

\subsection{Parameters Describing the Flux Power \NOTE{subsection\_c}}
\label{subsec:MeasP}

We measured the flux power at $k_m = 0.0011+0.0001m$, $m=1,2,..1480$ s/km.
We must average the power over many of these fine $k$ steps to increase 
the S/N enough to reveal the changes between the simulations. We found that
we obtained an excellent fit to all spectra in 7 bins from $-2 < log k < -0.8$
s/km, and also using a polynomial:
\begin{equation}
\log P(k) = A l + C l^2 + E l^3  + F l^4 + G l^5
\end{equation}
where $l = \log k  - B$.
We list the coefficients for this polynomial for all our simulations in
Table \ref{table:pfits}.
We ignore the increase in the S/N with $k$ that occurs because each 
spectrum samples more large $k$ modes.

To focus the discussion we use the polynomial fits to estimate the flux
power at three $k$ values:
\begin{itemize}
\item
\pl\ at $log k= -2.0$ s/km  % appears as $P_{-2}$
\item
\pmed\ at $log k= -1.5$ s/km, or $k \simeq 0.03162$ s/km % $P_{-1.5}$
\item
\ps\ at $log k= -1.0$ s/km. %            $P_{-1}$
\end{itemize}
\citet{viel03b} also show that high column density \lya\ lines with
\lnhi $>14$~\cmm\ dominate \pl , while lower column lines with
\lnhi\ 13 -- 14 \cmm\ dominate \ps .

The \pl\ is approximately the largest scale that we can estimate accurately
in our larger boxes.
The \pl\ includes modes $n \geq 1$ for a 9.6 Mpc box, $n \geq 2$ for a
19.2 Mpc box, $n \geq 4$ for a 38.4 Mpc box,
and $n \geq 16$ for our single largest box, $L=76.8$ Mpc.
It is customary not to place much faith in measurements at 
$n < 3$. Hence, the loss of power in \pmed\ is serious in the 9.6 Mpc
boxes, and important for the 19.2~Mpc boxes. 
We quantify this issue in \S \ref{subsec:Box}.

We prefer to keep \pl\ at scales that are relatively large for our boxes
because this helps us measure the effects of the loss of power, and to
provide overlap with measurements from intermediate resolution spectra. 
For example,
C02b use their LRIS sample at $log k < -1.85$ s/km ($k < 0.014$), while
\citet[M04a]{mcdonald04a} present the power of the SDSS QSO spectra at
$log k \leq -1.75$ s/km ($k \leq 0.01778$ s/km).

Our simulations with 75 kpc cells
contain power on scales six times smaller than \ps .
However, scales $k >0.1$ s/km are of less use today because
measurements of real spectra are poor. The metal lines are an increasing part
of the power, and many spectra are effected by photon noise in small scales.

\subsection{Errors on Measurement of the Simulated Spectra \NOTE{subsection\_c}}
\label{subsec:MeasErr}

In Table \ref{tbl:obsimathreefive}
we list observed parameters for different sets of spectra passing
through simulations A3 and A5. Set A3-1 is for the usual set of 25 spectra, 
starting from the point with the lowest density.
Set A3-2 starts at the same point as A3-1, but the directions are                
different because we used a different random seed for directions.           
In set A3-3 all 25 spectra start from different points, but they travel in      the same direction as 2, since they use the same seed for the directions.
Sets A3-4 to A3-10 use different starting points for each spectrum, with 25     spectra per set as usual.      

For each box we made 25 simulated spectra with a total path
length of 2.5 units of redshift. From Eqn. 30 of T04b the error from
a sample of real spectra with this size will be 0.0076 in \fbar .
Sets A3-3 to A3-10 have a mean \fbar $= 0.87479$ with $\sigma =  0.0028$.
They show less dispersion than we expect for real spectra because each sight 
line pass through the box about 8 times, different spectra all come from the 
same box, and the box contains less variety than portions of the universe 
with the same size. The error on the measurement of \fbar\ 
from a single set of 25 spectra will be at least 0.0028, with some dependence 
on \fbar\ that we will ignore.

We find that starting from the lowest point in the box was a mistake that
lead to systematic offsets that we correct.
\fbar\ = 0.87995 for A3-1 and A3-2. This is larger by 0.0051 than 
the mean from A3-3 to A3-10 of $0.8748 \pm 0.0010$
which is more than the error on a measurement using a set of 25
simulated spectra, but only half the error on the measurement of the 
real spectra. We will correct for this along with box and cell size effects
when we present scaling relations in \S \ref{sec:Scal}.
The size and sign of the effect are consistent with starting the spectra 
in the cell with the lowest density, where the gas will absorb less than usual.
We do not see any large effect from the directions in which the simulated 
spectra travel through the box.

The \bsig\ values for A3-3 to A3-10 have mean 24.95 \kms , with $\sigma = 0.18$
km/s. We see no effect here from the initial starting point because
A3-1 give \bsig\ very close to this mean, while A3-2 is $1.7 \sigma$ low.
However, the \bsig\ from A5 is over $2\sigma $ less than A3-1,
suggesting that the random seed has at least as large an effect as the
starting point and the size of the sample of 25 spectra.
These changes are all small compared to the measurement error for data
of approximately 1.5 km/s.

Simulation A5 is identical to A3 except that A5 used a different seed from our
other simulations, leading to a different pattern of density fluctuations.
The five measured parameters all differ by very small amounts:
0.0006 (0.07\%) for \fbar , 0.06 (1.1\%) for \pl , 0.04 (2.5\%) for \pmed ,
0.0008 (1.3\%) for \ps , and 0.44 (1.8\%) for \bsig .
The external errors that we should use comparing our simulations with data
or other simulations should be at least as large as these factors.

We do not give the corresponding discussion for the flux power, but we can
estimate the sense of the effects.
When we decrease the \fbar\ by 0.0051, adding back the typical absorption that
is underestimated at the start of each simulated spectrum, we expect the
large scale flux power to increase. The sizes of these changes can be
estimated from the Figures given in \S \ref{sec:Out}.

\section{Observed Data \NOTE{section\_d}}
\label{sec:ObSmpl}

We discuss the observational data before the simulations because the 
errors that we would like for the simulations should be less than those in the 
data.

\subsection{Mean Flux \NOTE{subsection\_d}}
\label{subsec:ObSmplFlux}

For the mean flux at redshift 1.95 we begin with our measurements in T04b.
We use the estimated amount of absorption from the low density IGM alone
because strong \lya\ lines from Lyman Limit Systems (LLS) and metal lines
have a significant effect on the 
absorption (\citet{viel03b, tytler04b, mcdonald04c})
but they are not included in our simulations.
We measure all absorption in the \lyaf\ then we subtract estimates of the
absorption from the \lya\ lines of LLS and also from metal lines.
We subtract in two ways, giving lower and upper bounds on the absorption
from the low density IGM. In T04b we used DA $= 1 - {\bar F}$.

Ideally we would measure the optical depth at each wavelength, find mean
the optical depths and subtract.
In practice we do not know the optical depth in regions with a lot of 
absorption. Such regions are a larger proportion of the absorption from
the \lya\ of LLS and the metal lines than the low density IGM.
If we convert DA values to effective optical depth before we subtract we
overestimate the absorption from the low density IGM. This gives
$\tau(DA8s)=\tau(DA7s)-\tau(DA6s)-\tau(DMs) = 0.130$, or 
$DA9s(z=1.90)=0.122 \pm 0.010$.
Here DA7s is the amount total absorption in the \lyaf\ at $z=1.90$, 
DA6s is the amount of absorption from \lya\ lines of LLS, 
DMs is the amount of absorption from metal lines, and
DA9s is defined by this equation.

We underestimate the DA from the low density IGM when we subtract the
DA values, without converting to optical depth.
This is because absorption is a division process, removing a proportion
of the photon flux, and not a subtraction.
Subtracting DA values, Eqn. (22) of T04b gives $DA8s(z=1.90)=0.118 \pm 0.010$.
Since a lot of the absorption from metal 
lines and \lya\ of LLS is saturated, we use the mean of
DA8s and DA9s, and we
scale this to $z=1.95$ using Eqn. (5) of T04b, $DA(z) \propto (1+z)^{2.57}$,
giving 
$DA(z=1.95) = 0.125 \pm 0.010$, or
\begin{equation}
{\bar F } = 0.875 \pm 0.010, ~~or ~~ \taueff = 0.1335 \pm 0.0115.
\end{equation}

\subsection{b-values \NOTE{subsubsection\_d}}
\label{subsec:ObSmplB}

As in T04b we use $b$-values from Figure 10 of \citet{kim01},
at redshifts $z = 1.61$, 1.98 and 2.13. We use their sample A that includes
lines with $12.5 < $ \lnhi\ $< 14.5$~\cmm\ and errors of $<25$\% in both \nhi\
and $b$. The sample has 286 lines, from $1.5 < z < 2.4$, and shows no evolution.
The mean redshift is 2.00. In Fig. 18 of T04b we showed that the distribution 
of these lines is very similar to that of lines in simulation B and
to the function in Eqn. (5) when
\begin{equation}
\bsig = 23.6 \pm 1.5 \rm ~\rm km/s.
\end{equation}

\subsection{Published Flux Power \NOTE{subsection\_d}}
\label{subsec:ObSmplP}

At least four papers present measurement of power on relevant scales:      
\citet[M00a]{mcdonald00a},
C02b, K04e and M04a.
The first three measure power on the scales of most interest to us,
$0.01 < k < 0.1$ s/km, but 
we have limited overlap with the SDSS spectra in the fourth reference
because our typical boxes are small compared to the smallest scale set
by the intermediate spectral resolution.

Unfortunately there is no agreement on a standard definition for
flux power in the literature. C02b discuss options and 
K04e label three quantities that we can take the power of:
F1, F2 and F3. F1 gives the power of the amount of 
the flux in units of the unabsorbed continuum level, or normalized flux.
This natural definition is very sensitive to the
mean amount of absorption.
F2 gives reduced (not zero) sensitivity to the mean absorption
because it is the power of the normalized flux, divided by the mean
of the normalized flux over some arbitrary range, measured from one
spectrum at a time, the average of many spectra together, or a model (see
C02b \S 2.2).
F3 avoids continuum fits, and is the power of the observed flux, in cgs
units, divided by its mean.
The differences are significant. From \citet[Table 3]{kim04e} we see that
P(F1)/P(F2) has a mean of 0.51, and it drifts systematically up and down from
around 0.45 to 0.53 over the listed $k$ range. 
P(F3)/P(F2) has a mean of 1.03 at $0.0056 < k < 0.075$ s/km,
and it rises to $>1.2$ at both smaller and larger $k$ values.

There are other significant problems with flux power. For real
spectra, flux calibration errors can be larger than the photon
noise, the continuum level errors are hard to estimate, and metal lines and
\Lya\ lines of LLS contribute significant power. Samples are often small 
and biased to have excess or too little metal lines absorption. 
At $k > 0.1$ s/km metals
are about 50\% of the signal, and there are no large published samples that 
are metal free. 
Moreover, there appear to be real differences between measured power values
even when the same definitions were apparently used.
With the simulated spectra the power is inaccurate because the
boxes are often many times too small.

For these reasons we can not yet make definitive contact between the 
simulations and data. The uncertainties in the data may be
larger than those in the simulations, including the external errors.

\subsubsection{Flux Power From Kim et al 2004 \NOTE{subsubsection\_d}}
\label{subsubsec:ObSmplPKim}

We have scaled the power measurements from K04e to match our $z=1.95$ and to
approximately remove the power from metal lines.
Our definition of flux power matches F2 in Eqn. 3 of K04e.  They list
revised F2 at various $k$ values for samples with mean redshifts of
$z=1.87$, 2.18 and 2.58 in Table 5 of \citet{kim04e}.
The F2 flux power is slightly lower at lower redshifts, with
F2(1.87)/F2(2.18) typically 0.9 with a range of 0.7 -- 1.1 for $-2 < k
< -0.8$ s/km. For each $k$ value that they list, we linearly interpolate
in $log (1+z)$ between log PF2(1.87) and log PF2(2.18) to find log PF2(1.95). 
We do the same with relative errors (not log relative errors) 
that range from 15 -- 50\% at $k=0.001$ to $0.03$ s/km,
and from 6 -- 11\% at larger $k$ values.
F2 values changes rapidly with $k$ near \ps : dropping a
factor of 2 from $k=-1.0$ to $-0.9$. 
We list these quantities in Table \ref{table:kim.tbl}.

Next we correct for metal lines. In Table \ref{table:kim.tbl} we list 
$f_m(k,z=2.36)=P_{metals}^{1D}(k)/P_{F1}^{1D}(k)$, the metal lines power
as a fraction of the F1 power at a mean $z=2.36$ from \citet[Fig. 3]{kim04a}.
We assume that a similar ratio would apply to F2, and we use a constant
value of $f_m = 0.0667$ for all $k < 0.01$ s/km because we suspect that
the error is larger than the fluctuations in their figure.

We expect that $f_m$ increases as $z$ drops, because the
total absorption from the low density IGM evolves faster than that
from metal lines (T04b). \citet[Fig. 16]{mcdonald04a} see a rapid change in
power at $z<2.3$ compared to higher $z$ that may have the same origin.
To correct for metal line power, we assume that the metal flux power 
does not change with redshift. 
First we find the total power at $z=2.36$ by interpolation in $log(1+z)$ 
between $log PF2(2.18)$ and $log PF2(2.58)$. 
Then we calculate the metal power $PF2M=f_m PF2(2.36)$. Finally we get the
power in the low density IGM $PF2I(1.95) = PF2(1.95)-PF2M$.
We list the PF2I(1.95) values in Table \ref{table:kim.tbl} 
and we list the power at the scales of \pl , \pmed\ and \ps\ in the
top row of Table 1.
The $1\sigma $ errors that we list ignore the error
in the removal of the metal lines, and hence they are much too small for
\ps .

We do not know whether the metal power in the 13 QSOs where K04e
fit metal lines is representative of their whole sample of 27 QSOs, 
and especially the QSOs that contribute to the $F2(1.85)$.
There are three reasons why the $f_m$ values
might underestimate the metal power on scales $k < 0.04$ s/km. 

First, the measurements in T04b indicate that $f_m$ is of order 
0.26 on scales around 0.0006 s/km in a larger sample at $z = 1.9$,
larger than the $f_m$ values of 0.06 -- 0.1 on scales $k<0.04$ s/km.

Second, at $z=1.9$ metal lines are 19\% of all absorption in the \lyaf .
It is likely that metals account for a larger fraction of the power
on all scales than they do of the absorption, because the clustering of metal 
lines on scales out to 600 km/s is extremely strong, and much easier to detect 
that the clustering of the \lya\ in the \lyaf\ \citep{sargent80}. 

Third, K04e state that their sample contained only 4 
sub-damped \lya\ absorbers (DLAs) that they removed. This is less than we 
might expect in a typical 
sample of 27 QSOs, and hence we expect that $f_m$ could be significantly 
larger than 
the values that we list.  A single LLS or DLA can contribute a lot of 
absorption and hence power.

\subsubsection{Flux Power from Other Papers \NOTE{subsubsection\_d}}
\label{subsubsec:ObSmplPO}

\citet[section 1]{mcdonald04b} noted possible systematic normalization 
errors and/or underestimation of the errors when they compared measurement
from different papers. We find that the power in K04e and C02b agrees at 
$z=2.13$, but that the power in both \citet{mcdonald00a} and 
\citet{mcdonald04b} is systematically lower, for no know reason.

\citet[Table 4]{mcdonald00a} list flux power at $z=2.41$ from complete
HIRES spectra of 1 QSO and partial spectra of 4 others.  In Fig. 5c
they show that metal lines have $f_m = 0.57$ at $k = 0.126$ s/km, more
than the $f_m =0.39$ from K04e at $z=2.36$ for $k=0.1334$, but
consistent considering the huge errors expected using so few QSOs.
\citet[Fig. 9]{kim04a} suggests that the M00 power is 
consistent with that of K04e at $z=2.35$.
Croft kindly points out that
M00 give the power of $F - \fbar $, where as $F2 = (F-1)/\fbar $.
Since power is proportional to the square of the signal, we should divide the 
power in M00 by \fbar $^2$ for comparison 
with F2. Using \fbar $=0.818$ for $z=2.41$ from M00 Table 1,
this increases the M00 power by a factor of 1.494.
We scaled the K04e F2 power, including metal lines but not the four sub-DLAs,
to $z=2.41$ for the comparison. We list values in Table \ref{tbl:m00k04}.
The two look extremely similar on a log $k$ -- log Power plot, but the 
systematic differences are much larger than the relative errors that we 
quote in Table \ref{table:kim.tbl}.
The M00 power is lower over most $k$ values, and the ratio PM00/PK04 is
1.04 at \pl , 0.84 at \pmed\ and 0.76 at \ps .
This result cast some doubt on our rescaling of the M00 power, because the
K04e power agrees with C02b.

In their Table 7 C02b give flux power of F2 from their subsample A of  9 
HIRES spectra that cover $1.7 < z < 2.3$ with a mean $z=2.13$. 
\citet[Fig 7]{kim04e} shows that the K04e power at $z=2.04$ 
is approximately 1.5 times that of C02b at $z=2.13$. However, 
when we compare to the K04e power at $z=2.18$, or a rescaling to 
$z=2.13$, we find excellent 
agreement for $-2.5 < k < -1.2$ s/km, with the exception of the C02b power
at $k=0.00437$ s/km that is obviously too low from Table 7 of C02b.
At $k > -1.2$ s/km we see more power in C02b because the C02b sample of
HIRES spectra had lower S/N, and perhaps also more metal lines than
the K04e sample.

M04a gives power at $k < 0.02$ for $z=2.2$ from SDSS
spectra.  The $P_F$ tabulated in their Table 3 is power of $F2$.
Their Fig. 23 shows that the power decreases smoothly with $z$, but
with some large deviations for the $2.2$ measurements at $k=$ 0.01413
and 0.01778 s/km, the two largest $k$ values.  For $k = 0.00891$ to
0.01122 s/km, we fit a line to their $\log P_F$ (values taken from
their Table 3) as a function of $\log k$ and extrapolated the result
to $z = 1.95$.  We found that a straight line 
produced an excellent fit to $\log P_F$ vs. $\log k$
between $2.2 < z < 3.8$, so we expect that
the small extrapolation from $z = 2.2$ to $z = 1.95$ is probably
valid.  We find $P_{F2}(k=0.00891, z=1.95) = 7.28$ \kms, and
$P_{F2}(k=0.01122, z=1.95) = 6.11$ \kms.  Doing a linear interpolation in
$k$ between these two points gives $P_{F2}(k=0.01, z=1.95) = 6.7$ \kms .
M04a quote relative errors of approximately 0.044 on the power around this
$k$ at $z=2.2$. The error on the power that we extrapolate to $z=1.95$
will be larger, by perhaps 30\% giving $\pm 0.38$.
This value, $6.7 \pm 0.38$ s/km includes metal lines, yet it is 
a factor of 0.76 smaller, $2\sigma $, than the value that we 
interpolated from K04e, $8.8 \pm 1.0$ s/km.
We give both values near the top of Table \ref{table:SimTable.tbl}.

\subsection{New Measurement of Flux Power \NOTE{subsubsection\_d}}
\label{subsubsec:ObSmplNewP}

We were tempted to make our own measurements of the flux power, to help
resolve definition and normalization uncertainty, and also because we can
remove from the small sample both the metal lines and the \Lya\ of LLS.

The data set we use consists of high resolution quasar spectra taken
with the Keck HIRES (FWHM $=8$ \kms ) and VLT UVES (FWHM$=4.3$ \kms ) 
spectrograph. The sample we use consists of
6 spectra, with QSO emission redshifts ranging from 1.71 to 2.65.  
The HIRES data were obtained as part of the effort to measure the mean
cosmic baryon density $\Omega_bh^2$ by comparing the ratio of
deuterium to hydrogen with the predictions of primordial
nucleosynthesis \citep[e.g.][]{kirkman03a}.  The details of
observational techniques, data reduction and continuum fitting are
described in more detail in the above paper. 

We fitted 
the metal and \lya\ lines in all these spectra, and we made artificial noise 
free spectra containing only the \lya\ lines with \lnhi $< 17.2$ \cmm .
We used these noise free fits because some of the spectra are low S/N, and we 
know from
\citet{viel03b} that line profiles contain most of the power information.
We do not claim that this will give an accurate power measurement, but we
were able to use exactly the same algorithms on these spectra and those
from simulations.

The region we use from each spectrum spans the wavelength range from 1000
\kms\ redward of Ly$\beta$ to 3000 \kms\ blueward of
Ly$\alpha$, to avoid the effect of any ionizing radiation arising due
to the close proximity to the quasar.  The redshifts of the Ly$\alpha$
forest data ranges from 1.52 to 2.61.  In general, signal to noise
ratio (S/N) increases with wavelength due to CCDs sensitivity and
atmospheric extinction.  The QSO redshift, signal to noise ratio (S/N)
and length of each spectra is shown in Table \ref{tbl:obs_table}.

As in Croft et al. (2002), we scale the optical pixel depths by a
factor of ${(1 + z)^{4.5}}$ to the mean z of the sample in order to
mitigate the effects of evolution.
To reduce evolution we used spectra covering only redshifts 1.7 -- 2.3
and to reduce differences in normalizations and windowing we divided
each spectrum into segments of length 0.1 in $z$, the same length as the
simulated spectra. We had a total of 34 segments, with slightly more at 
z greater than 2.0.  

We calculated the flux power
spectrum of each QSO segment and averaged them.  To help others to
compare with our results, we give the normalization now.  Given a
spectrum of length L and number of points N, we define the power
spectrum as the ${DFT \times L/N^2}$ where L is in units of km/s and
$DFT$ is the dimensionless output from the Discrete Fourier
Transform. This normalization matches that of \citet{croft02b}, and
the F2 of K04e. When we divided by the mean flux, we took the mean for
each QSO separately, as did K04e. 

We call this new power spectrum PJ05. In Table \ref{table:pfits} 
we give the coefficients of the polynomial representation
(\S \ref{subsec:MeasP}),
and we show this polynomial and compare it to simulations in \S \ref{sec:Conc}.
We list values from the polynomial fit near the top of 
Table \ref{table:SimTable.tbl}.

We again see systematic differences between power spectra. PJ05 is larger
than PF2I at all $log k < -1.1$ s/km, and smaller otherwise.
The ratio PJ05/PF2I is 1.27 at \pl , 1.33 at \pmed\ and 
0.90 at \ps ,
although the difference is only significant around \pmed\ where
PJ05-PF2I = $5\sigma $(PF2I). 
We understand the sense of the differences.
The spectra that we used for PJ05 often have lower S/N and larger continuum
level errors than those used in PF2I. This would tend to make PJ05 larger
on large scales. In addition, for PJ05 we used fits to spectra rather than
spectra, and this leads to a lack of power on small scales, from the lack
of photon noise and weak \lyaf\ lines.
For these reasons we do not consider PJ05 to be as accurate as PF2I, and
we do not know how to give errors on PJ05. However, PJ05 does suggest that 
the methods that we use on the simulated spectra should give results
similar to both PJ05 and PF2I.
\section{Convergence Tests \NOTE{section\_c1}}
\label{sec:ConvTst}

In this section we examine the effects of the box size and then the
cell size (resolution) on each of the parameters that we measure from
the simulated spectra: the mean flux, b-values and the flux power
spectrum.  We would like the parameters measured from the simulations
to have errors that are smaller than those of the data.
We find that surprisingly large boxes, $ > 100$ Mpc, and small cell size
($<30$ kpc) are required. While we have no single simulation with
these ideal parameters, we do explore a range that shows us how to scale
the results from practical smaller simulations to those we expect from such 
ideal simulations.

\subsection{Box Size \NOTE{subsection\_c1}}
\label{subsec:Box}

We need a large box size to contain large scale power.  Simulations
in the A series used identical inputs except for the box size. The
pair W1 and K2 also differ only in box size, but they have 
\gammahe $=3.3$ that makes them hotter
than the A series with less small scale power and wider lines.

In Fig. \ref{fig:boxFbarRaw.fig}
we plot the \taueff\ for these simulations.
As the box size doubles in the A series; from 9.6 to 19.2, 38.4 and to 76.8 Mpc;
the mean flux decreases by 
0.0040, then $-0.0023$, then 0.0066.
A2 shows a different behavior, but the size of the effect is
less than two times the random error for a single set of 25 spectra:
$\sigma =  0.0028$ (\S \ref{subsec:MeasErr}).
We can ignore A5 since it uses the same box as A3.
The effect of doubling the box size is smaller than the measurement error,
but not by much, and the convergence is slow.
The change in \fbar\ for K2W1 shows the same trend  as A3A.

From Table 1 and Fig. \ref{fig:boxbsigRaw.fig}
we see that the \bsig\ values increase with box size. The
increase is seen in both the A series and with W1K2, and it is a large 
increase of 1.7 km/s (6.7\%) from 19.2 to 76.8 Mpc.
As we double the box size starting from 9.2 Mpc in the A series, the 
\bsig\ increases by factors of 1.086, 1.030 and 1.036; again a slow convergence.

The 1D flux power spectra change shape as the box size increases.
In Fig. \ref{fig:powerbox2.fig} we show the 1D flux power from
simulations A3, A4, W1 and K2 in units of the power from simulation A.
The A series differ in box size alone. We see that the larger
boxes lead to reduced power at $ k > 0.05$ s/km for the A series with
75 kpc cells and \gammahe $=1.8$. We see a similar trend for W1K2,
shifted lower on the plot and tilted because we have divided by the flux 
power of A.
We note that simulation A2 seems to have flux power
that is out of order with their respective series. We suspect that this
may be some error, but we do not delete the entries because we could
not find any errors.

The changes in \ps\ as the box size doubles are factors of
0.772, 0.852 and 1.075. For the first two increases in box size, these changes 
are larger than the changes in \bsig . However going from the 38.4 Mpc to the
76.8 Mpc the \ps\ increases rather than decreases. 
The general inverse relation between \bsig\ and \ps\ follows from
\citet[Fig. 2]{viel03b}.

Simultaneously with the decrease in \ps\ we see \pl\ increase with
increasing box size. \pl\ increases with the box size because the larger 
boxes contain more long modes.
The W1K2 shows similar trends for \bsig\ and the flux power.

The precise effect of the
loss of large modes is complicated when we compare simulations.
Nearly all boxes have the same
seeds. The overall density pattern is similar in all such boxes, but the
$k$ ranges corresponding to the pattern change with the box size.  

The \ps\ drops as the box size increases because increased matter power
leads to decreased flux power on small scales. As we enlarge the box, keeping
the cell size and all cosmological and astrophysical parameters constant,
more long modes enter. The power increases on all scales, and especially those 
that were near the size of the smaller boxes.
As the simulation evolves the matter power increases on all scales leading to
larger matter power on all scales in larger boxes by $z=1.95$.
\citet[Fig. 10a]{mcdonald03}
and \citet[Fig. 13]{mcdonald04b} both show that an increase in matter power
on all scales leads to a decrease in the 3D flux power on scales
$k >$ 0.01 -- 0.02 s/km at $z = 2.2$.
According to these papers the reduction of small scale 3D flux power 
power is from
the increase in nonlinear peculiar velocities, a fingers-of-god effect.
The 1D flux power then also decreases on small scales as the matter power
increases.

\subsection{Cell Size or Resolution \NOTE{subsection\_c1}}
\label{subsec:Res}

The cell size has to be small enough to resolve the structures in the
IGM that change the output parameters.  The K series simulations differ only
in the cell size, as do the series A4, B and B2.
The K series reach our smallest cell size, 18.75~kpc, but they are all in
9.6~Mpc boxes. The A4BB2 series are in 19.2~Mpc boxes, with
\gammahe $=3.3$, but they stop at 37.5~kpc cells.

In Fig. \ref{fig:cellFbarRaw.fig} we show \taueff\ against cell size.
The mean flux increases by 0.006 and 0.005 as the cell decreases from
150, to 75 to 37.5~kpc.  These changes are larger than we would like,
and imply that cells of less than 37.5~kpc might be needed to measure
the mean flux to within 0.01. 

In Fig. \ref{fig:cellbsigRaw.fig} we show the decrease in \bsig\ with
decreasing cell size. The effect is large for cells $> 75$ kpc, but
small for smaller cells.

Fig. \ref{fig:powercell2.fig}
shows the flux power for the K series simulations with different cell size. 
The K series show  simple patterns: as the cell size decreases, from K3 to
K2 to K1, the power on small scales $k > 0.018$ s/km increases while that on
larger scales decreases. We did not expect the large scale
power to decrease, because these simulations all began with the same large
scale 3D matter power.
Perhaps this is an effect of long modes missing from all the boxes; modes 
that are better sampled with the smaller cells.

Fig. \ref{fig:powercell1.fig} shows the effect of cell size on flux
power for A4BB2.
The results looks rather different from the K series, and we suspect some 
error, perhaps in the power for simulation B.

For comparison, \citet[Section 2.2]{mcdonald04b} use 14.29~Mpc boxes 
with 55.8~kpc cells,
and they test for resolution convergence with a 7.14~Mpc box
and 55.80 and 27.90~kpc cells.
At $z=2.125$ the flux power is lower in the lower resolution simulation
by about 0.2\% from $0.01 < k < 0.05$ s/km, after they adjusted the redshift
of reionization, and 1\% larger before the correction.
The differences are several times larger at $k=0.1$ s/km.

While \citep{bryan99} found that 37.5 kpc is sufficient to resolve
the \lyaf\ at $z=3$ and we prefer smaller cells at $z=2$ because the
density contrast is larger. 

When we compare simulations with differing box and cell size we will
apply the systematic corrections that we derive in \S \ref{sec:Scal}.
\section{Correlation of Measured Output Parameters \NOTE{section\_c2}}
\label{sec:Out}

In this section we examine the correlations between the five output
parameters that we measure from the simulated spectra: the mean flux,
b-values and the flux power on three scales.  

In Fig. \ref{fig:bsigTau1.fig}
models with the large 150 kpc cells, O1, O3, P1, P2, Q, Q1 and K3 are on the 
upper right. They are joined by M that has broad lines because it has low \sig\ 
and high \gammahe\ values. We will see these same simulations stand apart from
others in many subsequent plots. 
The remaining simulations show larger \bsig\ values for smaller \taueff .
This is not the trend expected from line saturation that has the opposite
sense.

It is well known that for a given matter power the flux power is a 
strong function
of the mean flux (\citep{croft02b, gnedin02, viel04a}; also \S 1 of T04b).
The flux power increases as the mean flux decreases, by Parseval's theorem,
because a lower mean flux corresponds to more flux variance.
In Fig. \ref{fig:Pm2Fbar.fig}
we see that the effect remains strong for \pl , even though the flux 
power that we use, coming 
from F2, is less sensitive to the mean flux than other definitions.
The models on the upper left all have the large 150 kpc cells.
We see a similar trend for \pmed\ in Fig. \ref{fig:Pm15Fbar.fig}.
However, in Fig. \ref{fig:Pm1Fbar.fig} we see much more scatter for \ps ,
because other factors have a large effect on the power on small scales.

The trends in Figs. \ref{fig:Pm2Fbar.fig} and \ref{fig:Pm1Fbar.fig} suggest that
we should increase \pl\ by 6\% and \pmed\ by 5\%
when we the decrease the \fbar\ by 0.0051 to correct the absorption missing
at the start of each spectrum.

\citet[Fig. 1]{viel03b} show that the flux power
for all three $k$ values; \pl , \pmed\ and \ps ; is extremely similar to 
the power of a random arrangement of a sample of real absorption lines
fitted with Voigt profiles. When they artificially
increased all $b$-values, the power spectrum shifts to larger $k$ in their
Fig. 2.
In the band of \ps\ they find that doubling all $b$-values, making all
lines broader, gives about 8 times less power. Halving all $b$-values gives
5 times less power. We expect to readily see a strong correlation
between \ps\ and \bsig\ and we expect to be sensitive to much
smaller changes.
The band \pl\ is less sensitive to changes in the $b$-values, and in the
opposite sense to \ps , while the \pmed\ is insensitive to $b$-values.

In Fig. \ref{fig:Pm2bsig.fig},
\ref{fig:Pm15bsig.fig} 
and \ref{fig:Pm1bsig.fig}
we show how the flux power depends on \bsig . In all three Figs. simulations
O1, O3, P1, P2, Q and Q2; all with the 150 kpc cells; lie in the upper right
and can be ignored. We see that power decreases as \bsig\ increases for all
three power bands. The effect is small for \pl , intermediate for \pmed , and 
strong for \ps , as expected. 
Since our simulations differ in various input parameters, and we did not 
change $b$-values explicitly, we should not conclude that the
trends that we see for \pmed\ and \pl\ differ from the results 
of \citet[Fig. 2]{viel03b}.

\subsection{Extra Information in the Flux Power\NOTE{subsection\_c2}}
\label{subsec:OutP}

It is interesting to ask whether the flux power contains any information
that is not in the mean flux and b-values. We ask this because the
flux power is strongly correlated with both the mean flux and b-values,
and the dispersion in flux power seems to contain approximately
two degrees of freedom. We can think of one of these as controlling 
the amplitude and the second the shape, or the ratio \pl /\ps .
The answer will depend on the level of accuracy required. 

A result in \citet{mcdonald04b} also hints that the flux power is 
contained in the mean 
flux and \bsig . They state that they can derive the mean flux to better
than 1\% error when they match their observed flux power to simulations.
Since they use intermediate resolution spectra that do not resolve lines,
their flux power contains only the longer scale information on b-values, 
and their results will be sensitive to their spectral resolution.

While it is clearly the case that the flux power on the largest scales
is strongly correlated with the matter power that we parameterize with 
\sig\ and $n$,
on small scales the flux power is most strongly effected by the factors that
determine the mean flux and $b$-values. These factors
include the temperature which we parameterize with \gammahe\
and the combination of parameters that set the ionization, especially
\gammah . The power that we measure at $k > 0.01$ s/km is not a good 
measure of the matter power because of 
the importance of these other factors.

We have not tried to determine whether, to the level of accuracy of our 
simulations, the mean flux and \bsig\
contain essentially all the information that is in the flux power on small 
scales. This comparison would be difficult because we are most interested
in the largest scales, but our boxes are missing power on these
scales, and the power on these scales is effected by our choice to start
the simulated spectra at the cell with the lowest density in each box.

\section{Scaling Laws \NOTE{section\_e}}
\label{sec:Scal}

We now determine the scaling relationship between the two groups of
parameters, the input to the simulations and the outputs that we
measured from the simulated spectra.  We will give scaling
relationships between individual inputs and outputs.  We vary only one
input at a time, noting the values of all the others that we keep
fixed. We find that the cross terms are small, allowing us to vary
five different inputs simultaneously and obtain accurate predictions
for the outputs.

There has been much work on scaling in the literature, including
recent work by \citet{mcdonald03, mcdonald04b}; and \citet{bolton04b}. We will not
attempt to compare our simulation results with theirs because we
typically differ in numerous input parameters, including the
corrections for box and cell size.

In Table \ref{tbl:scalings} we list functional forms and the
parameters that describe the scalings. We list five input variables in
the first column, and two output variables, \taueff\ and \bsig .  The
table entries give the value of the output parameter that we expect as
a function of each of the five input parameters.  If a parameter is
not explicitly used, then these scalings give the values expected
using the default value for our standard model, listed in Table
\ref{tbl:standard}. This standard also includes the corrections for
starting the lines of sight at random positions in the box, rather
than the lowest density point.
As elsewhere in this paper the box size $L$ is in comoving Mpc and cell 
size $C$ in comoving kpc, both for $h=0.71$.

To scale an output to different input parameters, we multiply the output
by the relevant scaling factor. We give relations for two of the five
outputs, \taueff\ and \bsig .
For example, simulation A gave \taueff $= 0.13778$ for cell size $C=75$ kpc.
Let us rescale this \taueff\ to the value that we would expect from a
simulation with $C=18.75$ kpc.
We multiply 0.13778 by \taueff $(C=18.75)/$\taueff $(C=75)$ = 0.9352, where
\taueff $(C) =  0.132828 + 0.00016748 C$, giving \taueff = 0.12885. 
We can calculate similar correction factors for both
\taueff\ and \bsig , to scale using all five 
input parameters for which we give scaling relations.

The scalings work very well within the range of parameters given in
Table \ref{table:SimTable.tbl}, and we can use them to scale an
output by one input parameter, two parameters or even all five inputs.
To rescale by more than one input parameter we simply apply the product of the
relevant correction factors.

The scaling relations presented in Table \ref{tbl:scalings} are shown
in Figures \ref{fig:boxFbar.fig} -- \ref{fig:gamma228bsig.fig}.  
Each Figure shows either \taueff\ or \bsig\ as a
function of one of the input parameters: box size ($L$), cell size ($C$), 
\sig, \gammah, or \gammahe.  In each figure, \taueff\ and
\bsig\ have been rescaled to our standard model values for all of the
input parameters {\it except} the one being displayed as the
independent variable in that figure.  There are three points to take
away from our scaling figures: (1) Our scalings work remarkably well
over a wide range of input parameters.  (2) The scalings are self
consistent.  (3) Our procedure of applying multiplicative correction
factors in succession to rescale for more than one change in input
parameters works well, implying that the cross terms are small.

There is nothing fundamental about using the scaling relations to
produce multiplicative correction factors.  While the multiplicative
correction factors are defined such that they must give correct
results when only rescaling by a single input variable, we use the
multiplicative factors because they also work well when rescaling by
more than one input.  This need not have been the case, and the fact
that the multiplicative rescalings work so well indicates that the
full equations giving \taueff\ and \bsig\ are separable in $L, C,
\sig, \gammah, $ and \gammahe\ with very small cross terms between the
input parameters.

In Fig. \ref{fig:bsigTau2.fig} we show the values we predict
for \bsig\ and \taueff\ when we scale the output parameters from each
simulation to the standard model. A measure of the accuracy of the
scalings is the dispersions in these scaled values: 
the mean \taueff $=0.13539$ with standard deviation $\sigma ($\taueff
$) = 0.0027$, while the mean \bsig $=24.139$ \kms\ with
$\sigma ($\bsig $) = 0.33$ \kms. The typical error from the use of the
scaling relations is then 1.4\% for \bsig , 
and 2.0\% for \taueff , equivalent to an error on the mean flux
\fbar\ of 0.27\%. We can predict the \bsig\ and \fbar\ that we will obtain 
from simulations with errors
four times smaller than the errors on the measurements from the data
that we presented in \S \ref{subsec:ObSmplFlux} and \ref{subsec:ObSmplB}.
This suggests that the measurements from the simulations could be
more accurate than the data, provided the calibrations are reliable.

The scalings should not be used for extrapolations, because the
functional forms may be unreasonable outside the range of our models.
We also note that the values for the parameters in a given equation
are strongly correlated. A wide range is allowed for a scaling
parameter in Table \ref{tbl:scalings} when the range of the input
parameter in our simulations is small.  For example, for \gammah\ we
used 0.5 -- 1.2 in different simulations, and the range allowed for the
exponent in
the equation \taueff $= 0.0517116 + 0.0845752 \gammah ^{-1}$ is much
larger than you might expect. We choose $-1$ because this provides a
good fit, but $-0.8$ and $-1.2$ also fit well. We can neither confirm
nor refute the exponent of $-1.44$ that \citet[Eqn. 9]{bolton04b}
found in a different way, by post processing their simulations to
change the \taueff\ value, although we have more to say on this in 
\S \ref{subsec:ScalBandC}.

\subsection{Effect of Specific Scalings}
\label{subsec:ScalBandC}

We now discuss some of the results and compare to prior literature.
We do not repeat our discussion of box and cell size in \S \ref{sec:ConvTst}.

Both $h$ and \obh\ are well known from other measurements
\citep{freedman01,kirkman03a,spergel03a} and so the few simulations that we
ran to explore their effects cover only a small range.

We discussed how the \fbar\ and \taueff\ depend on \ob\ and $h$ in T04b
\S 12.1.2 and 12.1.4. In Table \ref{table:SimTable.tbl} we give output
parameters for simulations S2K2S3 that vary in \ob\ and $h$ simultaneously
such that \obh\ is approximately constant. Their output parameters vary by
approximately the errors on the relevant data.

Since with various assumptions
$\tau \propto h^{-1}(\Omega _b h^2)^2$ 
(\citealt[Eqn. 17]{rauch97}, \citealt{bolton04b}), 
we can use the quantity
\begin{equation}
Y = \taueff h/(\Omega _b h^2)^2 \propto  \taueff /\tau , 
\end{equation}
to make an approximate estimate of how $\tau$ depends on \taueff . 
We find that Y increases as \taueff\ drops, but by only 4\% from
simulation S2 to S3. This trend is fit with with 
$\tau \propto $\taueff $^{-1.32}$, consistent with \citet[Eqn. 9]{bolton04b}
who found an exponent of $-1.44$. Although the changes are very small,
and unlikely to be reliable, the relation does significantly decrease the 
dispersion in the Y parameter: $\sigma (Y)/Y = $2\% drops to 0.5\% if we use
\taueff\ $^{-1.32}$ in place of \taueff\ in the Y definition.
However, this result is illustrative and not definitive, since changes in 
\ob\ also lead to changes in the gas temperature 
\citep{gardner03a,tytler04b} and Jeans smoothing
\citep[Section 12.1.2]{tytler04b}, both leading to changes in \taueff\ that 
are not included in the equations that we used here.
The scaling of \taueff\ with \gammah\ in 
Table \ref{tbl:scalings} is derived from simulations that
include these astrophysical effects and should be more reliable.

We do have two sets of simulations with slightly differing \om\ and \ol\ 
values for flat models. Larger \om\ corresponds to larger \fbar , and
smaller \pl\ and \bsig . 
However, the two sets show different trends for \pmed\ and \ps . 
In Fig. \ref{fig:Pm1bsig.fig} we saw that \ps\ and \bsig\ are typically
anticorrelated, as for K3 and V, and contrary to the trend for T and K2.
Although the change in \bsig\ is larger than the measurement error,
we are not convinced that this difference is robust.

\subsection{Effect of ${\sigma_8}$}
\label{subsec:ScalS}

In Fig. \ref{fig:sigma8FbarRaw.fig} we show the raw \taueff\ for
three sets of simulations: MK2, L3L4L5 and P2Q. For each set the
\taueff\ drops with increasing \sig\ as we noted in T04b.

In Fig. \ref{fig:sigma8bsigRaw.fig} we show that larger \sig\ gives
smaller \bsig , as was noted by 
\citet[Fig. 7]{bryan00} and 
\citet[Fig. 5]{theuns00b}. \citet{theuns00b} saw the trend in
a portion of their simulated spectrum, but the VPFIT
software that they used to obtain the b-value distribution failed to
show the change because more lines were added instead of wider ones.

The set UQ1 is different because its \pl\ and \pmed\ values drop with 
increasing \sig , the opposite trend from the other sets, perhaps 
because the cell size is large.

\subsection{Effect of ${\gamma_{912}}$}
\label{subsec:ScalGh}

In Fig. \ref{fig:gamma912FbarRaw.fig} we show the \taueff\ for
three sets of simulations, each of which differ only in \gammah.
All three sets show less absorption and smaller \taueff\
for larger \gammah , as expected \citep{croft02b}.

In Fig. \ref{fig:gamma912bsigRaw.fig} we show the \bsig\ for
three sets of simulations, each of which differ only in \gammah.
We see a slight increase in \bsig\ with \gammah .

\subsection{Effect of \gammahe\ }
\label{subsec:ScalGhe}

In Fig. \ref{fig:gamma228FbarRaw.fig} we show the \taueff\ for
four sets of simulations, each of which differ only in \gammahe.
All sets show less absorption in hotter models, as we expect.

In Fig. \ref{fig:gamma228bsigRaw.fig} we show how \bsig\ depends on
\gammahe\ for the same four sets as in Fig. \ref{fig:gamma228FbarRaw.fig}.
We see that the hotter models have larger \bsig .

In Fig. \ref{fig:specIJ.fig} we show an example of the effect of changing 
\gammahe\ on simulated flux spectra. 
As we expect, increasing the \gammahe\ results 
in more ionization and thus in less absorption. Accordingly, the flux spectra 
shows lines closer to the unabsorbed value of 1.0. However, it is not
obvious from this segment of spectrum, of length $\Delta z = 0.01$, that
simulation J has larger \bsig\ than simulation I. 

\subsection{Comparison to Bolton \NOTE{section\_e2}} 
\label{subsec:Bolton}

In Fig. \ref{fig:g912s8constg228.fig}
we show the \gammah\ that gives the observed
\fbar\ as a function of \sig . As we expected, hotter models need less
\gammah . This Figure is similar to Fig. 16 of T04b and
Fig. 3 (right) of \citet{bolton04b}, with five exceptions.
First, T04b used \gammahe = 1.4, while \citet{bolton04b} used \gammahe = 3.3.
This difference explains why the curve in \citet{bolton04b} lies well below 
that in T04b for \sig = 0.84 and 0.7.
A second differences is with the scalings used to correct for box and cell 
size. In T04b we did not make any such corrections, and the simulations that 
we used had $L=19.2$ Mpc and 75 kpc cells.
\citet{bolton04b} did correct for box and cell size, as we do here for
Fig. \ref{fig:g912s8constg228.fig},
but we expect there to be some difference
in these corrections because they are hard to estimate.
A third difference is in the redshift. We now work at $z=1.95$,
where as T04b and \citet{bolton04b} were for $z=1.90$.
A fourth difference is that we derived the approximate scalings that we 
used in T04b from simulations with \sig\ = 0.7 -- 1.09. 
Outside this \sig\ range the scaling relations that we showed were 
extrapolated, and can readily have large errors.
We can not calibrate these errors because all our simulations still have
0.7 -- 1.1. Values for \sig\ outside this range are of little cosmological
interest.
Since the main purpose of Fig. \ref{fig:g912s8constg228.fig} is to aid 
comparison with past work, and we know that 
we should increase \gammahe\ as \sig\ varies, we again show scalings 
extrapolated to a wider range of \sig\ values.
We see that T04b was relatively accurate at \sig $>0.9$ where
the scalings were small, and that our new scaling relations
show that we overestimated the \gammah\ required at \sig $< 0.8$.
The fifth difference is with the other cosmological parameters.
Here and in T04b we used 
\obh $ = 0.02218$, \om $=0.27$, $h = 0.71$ and $n=1$, 
while \citet{bolton04b} used
\obh $ = 0.02400$, \om $=0.26$, $h = 0.72$ and $n=0.95$. 
The points from \citet{bolton04b} lie between our curves for
\gammahe $=3.3$ and 1.4 for \sig $< 1$. They are 1.13 times higher than
our \gammahe $=3.3$ curve for \sig $=0.5$, rising smoothly to
1.27 times higher by \sig $=1.2$.
This difference might be largely explained by the different $n$ values,
since for smaller $n$, a smaller \sig\ value has the same matter power
on relevant scales.

Fig. \ref{fig:g912s8constg228Error1.fig} is similar to Fig. 
\ref{fig:g912s8constg228.fig} but drawn for the three DA values given
in T04b. We see that the range of \gammah\ allowed by the error in the 
measurement of DA of 0.01 is larger than that shown in  Fig. 16 of T04b.
The present figure superseded that given in T04b.
\section{Joint Determination of Cosmological Parameters \NOTE{section\_f}}
\label{sec:JointDeterm}

In general we can not break the degeneracy between \sig\ and \gammahe\ using 
only mean flux, small scale flux power and b-values with the accuracies assumed in this paper. We can find models that fit all these observables 
simultaneously for a wide range of \sig\ from 0.7 -- 1.1. We did not explore a 
larger range, and we did not examine the small differences between these 
models.

We can break this degeneracy using some other information on \sig\ from the 
large scale flux power (T04b) or from some other measure of large scale 
clustering,
such as the galaxy distribution or the CMB. The large scale flux power is
helpful because it responds in a different way to changes in \gammahe\ than
does the small scale power  that we have discussed. The power spectrum changes 
shape with changes in \sig\ and \gammahe . \citet[Fig. 10a]{mcdonald03} and
\citet[Fig. 13]{mcdonald04b} both show how the flux power responds
to small changes in various parameters. Increasing \sig\ causes more 
3D flux power on large scales, less on small scales, and no change 
near $k = 0.03$ s/km at $z=2.2$. On large scales the flux power is proportional 
to the amplitude of the mass power spectrum, or 
$\sigma _8^2$, as \citet{croft98} assumed and
\citet[App. C]{mcdonald00a} proved.

We can readily distinguish models that have differing \sig\ because they will 
have different large scale power. 
The large scale power, or equivalently, the fluctuations in the 
mean flux on large scales, gives \sig\ directly, avoiding the complications 
found in small scales. 
Unfortunately we can not explore large scale power with our existing simulations
because most of our boxes are too small. However, in T04b we showed that the
variation in the mean flux in spectral segments with length
$\delta z = 0.1$ was similar to the variations seen in our largest simulation,
A, and consistent with \sig $=0.90^{+0.13}_{-0.16}$ for $n = 0.95$.

In the next section we give the sets of parameters that are consistent with 
data. We will treat \sig\ as a control variable, but we could have used
some other parameter.

\subsection{A Concordance Model for the Lyman-$\alpha$ forest at z = 1.9 
         \NOTE{section\_g}}
\label{sec:Conc}

For a given \sig\ and other cosmological parameters, we can
estimate \gammah\ and \gammahe . In Figs. \ref{fig:g912s8.fig},
\ref{fig:g228s8.fig} and \ref{fig:g912g228.fig} we show the pairs of values
that give the observed \fbar\ and \bsig\ at $z=1.95$ for our standard model.
We do this by starting at our standard simulation and using our scaling 
relations to calculate the change in the input parameters necessary to 
match the two observables. We then have two equations in 
two unknowns. We then solve for the pair of values that would have matched 
the observed data for some value for the third parameter; for example, we 
can find values for \gammah\ and \gammahe\ which result in a match for some 
value for \sig . We do not expect any significant error arising from the 
implicit assumption that the scaling laws between two of our input 
parameters do not depend on the third parameter as we have shown this to 
be true in \S \ref{sec:Scal}.
   
For \sig\ in the range 0.8 -- 1.0 the \gammah\ and \gammahe\ are both
similar to unity, implying that the photoionizing spectrum is similar to that
predicted by \citet{madau99a}. 
For \sig $=0.9$ and the other standard model parameters from Table
\ref{tbl:standard} we find \gammah $= 1.0$ and \gammahe $= 1.26$.
The \gammah $=1.0$ implies 
a mean ${\Gamma = 1.329 \times 10^{-12}}$ s$^{-1}$ and a minimum
$J_{912}(z=2) = 0.30 \times 10^{-21}$ \flux\ using \citet[Eqn. 3]{hui02}, or
$J_{912}(z=2) = 1.7 \times 10^{-21}$ \flux ,
if we assume a hard spectrum with a volume emissivity spectrum of 
slope $-0.5$.
This may be compared to recent measurements of
the ionizing rate.  A recent analysis of the proximity effect
gave ${\Gamma = 1.9_{-1.0}^{+1.2}} \times 10^{-12}$ s$^{-1}$ averaged
from z = 1.7 to 3.8 \citep{scott00b}.  \citet{steidel01} measured the specific
intensity to be $J_{912} =1.2 \pm 0.3 \times 10^{-21}$ \flux\
from Lyman Break Galaxies at $z \sim 3$ and
\citet{hui02} assume a functional form for the specific intensity to
convert this to $\Gamma = 1.89_{-0.99}^{+1.21} \times 10^{-12}$ s$^{-1}$. 
\citet{mcdonald01b} compare observed mean transmitted flux
with that of simulated spectra to infer the intensity of the ionizing
background. They find $\Gamma = 0.65 \pm 0.1  \times 10^{-12}$ s$^{-1}$ at z =
2.4.

In Fig. \ref{fig:powerdata1} we compare the power in simulations L4 and P5 to
the measurements from K04e and our own measurement that we call PJ05.
Both the measurements are for the low density IGM, with the signal from
the \lya\ lines of LLS and metal lines removed. 
The power in PJ05 is larger than that in K04e by approximately 
25\% at $log k < -1.3$ s/km then at $log k > -1.1$ s/km the PJ05 power 
becomes smaller.
These errors are smaller than the differences between other measurements
that we discussed in \S \ref{subsec:ObSmplP}.
We selected these two 
simulations because they gave unscaled \fbar\ and \bsig\ values, those in 
Table \ref{table:SimTable.tbl}, that are similar to measurements.
We see that the power in these simulations is indeed close to the data
for $log k > -1.4$ s/km, indicating that the flux power spectrum on these small 
scales is well described with two parameters, such as \fbar\ and \bsig .
There appears to be little additional information in the small scale flux
power spectrum.

On larger scales the power in the simulations is less than the measured power
of the data. This may simply be the lack of large scale
power put into the small 19.2 Mpc boxes. 
In Fig. \ref{fig:powerbox2.fig} we saw that simulations in 19.2 Mpc boxes
show less flux power than those in larger boxes on the same scales:
$k < -1.4$ s/km, but we have not determined whether the magnitude of the lack 
of power from the boxes is enough to match the data.
There will also be excess power in the data, especially at $log k < -2$ s/km,
because the error in the continuum fits has not been subtracted. 

The \lyaf\ is very sensitive to \sig , \gammah\ and \gammahe .
In Table \ref{tabcos} we show the expected error in our ability to
compute \taueff\ using our scaling relations. Here we assume that
the scaling relations are free of errors, which means that
we ignore errors in the corrections for box and cell size and we 
ignore the errors in the data that we used to derive the scalings.
We consider only the gradients of the scaling relations.
In the second column of Table \ref{tabcos} we give nominal prior values 
with errors, based on Table 7 of T04b.
The entries in Table \ref{tabcos} supersede the last three rows
in Table 7 of T04b.
The third column of Table \ref{tabcos} gives the derivatives of our
scaling relations evaluated at our standard model parameters, and the
fourth column gives that derivative times the error on the prior value. 
The values in the fourth column are similar in size to the error
in our measurement of DA in T04b, and given here in Eqn. (10).
If we knew all the cosmological parameters in Table \ref{tbl:standard}
without error, with the exception of one of \sig , \gammah\ or \gammahe ,
then we could measure that last parameter with an error similar to the error
listed on the prior. 

The main impediment to extracting cosmological information from the \lyaf\
is not the sensitivity, but the rather large uncertainties in the astrophysical
parameters \gammah\ and \gammahe .
Instead, the small scale information that we have discussed in this paper 
provide the best constraints on \gammah\ and \gammahe , and when we have 
optically thick simulations we will be able to convert \gammahe\ into
a constraint on the ionizing spectrum.

\section{Conclusion \NOTE{section\_h}}
\label{sec:Conclude}

We have explored the relationships between several cosmological and 
astrophysical parameters that effect the \lyaf\ in hydrodynamic simulations.
We have used 40 hydrodynamical cosmological simulations of the $z = 1.95$
\lyaf\ to measure the background radiation amplitude
and the matter power amplitude. Both the data and simulations now
support high accuracy. The HIRES and UVES spectra allow
better continuum fits and better removal of metal lines and Lyman limit
systems resulting in a mean flux measurement with an error of about
1\%. The simulations used include not only gravitational effects, but
also hydrodynamical effects as well as the effects of non-equilibrium
chemistry of the baryons and the effects of a background ionizing
radiation field. We did not rescale the output of the simulations to
different temperatures. Instead we ran complete simulations to explore 
different UV background intensity and heating rates.  We made corrections 
for the finite box size and resolution of our simulations.

We agree with other authors 
\citep{hui99d,mandelbaum03,mcdonald04a,seljak04a,viel04b}
that the careful comparison of
large, well calibrated sets of simulations and data allow the
extraction of both cosmological and astrophysical information from the 
\lyaf .

We found that the simulations required to give the required accuracy are
much larger than one might have expected:
box size $L > 100$ Mpc and cell size $C <50$ kpc, corresponding to
at least $2048^3$ cells, similar to, or more demanding than the recommendations
of others; eg. \citet{meiksin04a} who recommend $L > 35$ Mpc and $C < $ 42 -- 85 kpc at $z>3$. We find that 
boxes must be large enough to include modes two
orders of magnitude larger than the filaments in the \lyaf , much larger than
those that have been used in nearly all prior work, a point made by
\citet{barkana04a}.
These simulations must also simultaneously have small cell size because 
larger cells give inaccurate absorption statistics, as we saw for
our simulations with 150 kpc cells.
However, once we resolve the \lyaf , box size becomes the most
critical parameter.

We have found that different output parameters are correlated.
When we compare the \ps , b-values and mean flux we find that we have
only two independent degrees of freedom, to the level of accuracy of this 
study. We can predict the third parameter, in our case the power spectrum
at $log k > -1.5$ s/km, from the other two.

We find that the mean flux and \bsig\ are in many ways preferable to
the flux power. The flux power is hard to measure because it is sensitive
to continuum errors on intermediate and large scales,
photon noise, metal lines and the \Lya\ of LLS on all scales and especially
small scales for metals. The flux power is extremely sensitive to the mean flux
and to redshift. There is no standard definition, and there are disagreements
over values for apparently identical definitions. 

We find scaling relations between the input and output parameters for the 
simulations. For the input parameters we explore \sig , \gammah , \gammahe ,
box and cell sizes and for the output we explore \fbar\ and \bsig. We find that the 
scalings are well determined and allow us to predict the outputs for any of 
our simulations, with errors of 0.3 km/s (1.4\%) for \bsig , and 
0.0027 (2.0\%) for \taueff , equivalent to 0.0024 (0.27\%) for \fbar\ at
$z=1.95$.
We can also use the scalings to determine the input parameters that
would yield a given pair of output parameters.
The errors from the scalings are four times smaller than the errors on the 
measurements for the data that we presented in 
\S \ref{subsec:ObSmplFlux} and \ref{subsec:ObSmplB}.

We use the scaling relations to scale our outputs to a standard model with
\ob = 0.044,
\om = 0.27,
\om = 0.23,
$h = 0.71$,
\sig = 0.9,
$n=1.0$,
\gammah $=1.4$,
\gammahe $=1.0$,
box size 76.8 Mpc and
cell size 38.75 kpc.
We also correct for the our mistaken choice of starting the
simulated spectra at the lowest density point in the simulations.

We can break the degeneracy between \sig , \gammahe\ and \gammah\
using large scale power or other data to fix \sig .
When we pick a specific \sig\ value, the simulations tell us the   
the \gammahe\ and hence the temperature that we need to match
the observed small scale flux power and b-values.
We can then also find the H ionizing rate required to match the mean flux
for that combination of \sig\ and \gammahe .
Our scaling relations give the parameters for various \sig\ values.         

This work demonstrates that precision measurements from the \lyaf\ have
excellent promise. We can make simulations and observations 
that are each accurate to several percent and we can make significant 
improvements with both.
\acknowledgments

We thank Rupert Croft, Pat McDonald, Kev Abazajian,
Carlos Frenk and Daniel Eisenstein for helpful comments
and discussions. This work was supported in part by NSF grant AST-9803137
under the auspices of the Grand Challenge Cosmology Consortium,
NSF grant AST-0098731, NAG5-13113 from NASA, and HST-AR-10288.01-A from the
Space Telescope Science Institute. The spectra were obtained from the
the W. M. Keck Observatory that is a joint facility of the University of
California, the California Institute of Technology and NASA,
and from Paranal Observatory of the European Southern Observatory
for programs No. 166.A-0106 and 066.A-0212.
\clearpage
\bibliographystyle{apj}
%\bibliography{../../database/archive}
\bibliography{archive,mike,confproceed}
\clearpage
\begin{deluxetable}{cccccccccccccccc}
\rotate 
\tablecaption{Input Parameters of the Hydrodynamic Simulations 
\label{table:SimTable.tbl}} 
\tablewidth{0pt}
\tablehead{ 
\colhead{Name} & 
\colhead{$\Omega_b$} & \colhead{$\Omega_m$} & \colhead{$\Omega_{\Lambda}$} & \colhead{$h$} & 
\colhead{$\sigma_8$} & \colhead{${\gamma_{912}}$} & \colhead{${X_{228}}$} & \colhead{L} & 
\colhead{N} & \colhead{Cell} &
\colhead{$\bar{F}$} & \colhead{$P_{-2}$} & \colhead{$P_{-1.5}$} & \colhead{$P_{-1}$} & \colhead{${b_{\sigma}}$} 
} 
\startdata 
PJ05 &&&&&&&&&&& ... & 11.1 & 2.76 & 0.0755 & ...\\
M04a &&&&&&&&&&& ... & 6.7 & ... & ... & ...\\
Data &&&&&&&&&&& 0.875 & 8.8 & 2.07 & 0.0841 & 23.6\\
$\sigma (Data)$ &&&&&&&&&&& 0.10 & 1.0 & 0.14 & 0.0067 & 1.5\\
\\
A&0.044&0.27&0.73&0.71& 0.90& 1.0& 1.8&{\bf 76.8}& ${1024^3}$&75& 0.8713 & 6.22448 & 2.00096 & 0.0580288 & 26.662\\
A2&0.044&0.27&0.73&0.71& 0.90& 1.0& 1.8&{\bf 38.4}& ${512^3}$&75& 0.8779 & 5.26255 & 1.68276 & 0.0539609 & 25.741\\
A3&0.044&0.27&0.73&0.71& 0.90& 1.0& 1.8&{\bf 19.2}& ${256^3}$&75& 0.8756 & 5.29052 & 1.80656 & 0.0633351& 24.995\\
A5&0.044&0.27&0.73&0.71& 0.90& 1.0& 1.8&{\bf 19.2}& ${256^3}$&75& 0.8762 & 5.23220  & 1.76246 & 0.0625227& 24.560\\
A4&0.044&0.27&0.73&0.71& 0.90& 1.0& 1.8&{\bf  9.6}& ${128^3}$&75& 0.8796 & 4.51003 & 1.73439 & 0.0820087& 23.007\\
\\
W1&0.044& 0.27& 0.73& 0.71& 0.94& 1.0& 3.3&{\bf 38.4}& ${512^3}$&75  & 0.896& 4.59288 & 1.2719 & 0.025351 & 29.212\\ 
K2&0.044& 0.27& 0.73& 0.71& 0.94& 1.0& 3.3&{\bf 19.2}& ${256^3}$&75  & 0.897& 4.17679 & 1.2170 & 0.028068 & 27.778\\
%W3&0.044& 0.27& 0.73& 0.71& 0.94& 1.0& 3.3&{\bf 9.6}& ${128^3}$&75   & 0.889& 4.86138 & 1.4695 & 0.035802 & 27.760\\
\\
B2&0.044&0.27&0.73&0.71& 0.90& 1.0& 1.8&9.6& ${512^3}$ &{\bf 18.75}& 0.8925& 3.52847 & 1.47406 & 0.0770033& 22.7964\\
B&0.044&0.27&0.73&0.71& 0.90& 1.0& 1.8&9.6& ${256^3}$ &{\bf 37.5}& 0.8819 & 1.71394 & 4.62606 & 0.0789581 & 22.8300\\
A4&0.044&0.27&0.73&0.71& 0.90& 1.0& 1.8&9.6& ${128^3}$&{\bf 75}& 0.8796 & 4.51003 & 1.73439 & 0.082008 & 23.007\\
\\
K1&0.044& 0.27& 0.73& 0.71& 0.94& 1.0&3.3&19.2&${512^3}$&{\bf 37.5} & 0.902& 3.85064 & 1.29774 & 0.0375323 & 27.110\\
K2&0.044& 0.27& 0.73& 0.71& 0.94& 1.0&3.3& 19.2& ${256^3}$&{\bf 75} & 0.897& 4.17679 & 1.21702 & 0.0280679 & 27.778\\
K3&0.044& 0.27& 0.73& 0.71& 0.94& 1.0&3.3& 19.2& ${128^3}$&{\bf 150}& 0.891& 4.46281 & 1.07452 & 0.0166946 & 32.892\\
\\
L4&0.044& 0.27& 0.73& 0.71& 1.00& 0.8& 1.4&{\bf 19.2}& ${256^3}$&{\bf 75}&0.8664&5.82605&2.0934&0.0890899&22.577\\
U&0.044& 0.27& 0.73& 0.71& 1.00& 0.8& 1.4&{\bf 38.4}& ${256^3}$&{\bf 150}&      &6.36059&1.66403&0.0470178&\\
\\
L5&0.044& 0.27& 0.73& 0.71& 1.09& 0.8& 1.4&{\bf 19.2}& ${256^3}$&{\bf 75} & 0.8753&5.55675&1.89446 & 0.0871944 & 22.1579 \\
Q1&0.044& 0.27& 0.73& 0.71& 1.09& 0.8& 1.4&{\bf 38.4}& ${256^3}$&{\bf 150} & 0.86 & 6.93897&1.82575&0.0477104& 27.516 \\
\\
A&0.044&0.27&0.73&0.71& 0.90& 1.0& 1.8&{\bf 76.8}& ${1024^3}$&{\bf 75}& 0.8713 & 6.22448 & 2.00096 & 0.0580288 & 26.662\\
A2&0.044&0.27&0.73&0.71& 0.90& 1.0& 1.8&{\bf 38.4}& ${512^3}$&{\bf 75}& 0.8779 & 5.26255 & 1.68276 & 0.0539609 & 25.741\\
A3&0.044&0.27&0.73&0.71& 0.90& 1.0& 1.8&{\bf 19.2}& ${256^3}$&{\bf 75}& 0.8756 & 5.29052 & 1.80656 & 0.0633351 & 24.995\\
B2&0.044&0.27&0.73&0.71& 0.90& 1.0& 1.8&{\bf 9.6}& ${512^3}$ &{\bf 18.75}& 0.8925&3.52847 & 1.47406 & 0.0770033& 22.7964\\
B&0.044&0.27&0.73&0.71& 0.90& 1.0& 1.8&{\bf 9.6}& ${256^3}$ &{\bf 37.5}& 0.8819 & 1.71394 & 4.62606 & 0.0789581 & 22.8300\\ 
\\
F& 0.044& 0.27& 0.73& 0.71&{\bf 0.90}&{\bf 1.04}&{\bf 1.8} &19.2& ${512^3}$&37.5&0.884&4.65843&1.7522&0.074021&23.325\\
K1&0.044& 0.27& 0.73& 0.71&{\bf 0.94}&{\bf 1.0}&{\bf  3.3}& 19.2& ${512^3}$&37.5&0.902&3.85064&1.29774&0.0375323&27.110\\
G& 0.044& 0.27& 0.73& 0.71&{\bf 1.10}&{\bf 0.63}&{\bf 4.2}&19.2& ${512^3}$&37.5&0.893& 4.48832&1.43529&0.0451349&26.246\\
\\
I& 0.044& 0.27& 0.73& 0.71&0.70&1.50&{\bf 0.40}&19.2& ${512^3}$&37.5&0.862&4.77332&2.10754&0.157915&20.174   \\
H& 0.044& 0.27& 0.73& 0.71&0.70&1.50&{\bf 0.66}&19.2& ${512^3}$&37.5&0.871&4.44712&1.85006&0.11792&21.531   \\
J& 0.044& 0.27& 0.73& 0.71&0.70&1.50&{\bf 0.90}&19.2& ${512^3}$&37.5&0.877&4.24435&1.68075&0.0928399&22.707   \\
\\
C&0.044& 0.27& 0.73& 0.71& 0.94& 1.2&{\bf 1.4}& 19.2& ${256^3}$&75  & 0.893 &4.48559&1.53054&0.0571994& 23.842\\
%N&0.044& 0.27& 0.73& 0.71& 0.94& 1.2&{\bf 1.8}& 19.2& ${256^3}$&75  &      &      &&&        \\
D&0.044& 0.27& 0.73& 0.71& 0.94& 1.2&{\bf 3.4}& 19.2& ${256^3}$&75  & 0.910 &3.67275&1.00311&0.0207897& 28.819 \\
E&0.044& 0.27& 0.73& 0.71& 0.94& 1.2&{\bf 5.4}& 19.2& ${256^3}$&75  & 0.919 &3.28659&0.771793&0.0113551& 32.806 \\
\\
P5&0.044& 0.27& 0.73& 0.71& 0.94& 1.0&{\bf 1.4}& 19.2& ${256^3}$&75  & 0.8774 & 5.08876&1.79106&0.0721239& 23.542 \\
K2&0.044& 0.27& 0.73& 0.71& 0.94& 1.0&{\bf 3.3}& 19.2& ${256^3}$&75   & 0.897  & 4.17679 & 1.21702 & 0.0280679 & 27.778\\
\\
O1&0.044& 0.27& 0.73& 0.71& 0.84& 0.5&{\bf 1.0}& 38.4& ${256^3}$&150 & 0.7617 & 11.6701 &3.38596&0.147094& 27.803 \\
O2&0.044& 0.27& 0.73& 0.71& 0.84& 0.5&{\bf 1.4}& 38.4& ${256^3}$&150 & 0.7741 & 10.8993 &3.07758&0.108957&30.1342  \\
O3&0.044& 0.27& 0.73& 0.71& 0.84& 0.5&{\bf 1.8}& 38.4& ${256^3}$&150 & 0.7839 & 9.07688 &2.71671&0.0826093& 31.927 \\
\\
L1&0.044& 0.27& 0.73& 0.71&{\bf 0.70}& 0.8& 1.4& 19.2& ${256^3}$&75  &        &      &&& 25.426 \\
L2&0.044& 0.27& 0.73& 0.71&{\bf 0.79}& 0.8& 1.4& 19.2& ${256^3}$&75  & 0.83   & 6.20307&2.51307&0.111913& \\
% incorrect bsig for L2 30.693 \\
L3&0.044& 0.27& 0.73& 0.71&{\bf 0.94}& 0.8& 1.4& 19.2& ${256^3}$&75  & 0.8626 & 5.52899 & 2.11876 & 0.0947183 &  23.1608\\
L4&0.044& 0.27& 0.73& 0.71&{\bf 1.00}& 0.8& 1.4& 19.2& ${256^3}$&75  & 0.8664 & 5.82605&2.0934&0.0890899& 22.577 \\
L5&0.044& 0.27& 0.73& 0.71&{\bf 1.09}& 0.8& 1.4& 19.2& ${256^3}$&75  & 0.8753&5.55675&1.89446 & 0.0871944 & 22.1579 \\
\\
M&0.044& 0.27& 0.73& 0.71&{\bf 0.70}& 1.0& 3.3& 19.2& ${256^3}$&75   & 0.8722 & 4.62589&1.36509&0.0322525& 31.015\\
K2&0.044& 0.27& 0.73& 0.71&{\bf 0.94}& 1.0&3.3& 19.2& ${256^3}$&75   & 0.897  & 4.17679 & 1.21702 & 0.0280679 & 27.778\\
\\
U&0.044& 0.27& 0.73& 0.71&{\bf 1.00}& 0.8& 1.4& 38.4& ${256^3}$&150&      &6.36059&1.66403&0.0470178&\\
Q1&0.044& 0.27& 0.73& 0.71&{\bf 1.09}& 0.8& 1.4& 38.4& ${256^3}$&150 & 0.86 & 6.93897&1.82575&0.0477104& 27.516 \\
\\
P2&0.044& 0.27& 0.73& 0.71&{\bf 0.84}& 0.7& 1.4& 38.4& ${256^3}$&150 & 0.82   & 8.30225&2.28294&0.069514&30.5494\\
Q&0.044& 0.27& 0.73& 0.71&{\bf 1.09}& 0.7& 1.4& 38.4& ${256^3}$ &150  & 0.85 &7.55325&2.04882&0.0565825& 27.159 \\
\\
O2&0.044& 0.27& 0.73& 0.71& 0.84&{\bf 0.5}& 1.4& 38.4& ${256^3}$&150 & 0.7741 & 10.8993 &3.07758&0.108957&30.1342  \\
P1&0.044& 0.27& 0.73& 0.71& 0.84&{\bf 0.6}& 1.4& 38.4& ${256^3}$&150 & 0.80   & 9.42176&2.61947&0.0854846&30.3377\\
P2&0.044& 0.27& 0.73& 0.71& 0.84&{\bf 0.7}& 1.4& 38.4& ${256^3}$&150 & 0.82   & 8.30225&2.28294&0.069514&30.5494\\
\\
%P3&0.044& 0.27& 0.73& 0.71&0.94&{\bf 0.7}& 1.4& 19.2& ${256^3}$&75  & 0.84 & 4.7575 & 0.587824 & 0.00215126 & 29.023 \\
L3&0.044& 0.27& 0.73& 0.71&0.94&{\bf 0.8}& 1.4& 19.2& ${256^3}$&75  & 0.8626 & 5.52899 & 2.11876 & 0.0947183 &  23.1608\\
P5&0.044& 0.27& 0.73& 0.71& 0.94&{\bf 1.0}& 1.4& 19.2& ${256^3}$&75  & 0.8774 & 5.08876&1.79106&0.0721239& 23.542 \\
C&0.044& 0.27& 0.73& 0.71& 0.94&{\bf 1.2}&1.4& 19.2& ${256^3}$&75  & 0.893 & 4.48559&1.53054&0.0571994& 23.842\\
\\
Q&0.044& 0.27& 0.73& 0.71& 1.09&{\bf 0.7}& 1.4& 38.4& ${256^3}$&150 & 0.85 & 7.55325&2.04882&0.0565825& 27.159 \\
Q1&0.044& 0.27& 0.73& 0.71& 1.09&{\bf 0.8}& 1.4& 38.4& ${256^3}$&150 & 0.86 & 6.93897&1.82575&0.0477104& 27.516 \\
\\
S2&{\bf 0.051}& 0.27& 0.73&{\bf 0.66}& 0.94& 1.0& 3.3& 19.2& ${256^3}$&75  & 0.892& 4.42602&1.29656&0.031181& 28.313\\
K2&{\bf 0.044}& 0.27& 0.73&{\bf 0.71}& 0.94& 1.0& 3.3& 19.2& ${256^3}$&75  & 0.897&4.17679 & 1.21702 & 0.0280679&27.778\\
S3&{\bf 0.038}& 0.27& 0.73&{\bf 0.76}& 0.94& 1.0& 3.3& 19.2& ${256^3}$&75  & 0.904& 4.03904&1.16607&0.0262167& 27.536 \\
\\
T&0.044&{\bf 0.22}&{\bf 0.78}& 0.71& 0.94& 1.0& 3.3& 19.2& ${256^3}$&75  & 0.887&4.48482&1.36005&0.0315291& 28.671\\
K2&0.044&{\bf 0.27}&{\bf 0.73}& 0.71& 0.94& 1.0& 3.3& 19.2& ${256^3}$&75  & 0.897&4.17679 & 1.21702 & 0.0280679&27.778\\
\\
K3&0.044&{\bf 0.27}&{\bf 0.73}& 0.71& 0.94& 1.0& 3.3& 19.2& ${128^3}$& 150 & 0.891&4.46281 & 1.07452 & 0.0166946 &32.892\\
V&0.044&{\bf 0.32}&{\bf 0.68}& 0.71& 0.94& 1.0& 3.3& 19.2& ${128^3}$& 150 & 0.905&4.11799&1.15153&0.026181 & 27.318\\
\enddata 
\end{deluxetable}

\clearpage
\begin{deluxetable}{ccccccc}        
\tablecaption{Polynomial Fit Coefficients to Simulation Flux Power Spectrum
\label{table:pfits}} 
\tablewidth{0pt}
\tablehead{ 
\colhead{Name} & 
\colhead{$B$} & \colhead{$A$} & \colhead{$C$} & \colhead{$E$} & \colhead{$F$} & \colhead{$G$}
} 
\startdata 
PJ05 & 1.726 &  43.709 & 42.008 &  15.075 &  2.407 &  0.144\\
A &   2.048 &  55.103 &  49.214 &  16.433 &   2.440 &   0.136\\
A2 &   2.652 &  81.246 &  64.923 &  19.468 &   2.600 &   0.130\\
A3 &   2.635 &  77.822 &  62.260 &  18.678 &   2.495 &   0.125\\
A4 &   2.615 &  80.807 &  66.037 &  20.249 &   2.765 &   0.142\\
A5 &   2.006 &  57.992 &  53.459 &  18.466 &   2.839 &   0.164\\
B &   1.916 &  56.461 &  54.123 &  19.469 &   3.122 &   0.188\\
B2 &   2.346 &  72.765 &  63.377 &  20.715 &   3.015 &   0.165\\
C &   1.874 &  54.883 &  52.543 &  18.871 &   3.020 &   0.182\\
D &   2.697 &  90.751 &  71.484 &  21.148 &   2.787 &   0.138\\
E &   2.207 &  69.003 &  59.082 &  18.988 &   2.719 &   0.146\\
F &   1.888 &  54.335 &  51.927 &  18.599 &   2.967 &   0.178\\
G &   1.954 &  56.939 &  52.984 &  18.500 &   2.880 &   0.168\\
H &   2.506 &  80.153 &  68.218 &  21.783 &   3.095 &   0.165\\
I &   1.859 &  54.190 &  53.822 &  20.029 &   3.316 &   0.206\\
J &   2.423 &  78.646 &  67.938 &  22.023 &   3.178 &   0.172\\
K1 &   1.949 &  57.636 &  53.391 &  18.538 &   2.866 &   0.166\\
K2 &   2.034 &  65.849 &  60.151 &  20.619 &   3.149 &   0.181\\
K3 &   1.617 &  54.038 &  54.058 &  20.286 &   3.394 &   0.213\\
L2 &   1.909 &  62.159 &  60.186 &  21.805 &   3.511 &   0.212\\
L3 &   2.078 &  56.175 &  50.878 &  17.243 &   2.599 &   0.147\\
L4 &   2.145 &  63.071 &  56.947 &  19.279 &   2.907 &   0.165\\
L5 &   3.375 &  89.469 &  61.666 &  15.949 &   1.837 &   0.079\\
M &   2.835 & 116.928 &  92.016 &  27.194 &   3.579 &   0.177\\
O1 &  -1.261 &  -2.682 &  -2.017 &  -0.046 &   0.788 &   0.307\\
O2 &   0.566 &  21.158 &  29.526 &  15.214 &   3.498 &   0.303\\
O3 &   2.738 &  66.863 &  50.661 &  14.339 &   1.801 &   0.085\\
P1 &  -1.315 &  -2.664 &  -2.285 &  -0.359 &   0.697 &   0.316\\
P2 &  -1.340 &  -2.602 &  -2.311 &  -0.402 &   0.713 &   0.334\\
%P3 &   0.112 &  28.200 &  45.711 &  27.535 &   7.385 &   0.743\\
P5 &   1.984 &  57.283 &  53.473 &  18.719 &   2.920 &   0.171\\
Q &   0.649 &  24.102 &  32.128 &  15.876 &   3.496 &   0.289\\
Q1 &   0.688 &  24.906 &  32.597 &  15.841 &   3.432 &   0.280\\
R &  -1.479 &  -2.143 &  -2.343 &  -0.913 &   0.320 &   0.257\\
S2 &   2.007 &  65.174 &  60.103 &  20.804 &   3.209 &   0.186\\
S3 &   1.839 &  57.767 &  54.973 &  19.623 &   3.121 &   0.186\\
T &   1.989 &  68.115 &  63.402 &  22.141 &   3.444 &   0.201\\
U &   1.980 &  44.419 &  39.218 &  12.948 &   1.904 &   0.105\\
V &   1.812 &  55.234 &  52.737 &  18.892 &   3.017 &   0.181\\
W1 &   2.050 &  65.009 &  58.622 &  19.827 &   2.987 &   0.169\\
%W3 &   2.070 &  57.687 &  51.092 &  16.937 &   2.498 &   0.138\\
\enddata
\end{deluxetable}

%------------------------------------------------------------------------
% source dtpc d:\projects\jena1\tablekim.txt from kim04etable5c.xls
\clearpage
\begin{deluxetable}{ccccc}
\tablecaption{Flux Power of F2 Scaled from K04e\label{table:kim.tbl}}
\tablewidth{0pt}
\tablehead{
\colhead{$k$}& \colhead{$PF2(k,z=2.36)$}&\colhead{$f_m$}
&\colhead{$PF2(k,z=1.95)$}&\colhead{$PF2I(k,z=1.95)$} \\
\colhead{(s/km)}&\colhead{(s/km)}&&\colhead{(s/km)}& \colhead{(s/km)} \\
}
\startdata
0.001	&	12.1020	&	0.0667	&	22.3768	$\pm $	6.8702	&	21.5696	$\pm $	6.6223	\\
0.0013	&	26.8893	&	0.0667	&	25.1011	$\pm $	8.3467	&	23.3076	$\pm $	7.7503	\\
0.0018	&	27.1282	&	0.0667	&	17.7488	$\pm $	4.2408	&	15.9394	$\pm $	3.8084	\\
0.0024	&	18.7607	&	0.0667	&	18.1680	$\pm $	5.0354	&	16.9166	$\pm $	4.6885	\\
0.0032	&	24.4826	&	0.0667	&	18.5827	$\pm $	4.0141	&	16.9497	$\pm $	3.6613	\\
0.0042	&	18.8581	&	0.0667	&	14.6722	$\pm $	2.4618	&	13.4143	$\pm $	2.2508	\\
0.0056	&	17.8765	&	0.0667	&	13.2725	$\pm $	1.4887	&	12.0801	$\pm $	1.3549	\\
0.0075	&	16.3082	&	0.0667	&	11.6026	$\pm $	1.8883	&	10.5149	$\pm $	1.7113	\\
0.01	&	11.8373	&	0.0667	&	9.6215	$\pm $	1.1073	&	8.8319	$\pm $	1.0164	\\
0.0133	&	11.0057	&	0.0823	&	7.2006	$\pm $	0.6441	&	6.2948	$\pm $	0.5631	\\
0.0178	&	8.0428	&	0.0776	&	5.2561	$\pm $	0.5319	&	4.6320	$\pm $	0.4688	\\
0.0237	&	5.6733	&	0.0770	&	3.6068	$\pm $	0.3591	&	3.1700	$\pm $	0.3156	\\
0.0316	&	3.2207	&	0.0849	&	2.3466	$\pm $	0.1634	&	2.0732	$\pm $	0.1443	\\
0.0422	&	1.9670	&	0.1020	&	1.2188	$\pm $	0.1059	&	1.0182	$\pm $	0.0884	\\
0.0562	&	0.9475	&	0.1230	&	0.6684	$\pm $	0.0404	&	0.5519	$\pm $	0.0334	\\
0.075	&	0.4110	&	0.1370	&	0.2969	$\pm $	0.0222	&	0.2406	$\pm $	0.0180	\\
0.1	&	0.1531	&	0.2350	&	0.1201	$\pm $	0.0096	&	0.0841	$\pm $	0.0067	\\
0.1334	&	0.0625	&	0.3900	&	0.0523	$\pm $	0.0047	&	0.0279	$\pm $	0.0025	\\
0.1778	&	0.0253	&	0.5050	&	0.0229	$\pm $	0.0025	&	0.0101	$\pm $	0.0011	\\
0.2371	&	0.0111	&	0.6570	&	0.0109	$\pm $	0.0010	&	0.0036	$\pm $	0.0003	\\
\enddata 
\end{deluxetable}
%------------------------------------------------------------------------

% source dtpc d:\projects\jena1\tablekim.txt from kim04etable5c.xls
\clearpage
\begin{deluxetable}{cccc}
\tablecaption{Flux Power of F2 at $z=2.41$ Scaled from K04e and M00
\label{tbl:m00k04}}
\tablewidth{0pt}
\tablehead{
\colhead{$k$}& \colhead{$PK04(k)$}&\colhead{$k$}&\colhead{PM00(k)}\\
\colhead{(s/km)}&\colhead{(s/km)}&\colhead{(s/km)}& \colhead{(s/km)} \\
}
\startdata
0.001&11.5733&...&...\\
0.0013&28.2664&0.00284&13.7045\\
0.0018&30.8258&0.00358&14.4667\\
0.0024&19.7370&0.0045&19.8767\\
0.0032&26.3861&0.00566&17.0372\\
0.0042&19.3851&0.00713&17.0372\\
0.0056&19.0843&0.00898&15.2438\\
0.0075&17.3480&0.0113&11.2087\\
0.01&12.6588&0.0142&8.8922\\
0.0133&11.5484&0.0179&6.1424\\
0.0178&8.5297&0.0225&5.3503\\
0.0237&6.1273&0.0284&3.7362\\
0.0316&3.4751&0.0357&2.226\\
0.0422&2.1282&0.045&1.5692\\
0.0562&1.0199&0.056&0.8294\\
0.075&0.4357&0.0713&0.4185\\
0.1&0.1630&0.0898&0.2033\\
0.1334&0.0667&0.113&0.0707\\
0.1778&0.0270&0.142&0.0324\\
0.2371&0.0118&...&...\\
\enddata 
\end{deluxetable}
%------------------------------------------------------------------------
%\clearpage
%\begin{deluxetable}{ccc}
%\tablecaption{Resolution and Box size effects on the Mean Flux \label{table:BoxResF.tbl}}
%\tablewidth{0pt}
%\tablehead{
%\colhead{Box Size(comoving Mpc)} & \colhead{Number of Cells} & \colhead{Mean Flux}
%}
%\startdata
%19.2 & $128^3$ &  0.891\\
%19.2 & $256^3$ &  0.897\\
%19.2 & $512^3$ &  0.902\\
%\\
%9.6  & $128^3$ &  0.889\\
%19.2 & $256^3$ &  0.897\\
%38.4 & $512^3$ &  0.896\\
%\enddata
%\end{deluxetable}
%--------------------------------------------------------------------------------------
%\clearpage
%\begin{deluxetable}{cccccc}
%\tablecaption{Effect on the Mean Flux of $h$ and ${\Omega_{\Lambda}}$.  
%\label{table:Obl.tbl}}
%\tablewidth{0pt}
%\tablehead{
%\colhead{${\Omega_b}$} & \colhead{h} & \colhead{${\Omega_{\Lambda}}$} & \colhead{${\Omega_M}$} & \colhead{$\bar{F}$} & \colhead{${kP/Pi}$}  
%}
%\startdata
%0.051& 0.66 &  0.73&  0.27&  0.892&  0.012\\
%0.044& 0.71 &  0.73&  0.27&  0.897&  0.011\\
%0.038& 0.76 &  0.73&  0.27&  0.904&  0.011\\
%\\
%0.044& 0.71 &  0.68&  0.32&  0.905&  0.011\\
%0.044& 0.71 &  0.73&  0.27&  0.897&  0.011\\
%0.044& 0.71 &  0.78&  0.22&  0.887&  0.012\\
%\enddata
%\end{deluxetable}

%------------------------------------------------------------------------------------------- 
\begin{deluxetable}{lccccl}
\tablecaption{QSOs Observed for Flux Power Measurement at $z=1.95$
\label{tbl:obs_table}}
%\tabletypesize{\tiny}
\tablewidth{0pt}
\tablehead{
\colhead{QSO}&
\colhead{$z_{em}$}&
\colhead{$z_{Ly\alpha}$}&
\colhead{$\lambda_{Ly\alpha}$ (\AA)}&
\colhead{S/N per pixel}&
\colhead{Instrument}
}

\startdata
HS0105+1619  & 2.65 & 2.09-2.61 & 3753-4389 & 50-90 & Keck HIRES\\
Q0450--1310  & 2.25 & 1.79-2.22 & 3392-3918 &  5-20 & Keck HIRES\\
HE0515--4414 & 1.71 & 1.52-1.69 & 3060-3265 &  5-25 & VLT UVES\\
Q0747+4259   & 1.90 & 1.55-1.87 & 3100-3487 &  5-15 & Keck HIRES\\
Q1243+3047   & 2.56 & 2.02-2.53 & 3666-4287 & 10-70 & Keck HIRES\\
HE2217--2818 & 2.41 & 1.89-2.38 & 3514-4109 & 35-50 & VLT UVES
\enddata

\end{deluxetable}

\begin{deluxetable}{ccc}
\tablecaption{Scaling Relations Between Input and Output Parameters
\label{tbl:scalings}
}
\tablewidth{0pt}
\tablehead{
\colhead{Input Parameter} &
\colhead{\taueff} &
\colhead{\bsig}
}
\startdata
Box Size ($L$)  &  $0.124447 + 0.00015187 L$   &   $-6.43176 + 24.9593 L^{0.0465855}$\\
Cell Size ($C$) &  $0.132828 + 0.00016748 C$   &   $24.0341 + 0.000244627 C^{2.03635}$\\
\gammahe        &  $0.339854 - 0.194975 \gammahe^{0.128823}$ & $ 16.4342 + 6.31086 \gammahe^{0.589379}$ \\
\gammah         &  $0.0517116 + 0.0845752 \gammah^{-1}$ & $ 23.9792 + 0.150065 \gammah$\\
\sig            &  $0.243733 - 0.119478 \sig$ & $ 41.1189 - 17.9121 \sqrt{\sig} $ \\
\enddata
\end{deluxetable}

\begin{deluxetable}{cc}
\tablecaption{Standard Model Parameters for our Scaling Relations
\label{tbl:standard}
}
\tablewidth{0pt}
\tablehead{
\colhead{Input Parameter} &
\colhead{Value} 
}
\startdata
\ob\ & 0.044\\
\om\ & 0.27\\
\ol\ & 0.73\\
$h$  & 0.71\\
\sig\ & 0.9\\
$n$  & 1.0\\
\gammah\ & 1.0\\
\gammahe\ & 1.4\\ 
box size $L$ & 76.8 Mpc\\
cell size $C$ & 18.75 kpc\\
redshift $z$ & 1.95\\
\enddata
\end{deluxetable}

\begin{deluxetable}{lccc}
\tablecaption{\label{tabcos} Cosmological Parameters Derived from our 
Measurement of DA at $z=1.9$}
\tablewidth{0pt}
\tablehead{
\colhead{Parameter} &
\colhead{Nominal Prior } &
\colhead{${\partial \tau \over \partial X}$} &
\colhead{$\sigma(\tau_{eff})$} 
}
\startdata
Power spectrum amplitude $\sigma _8$ ($n=1$)   & $0.9 \pm 0.1$        & -0.118  & 0.0118 \\
H~I ionization rate $\gamma_{912}$            & $1.0 \pm 0.1$        & -0.083  & 0.0083 \\
Enhanced He~II heating rate $X_{228}$ & $1.4 \pm 0.6$        & -0.019  & 0.0114 \\
\enddata
\end{deluxetable}

\begin{deluxetable}{ccc}
\tablecaption{Measured parameters for Spectra Passing through Simulations A3 and A5
\label{tbl:obsimathreefive}}
%\tabletypesize{\tiny}
\tablewidth{0pt}
\tablehead{
\colhead{Simulation ID}&
\colhead{$<F>$}&
\colhead{\bsig}
}

\startdata
A5    & 0.8762 & 24.560 \\
A3-1  & 0.8799 & 24.936 \\ 
A3-2  & 0.8800 & 24.650 \\
A3-3  & 0.8750 & 24.970 \\
A3-4  & 0.8714 & 25.229 \\
A3-5  & 0.8700 & 25.217 \\
A3-6  & 0.8788 & 24.934 \\
A3-7  & 0.8765 & 24.843 \\
A3-8  & 0.8753 & 24.824 \\
A3-9  & 0.8752 & 24.788 \\
A3-10 & 0.8761 & 24.793 \\
\enddata

\end{deluxetable}

%---------------------------------------------------
%  fig1
\clearpage

\begin{figure}
{\plotone{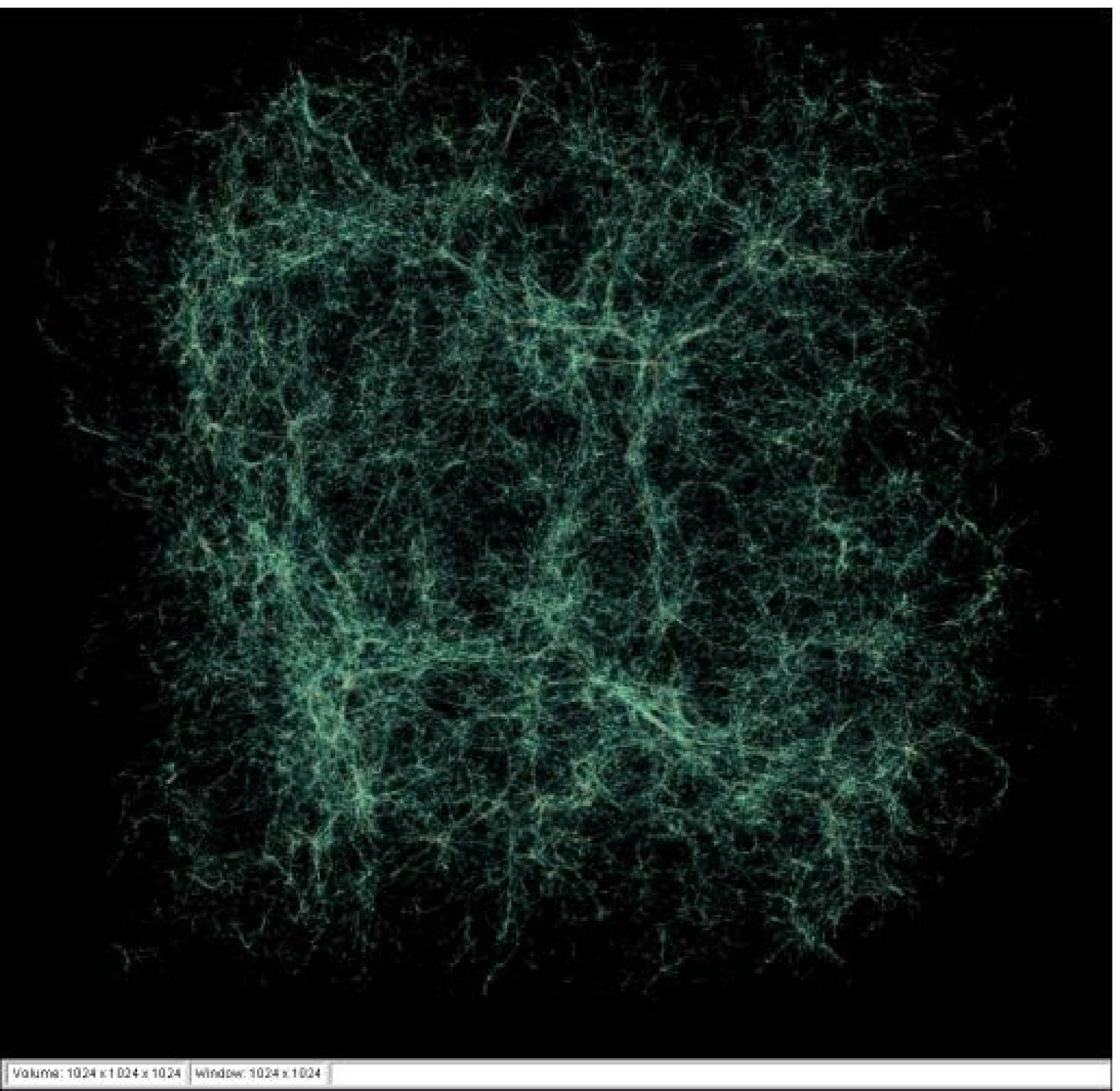}}
\caption{\NOTE{imageA.ps} Volumetric rendering of the log baryon density from
simulation A at $z=2.0$.}
\label{fig:imageA.fig}
\end{figure}

% output against box, then cell size in sc1.tex convergence

\clearpage
\begin{figure}
{\plotone{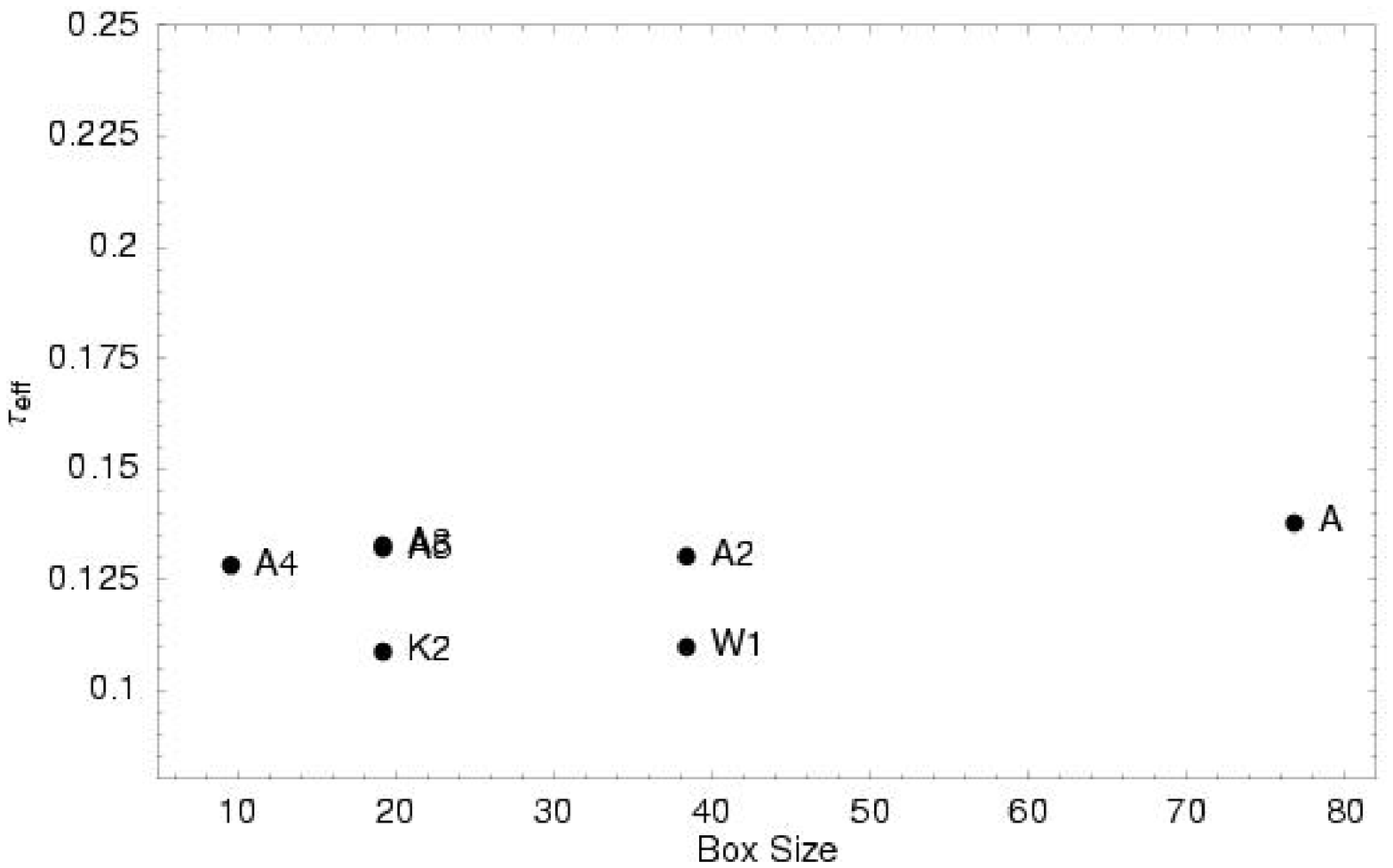}}
\caption{\NOTE{boxFbarRaw2.ps}
The change in \taueff\ with box size. The A simulations (\gammahe $=1.8$)
all have all the same parameters except for box size, as do the pair K2W1 
(\gammahe $=3.3$). The points and labels for A3 and A5 lie on top of
each other. In all Figures the box size is in comoving Mpc for $h = 0.71$.
}
\label{fig:boxFbarRaw.fig}
\end{figure}

\clearpage
\begin{figure}
{\plotone{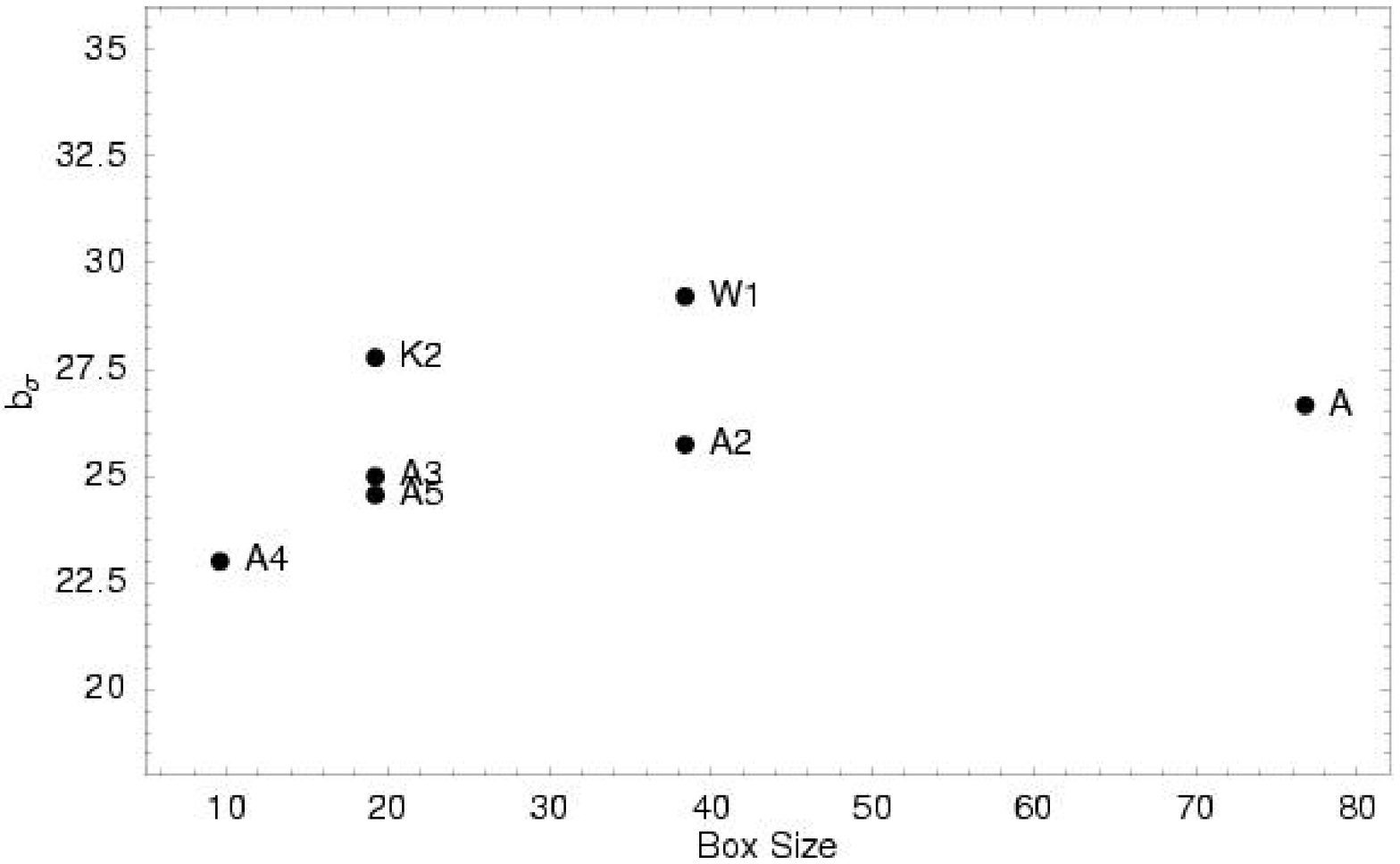}}
\caption{\NOTE{boxbsigRaw.ps}
The change in \bsig\ with box size.
The series K2W1 (\gammahe $=3.3$) are hotter and hence lie above the A 
series (\gammahe $=1.8$).
}
\label{fig:boxbsigRaw.fig}
\end{figure}

\clearpage
\begin{figure}
{\plotone{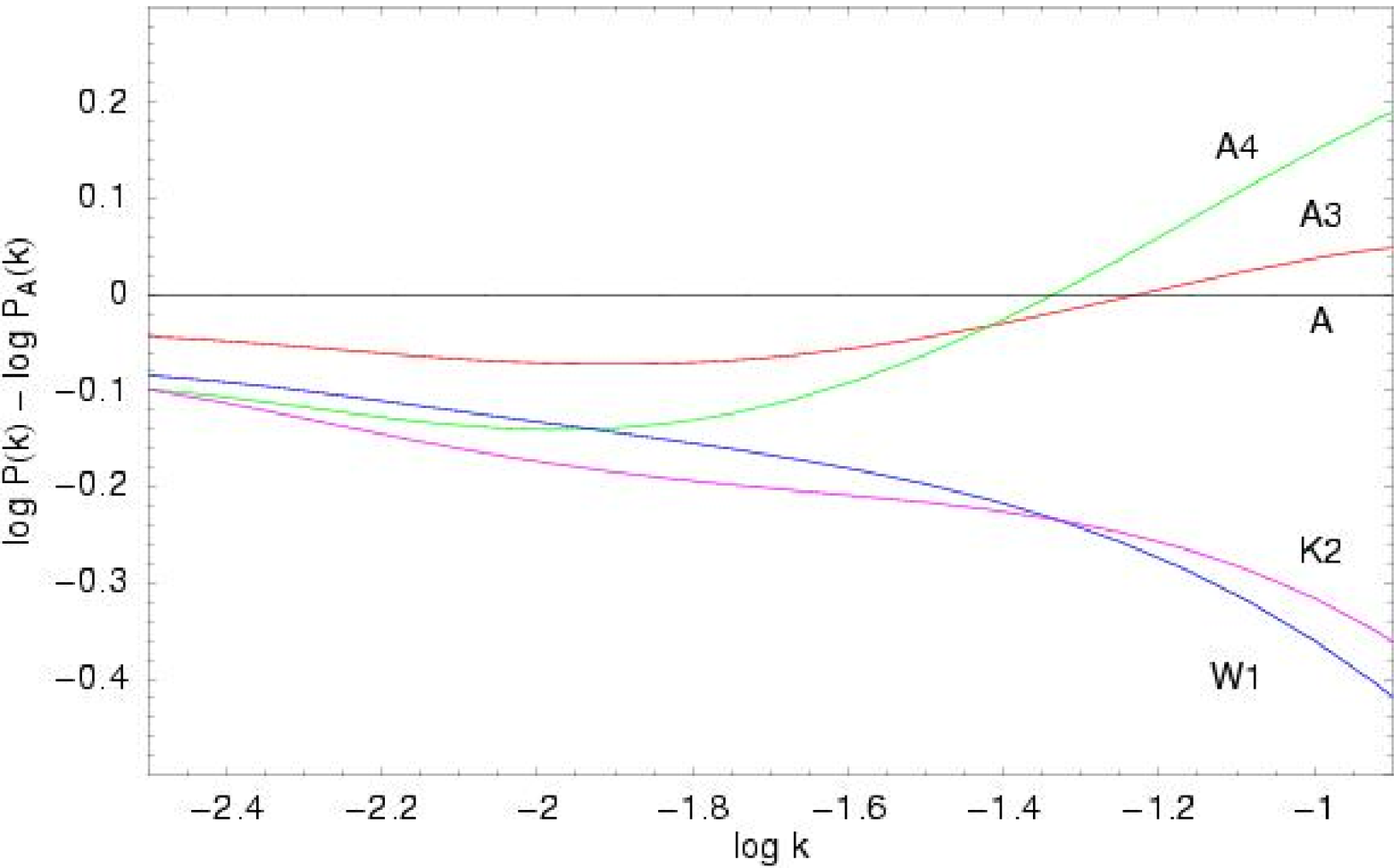}}
\caption{\NOTE{powerbox2.ps}
The effect of box size on flux power.
The 1D flux power as a function of wave number k in s/km.
We have divided the power spectra by that for the simulation with the
largest box, A.
As box size increases, from A4 (9.6 Mpc) to A3 (19.2 Mc) to A (76.8 Mpc), 
the power at $log k > -1.3$ s/km
(small scales) decreases, while that
on larger scales increases as the boxes contain
more large scale matter power (\S \ref{sec:CosSimSeeds}). 
K2 (19.2 Mpc) and W1 (38.4 Mpc) show the same trend, but 
tilted because we divided their flux power by A.
We show the power at smaller 
$k$ values than is justified by the box sizes to help illustrate the trends 
and differences.  The 19.2 Mpc boxes include only modes $n > 2$
for $log k = -2$ s/km (\S \ref{subsec:MeasP}).
In all plots involving power, $k$ is in units of s/km for our standard model, 
and the power is the 1D flux power of the quantity that 
\citet[Eqn. 3]{kim04a} call F2.
}
\label{fig:powerbox2.fig}
\end{figure}

\clearpage
\begin{figure}
{\plotone{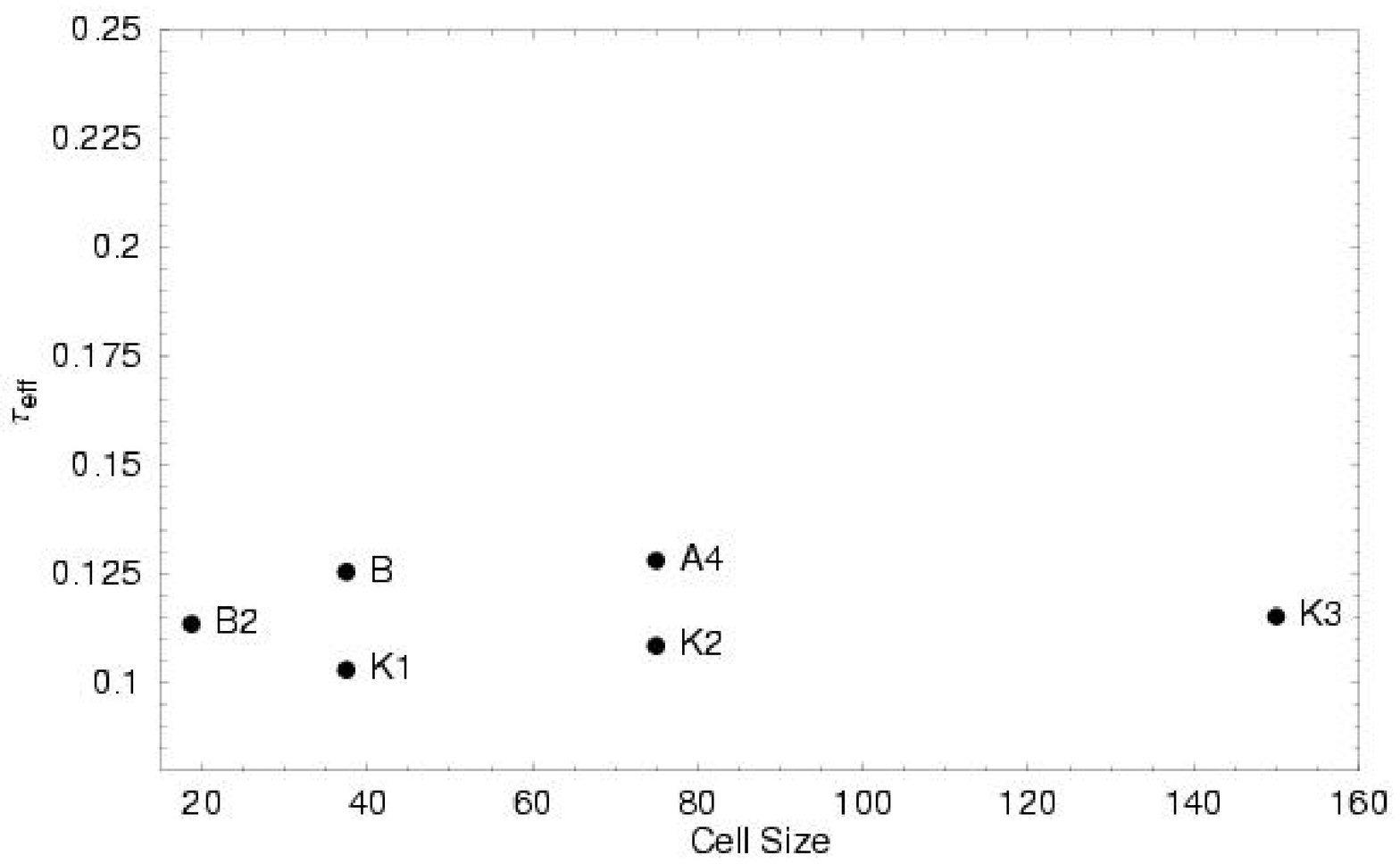}}
\caption{\NOTE{cellFbarRaw.ps}
The variation of \taueff\ with cell size.
Simulations B2BA4 (\gammahe $=1.8$) all have the same parameters except for 
cell size, as do the K series (\gammahe $=3.3$).
In all Figures the cell size is in comoving kpc for $h = 0.71$.
}
\label{fig:cellFbarRaw.fig}
\end{figure}

\clearpage
\begin{figure}
{\plotone{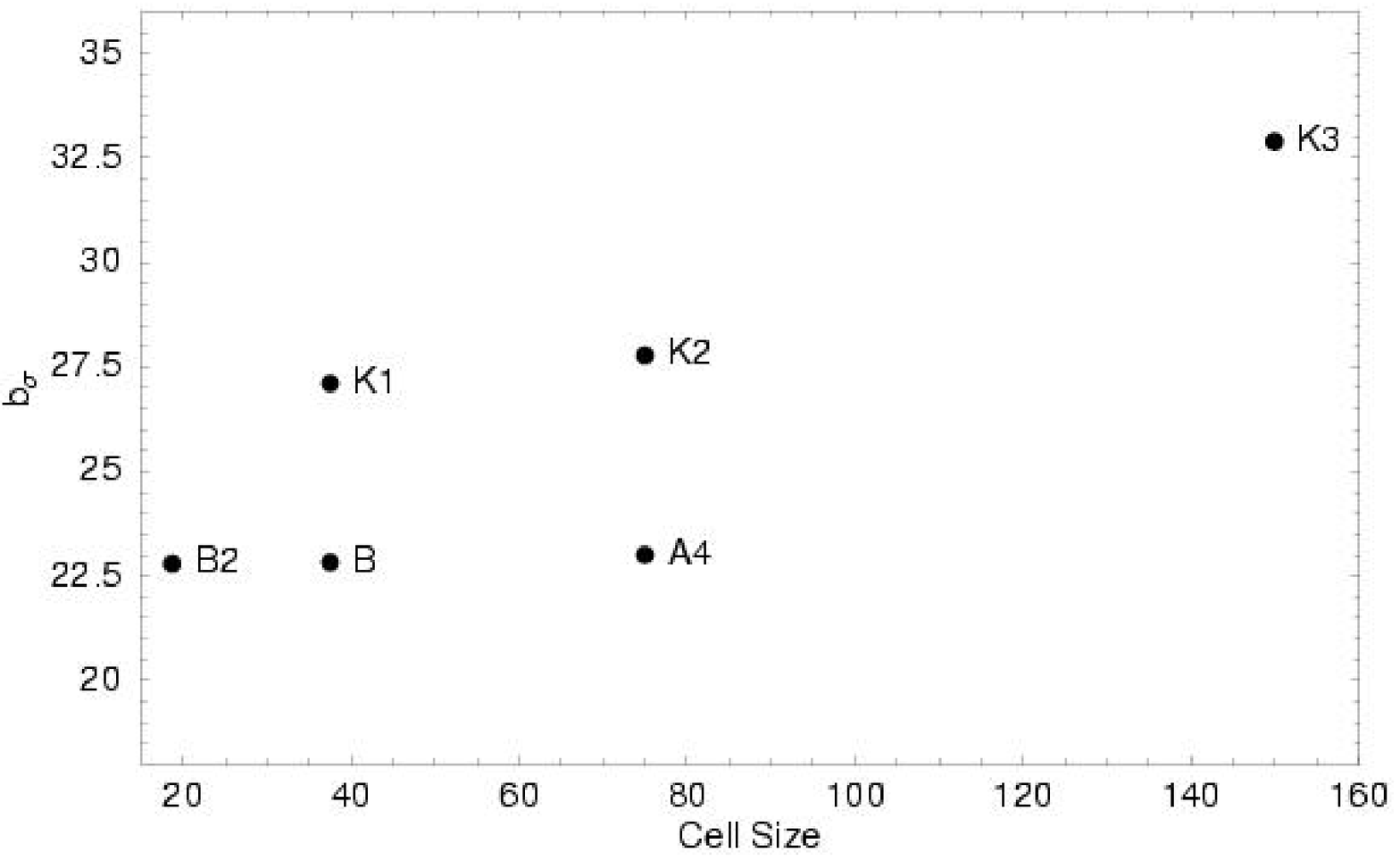}}
\caption{\NOTE{cellbsigRaw.ps}
The variation of \bsig\ with cell size.
Simulations B2BA4 (\gammahe $=1.8$) all have the same parameters except for
cell size, as do the K series (\gammahe $=3.3$).
}
\label{fig:cellbsigRaw.fig}
\end{figure}

\clearpage
\begin{figure}
{\plotone{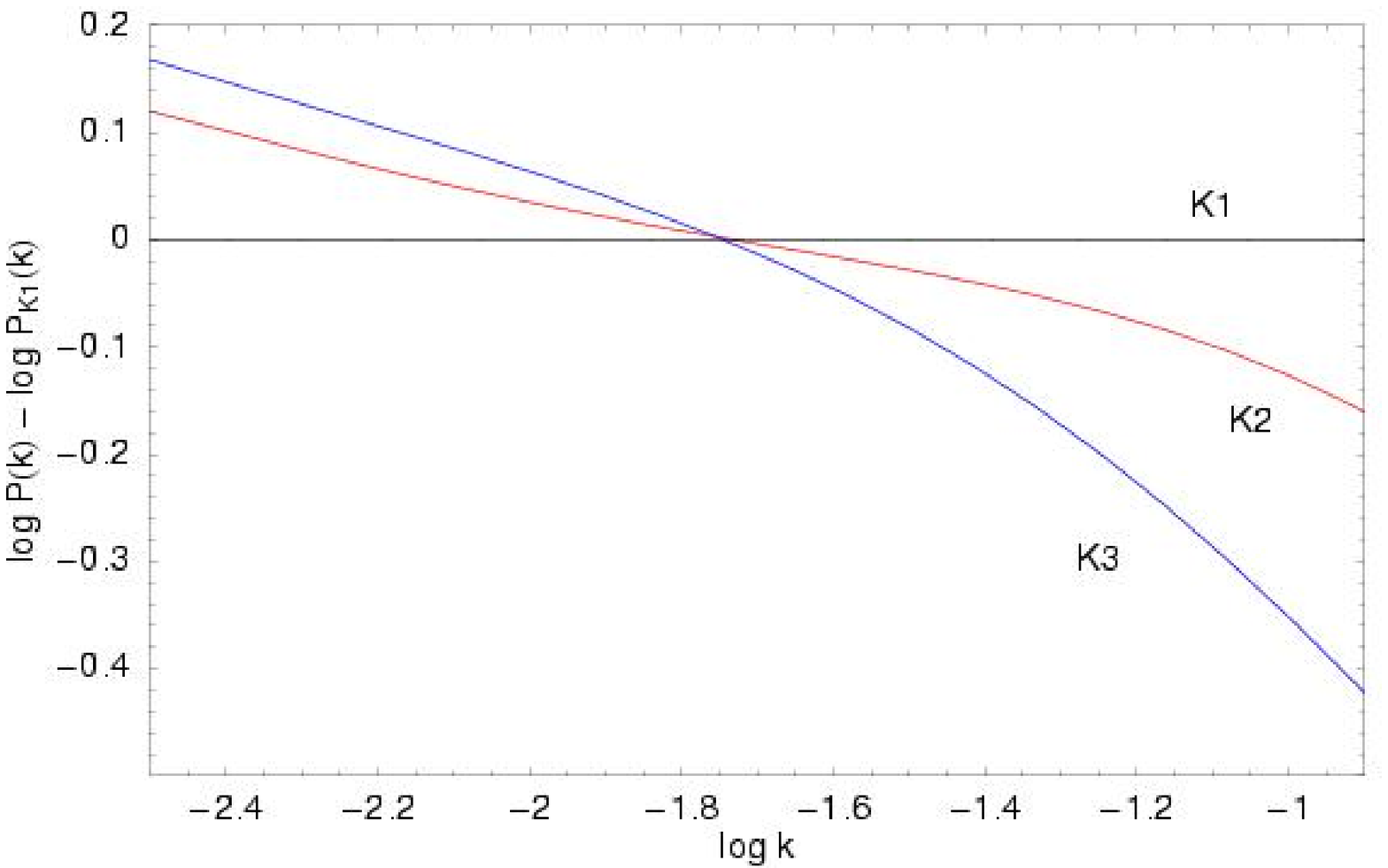}}
\caption{\NOTE{powercell2.ps}
The effect of cell size on flux power.
The 1D flux power as a function of wave number k in s/km.
The simulations shown have identical input parameters except for cell size.
We have divided the power by that for simulation K1 that has the smallest
cell size (37.5 kpc). With smaller cell size the small scale power increases
while that on large scales decreases.
}
\label{fig:powercell2.fig}
\end{figure}

\clearpage
\begin{figure}
{\plotone{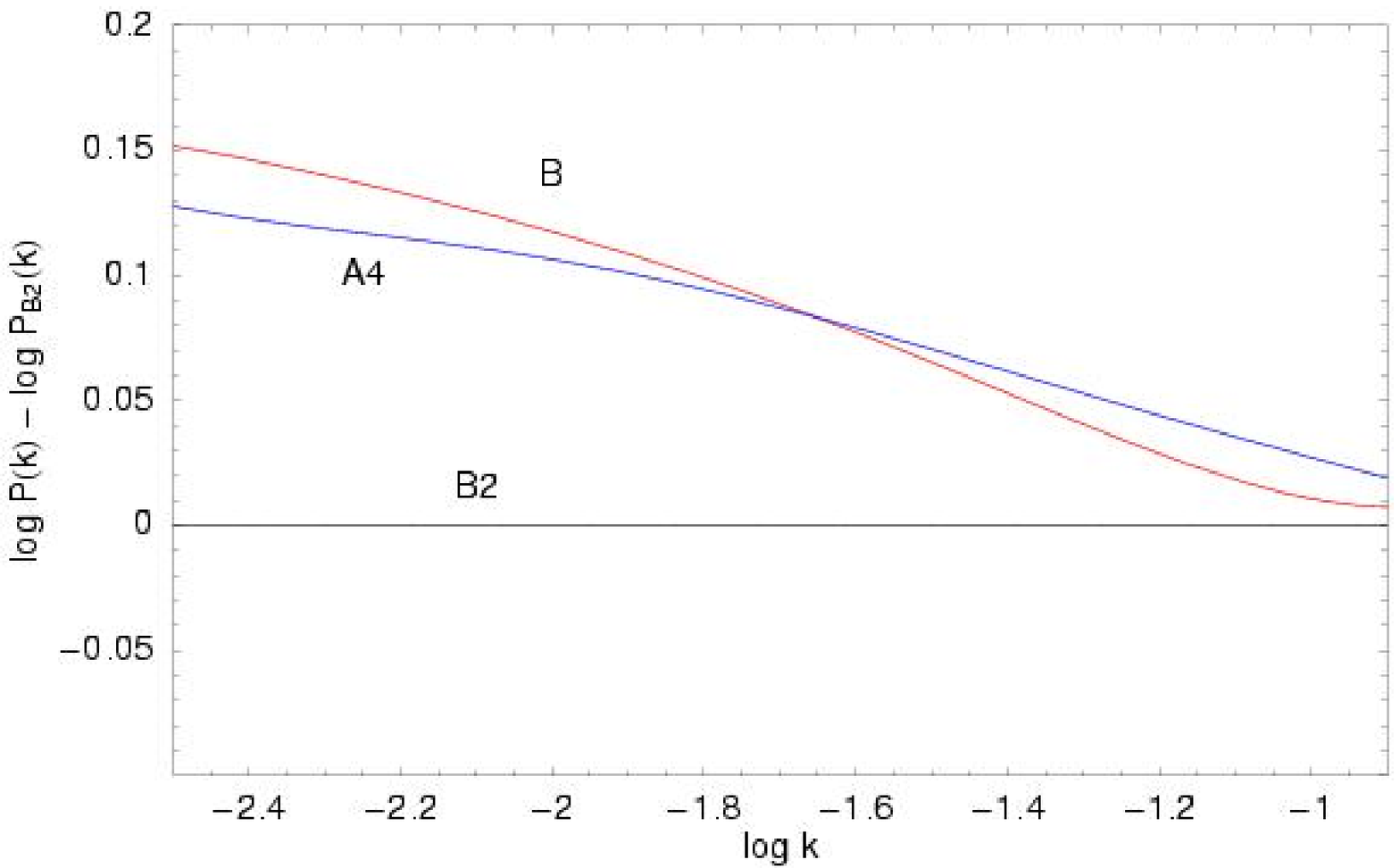}}
\caption{\NOTE{powercell1.ps}
The effect of cell size on flux power.
The 1D flux power as a function of wave number k in s/km.
The simulations shown have identical input parameters except for cell size.
We have divided the power by that for simulation B2 that has the smallest
cell size (18.75 kpc). 
}
\label{fig:powercell1.fig}
\end{figure}

% out-out in sc2.tex

\clearpage
\begin{figure}
{\plotone{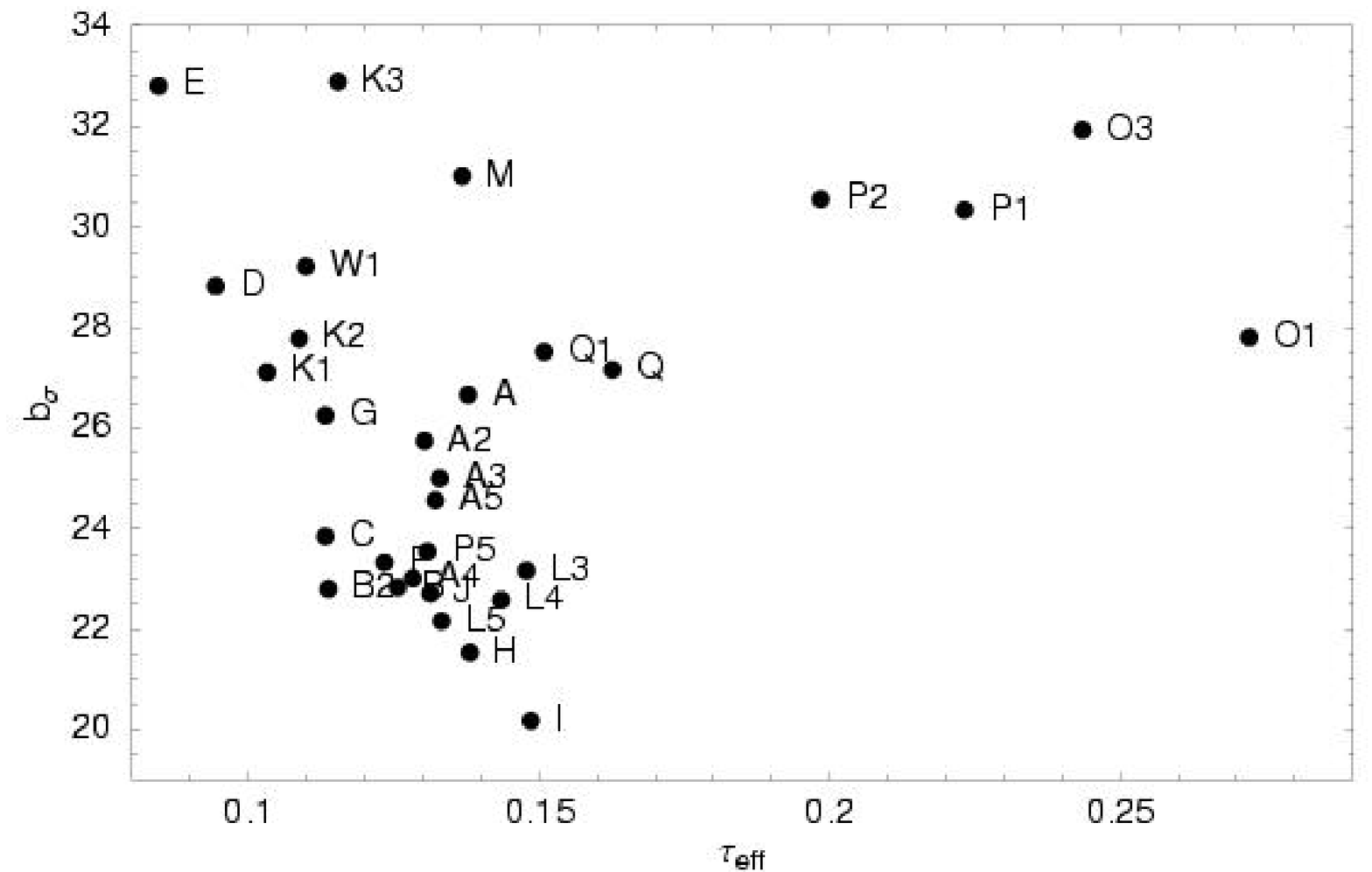}}
\caption{\NOTE{bsigTau1.ps}
A scatter plot showing the correlation of a pair of output parameters,
Line width parameter \bsig\ against effective optical depth for various
simulations. Cell size decreases to the lower left.
Simulations O1, O3, P1, P2, Q1, Q and K3 all have 150 kpc cells.
They are joined by M that has broad lines because it has low \sig\
and high \gammahe\ values.
K1, G, B, H and I all have 37.5 kpc cells while B2 has
18.75 kpc cells. For a given cell size, \bsig\ increases as \taueff\ decreases.}
\label{fig:bsigTau1.fig}
\end{figure}

\clearpage
\begin{figure}
{\plotone{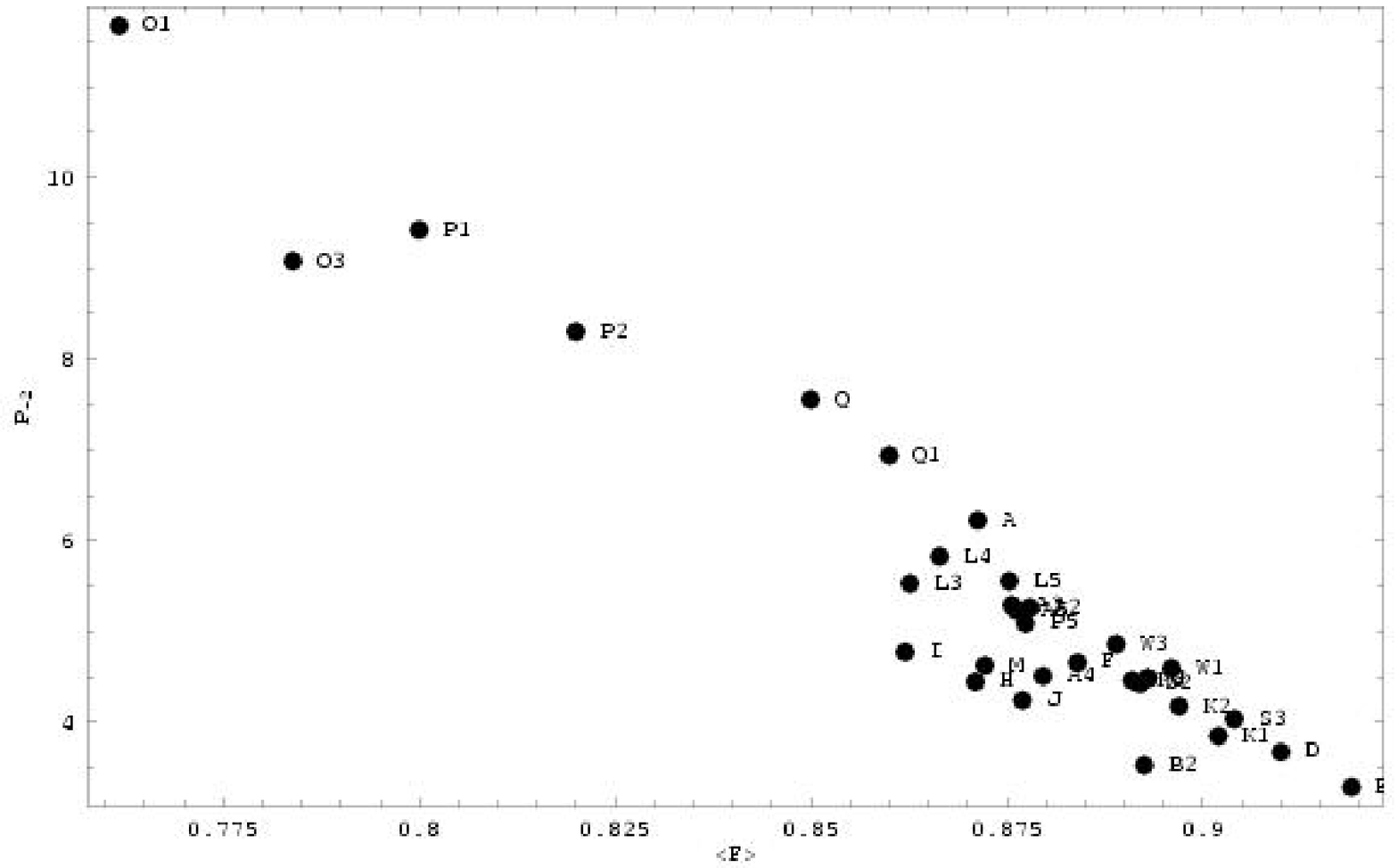}}
\caption{\NOTE{Pm2Fbar.ps}
A scatter plot showing the correlation of a pair of output parameters,
\pl\ against \fbar\ for various simulations. 
Simulations O1, O3, P1, P2, Q1, Q and K3 all have 150 kpc cells.
Simulations with small boxes, such as A4, B2 and B in the lower right,
will have \pl\ values that are significantly too small.
The simulations show a strong correlation, with larger \pl\
for more absorption.
}
\label{fig:Pm2Fbar.fig}
\end{figure}

\clearpage
\begin{figure}
{\plotone{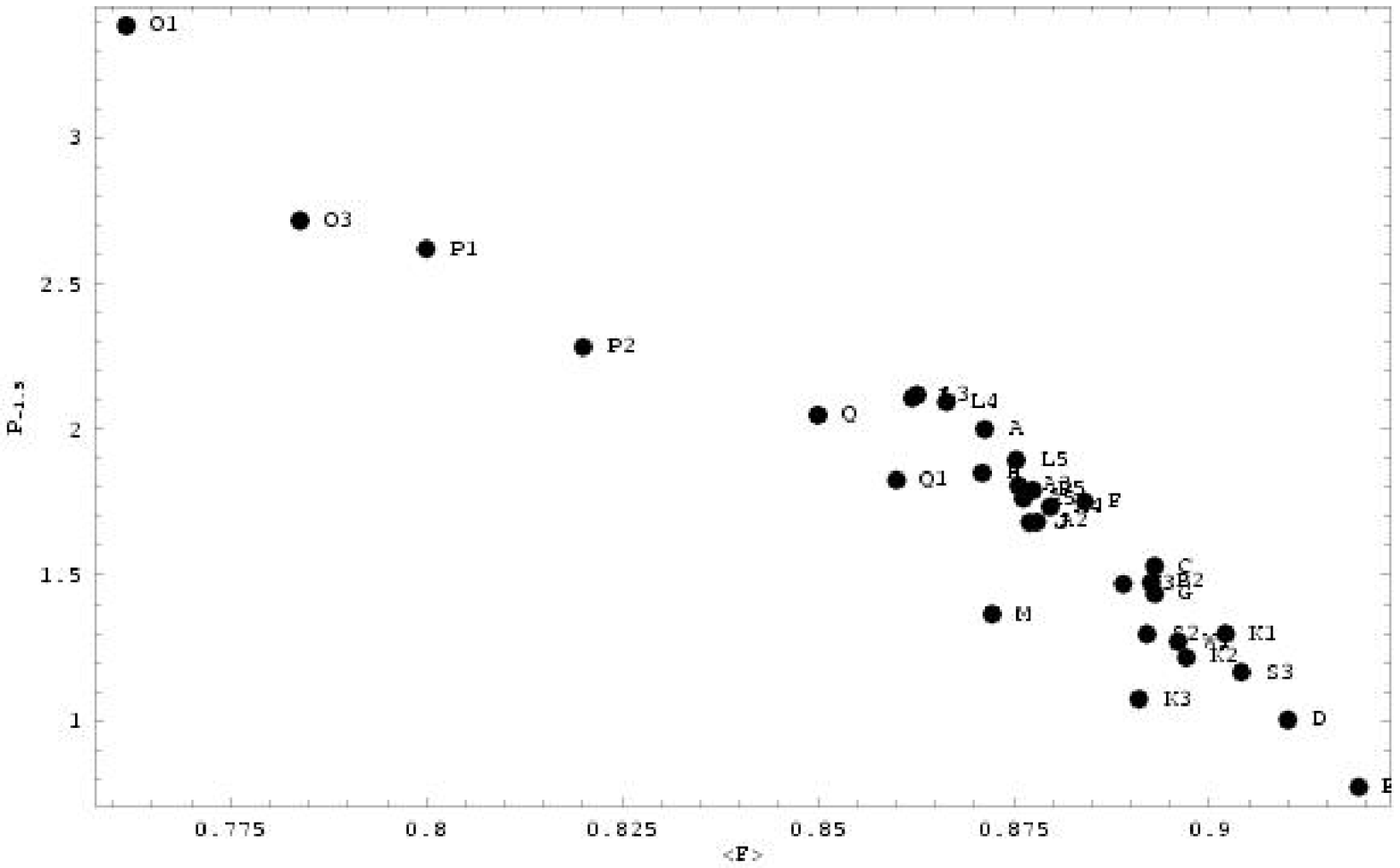}}
\caption{\NOTE{Pm15Fbar.ps}
A scatter plot showing the correlation of a pair of output parameters,
\pmed\ against \fbar\ for various simulations. 
Simulations O1, O3, P1, P2, Q1, Q and K3 all have 150 kpc cells and lie on
a diagonal line below and to the left of the other simulations.
They are joined by M that has broad lines because it has low \sig\
and high \gammahe\ values.
The remaining simulations show a strong correlation, with larger \pmed\
for more absorption.
}
\label{fig:Pm15Fbar.fig}
\end{figure}

\clearpage
\begin{figure}
{\plotone{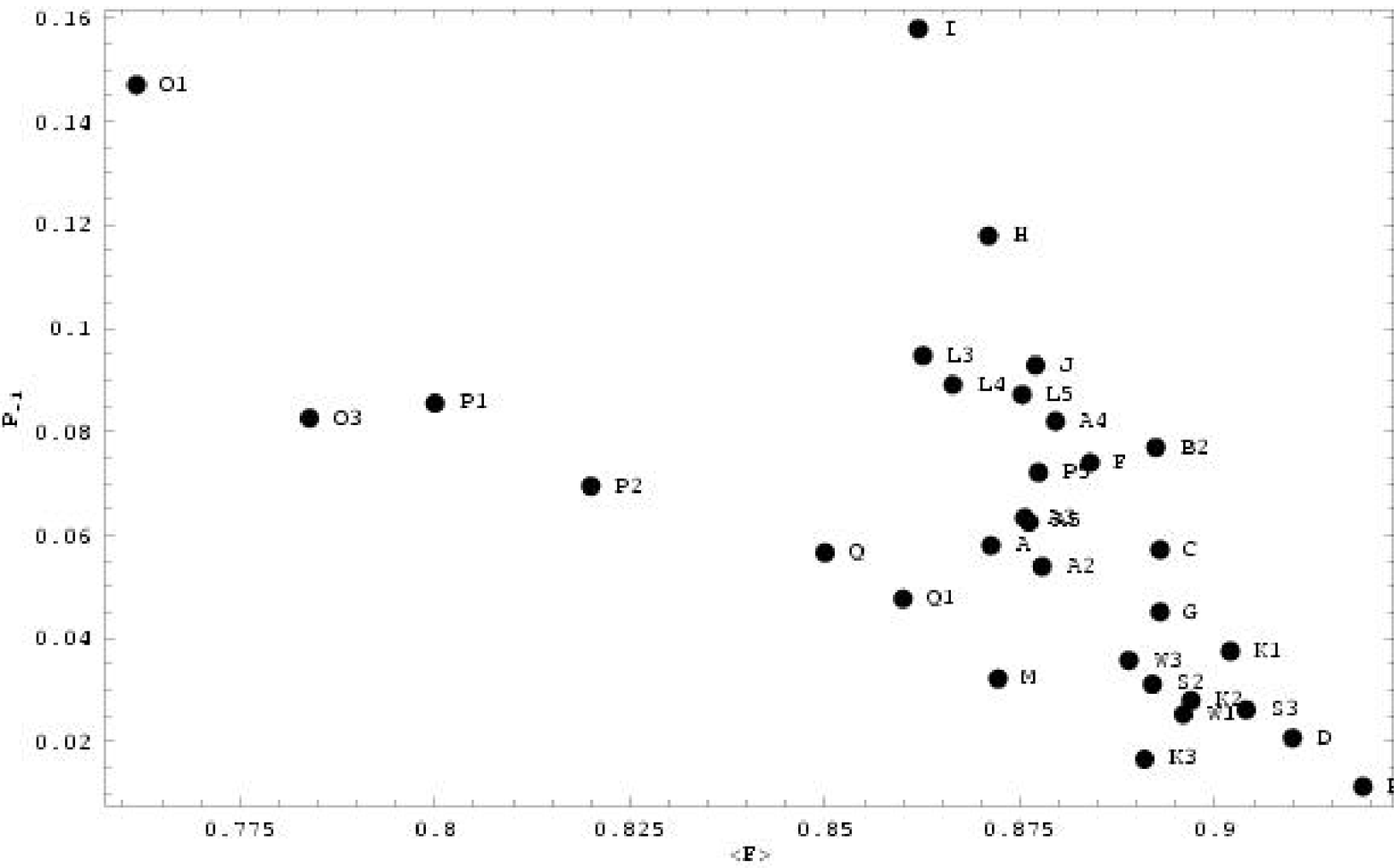}}
\caption{\NOTE{Pm1Fbar.ps}
As Fig. \ref{fig:Pm15Fbar.fig} but for \ps .
Simulations O1, O3, P1, P2, Q1, Q and K3 all have 150 kpc cells and lie on
a diagonal line below and to the left of the other simulations.
They are joined by M that has broad lines because it has low \sig\
and high \gammahe\ values.
Simulations in the upper right (I, H, J, F, C, G, K1) all have 37.5 kpc
cells, while B2 has 18.75 kpc cells.
For a given cell size the simulations show a correlation, with larger \ps\
for more absorption. There is more scatter than for \pl\ and \pmed\
because factors such as cell
size and \gammahe\ have a larger effect on \ps\ than on \pl\ and \pmed .
}
\label{fig:Pm1Fbar.fig}
\end{figure}

\clearpage
\begin{figure}
{\plotone{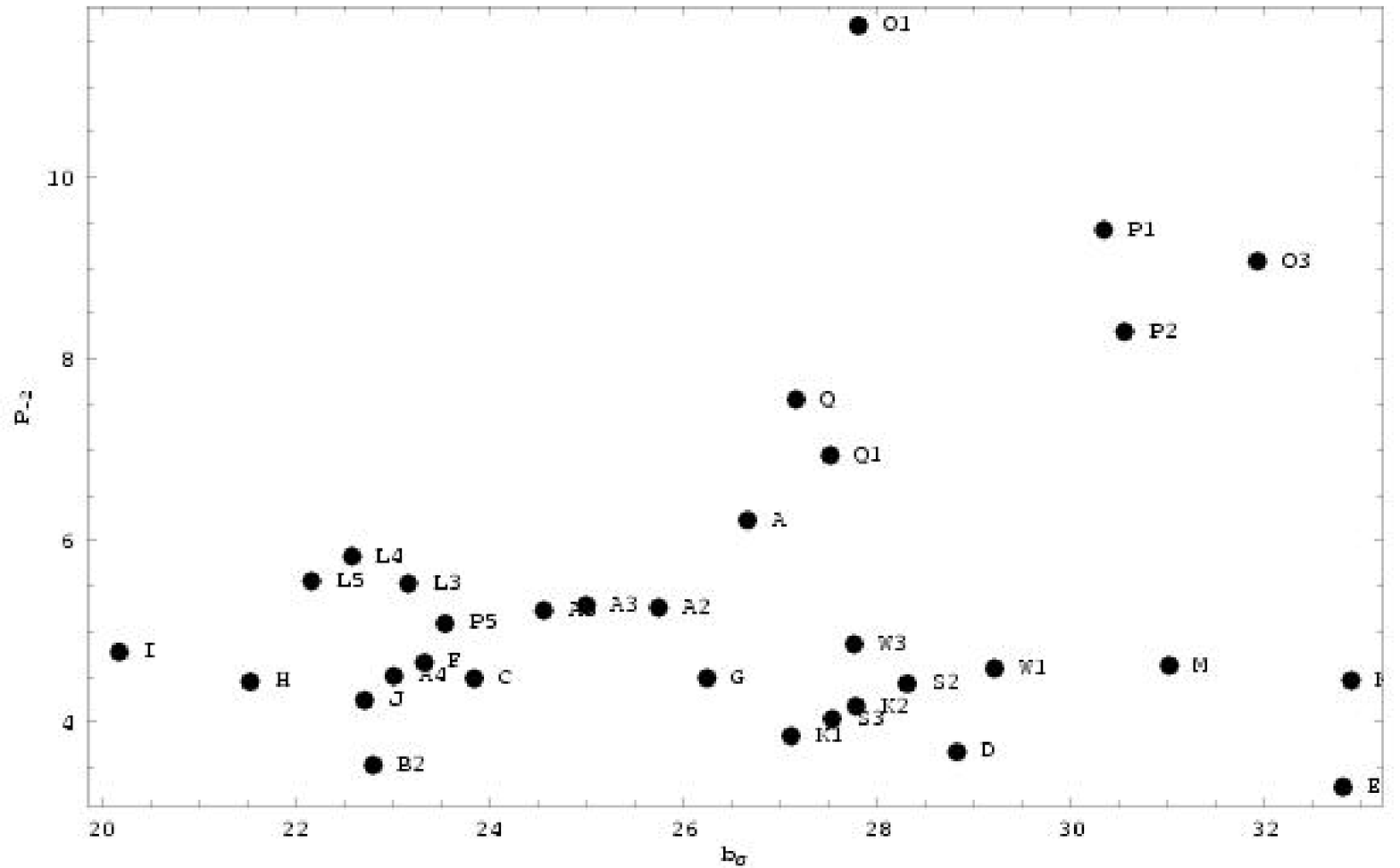}}
\caption{\NOTE{Pm2bsig.ps}
A scatter plot showing the correlation of a pair of output parameters,
\pl\ against \bsig\ for various simulations. 
Simulations O1, O3, P1, P2, Q1, Q and K3 all have 150 kpc cells and have
larger \pl\ values, above the other simulations on the plot.
The remaining simulations show a slight decrease of \pl\ with
increasing \bsig .
}
\label{fig:Pm2bsig.fig}
\end{figure}

\clearpage
\begin{figure}
{\plotone{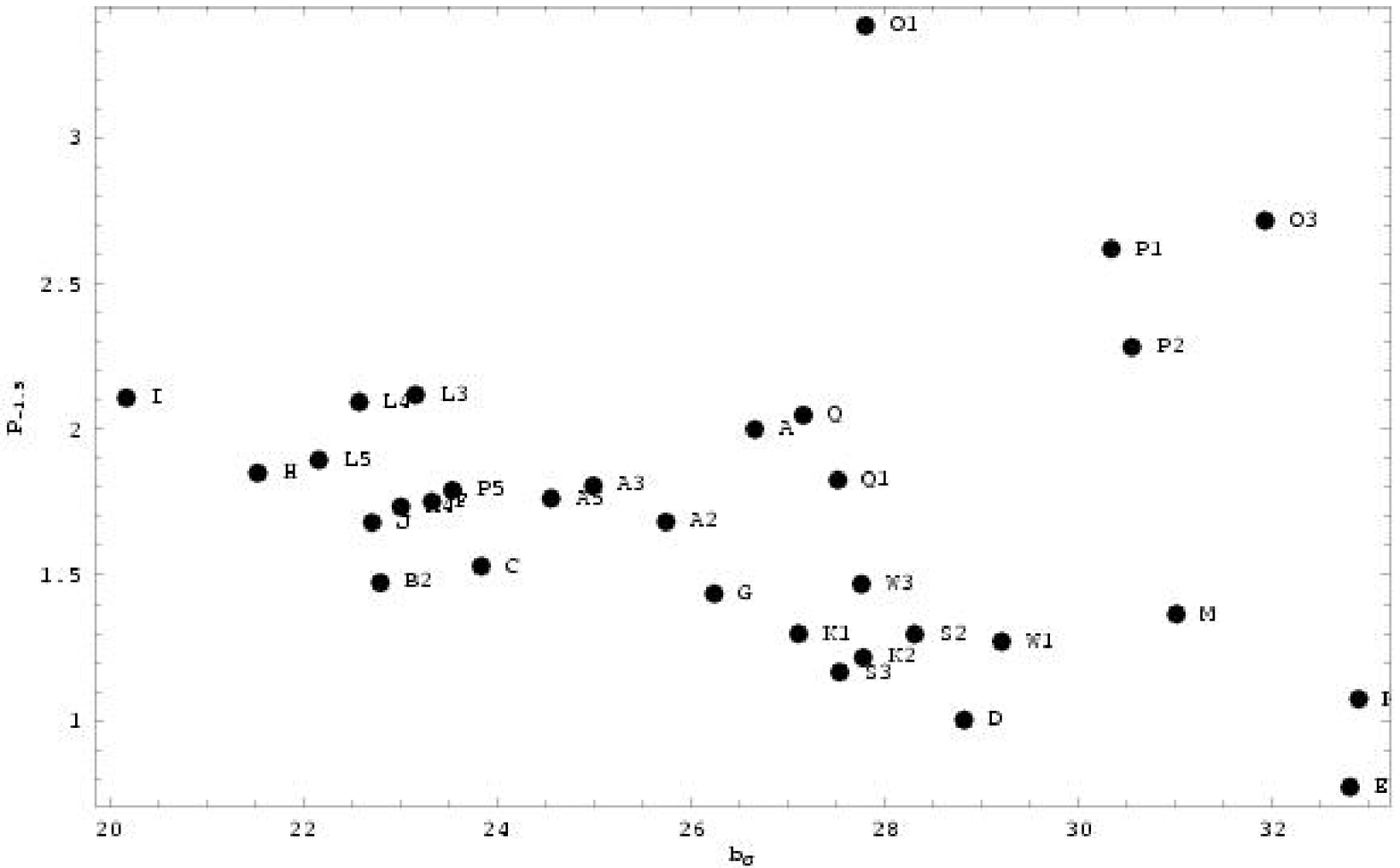}}
\caption{\NOTE{Pm15bsig.ps}
As Fig. \ref{fig:Pm2bsig.fig} but for \pmed . 
}
\label{fig:Pm15bsig.fig}
\end{figure}

\clearpage
\begin{figure}
{\plotone{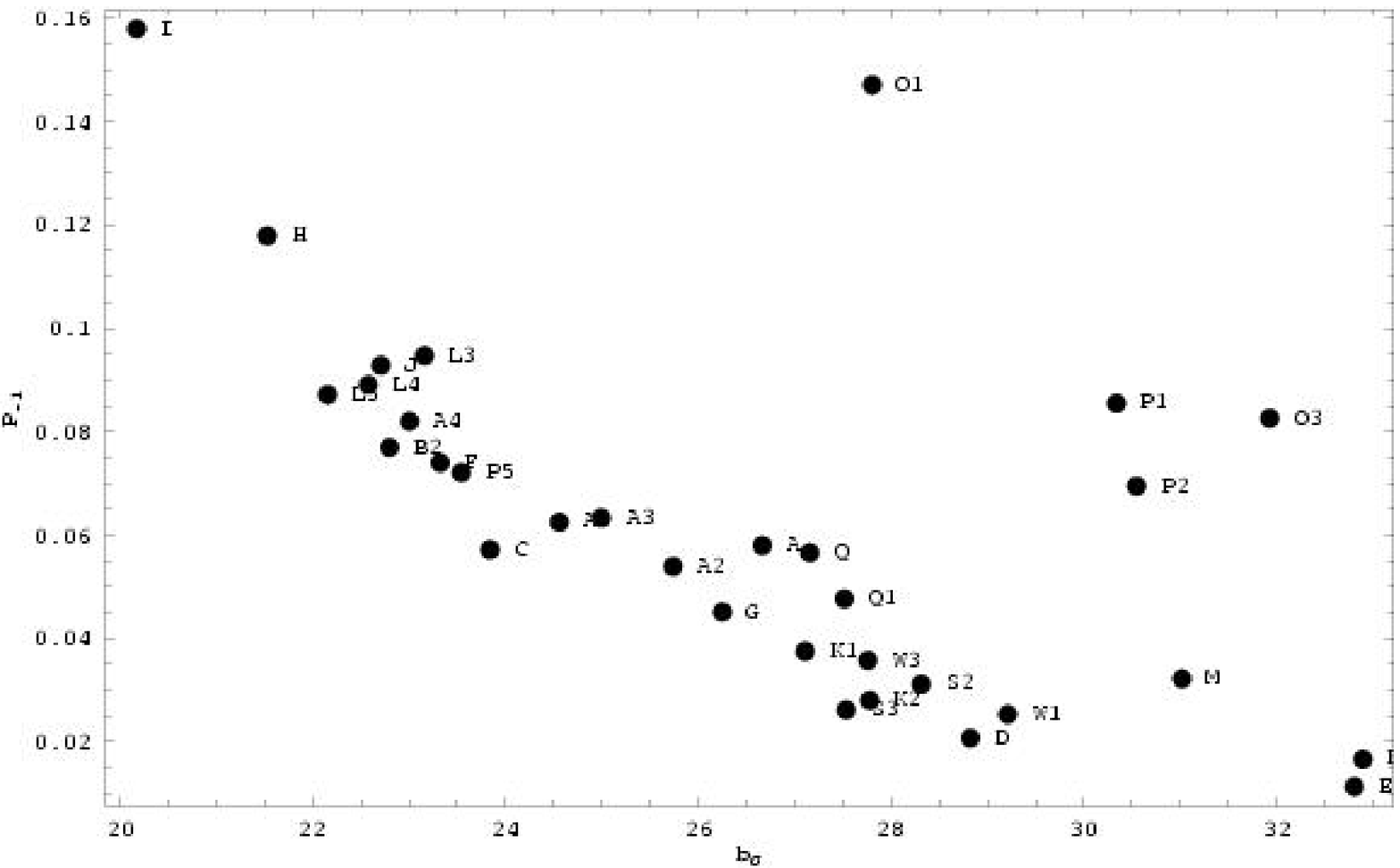}}
\caption{\NOTE{Pm1bsig.ps}
As Fig. \ref{fig:Pm2bsig.fig} but for \ps . 
}
\label{fig:Pm1bsig.fig}
\end{figure}

\begin{figure}
{\plotone{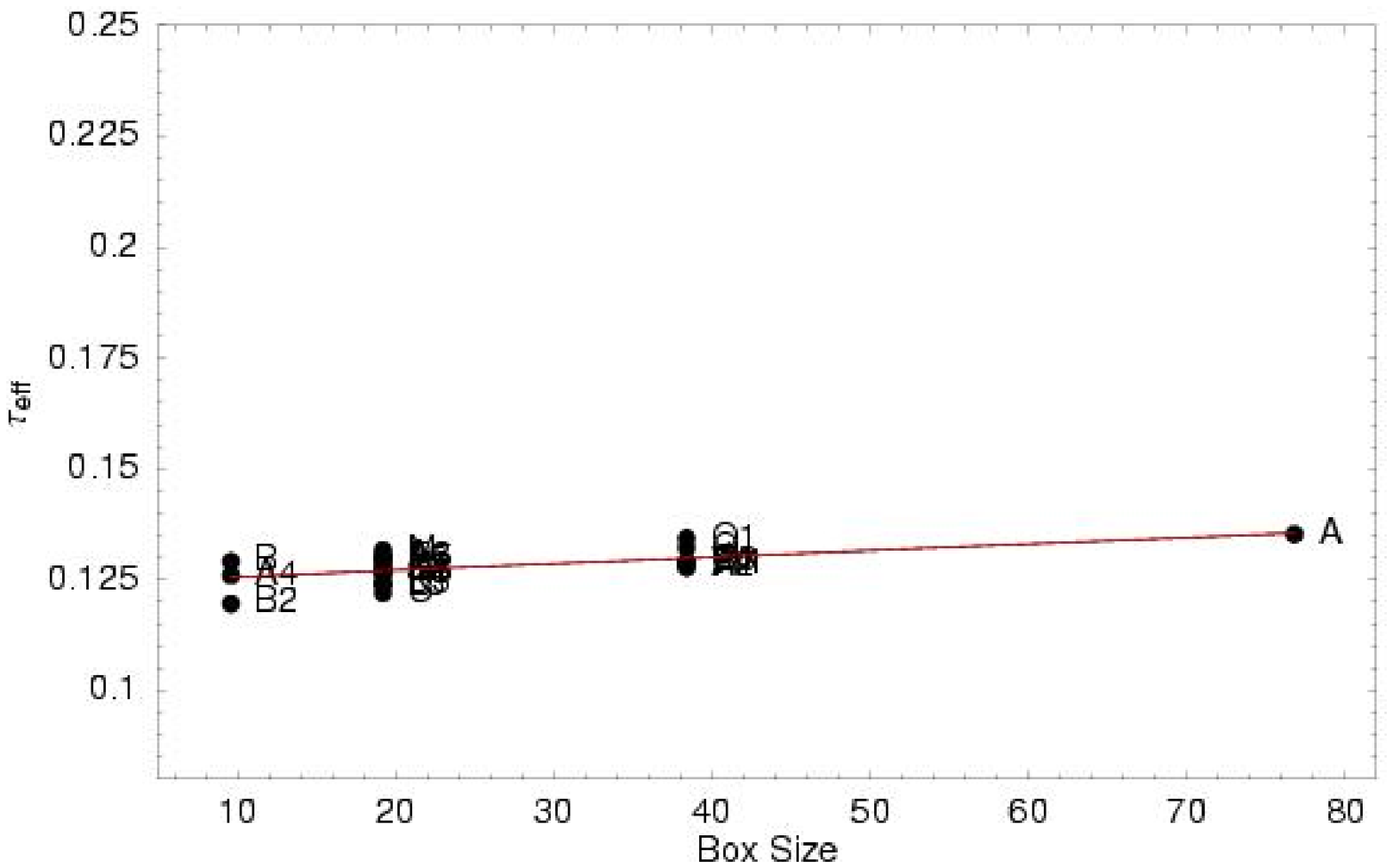}}
\caption{\NOTE{boxFbar.ps}
The output parameter \taueff\ after scaling using the relations in
Table \ref{tbl:scalings} to scale to our standard parameters, listed in
Table \ref{tbl:standard}. We do not scale by the quantity shown on the
horizontal axis, in this case box size.
}
\label{fig:boxFbar.fig}
\end{figure}

\clearpage
\begin{figure}
{\plotone{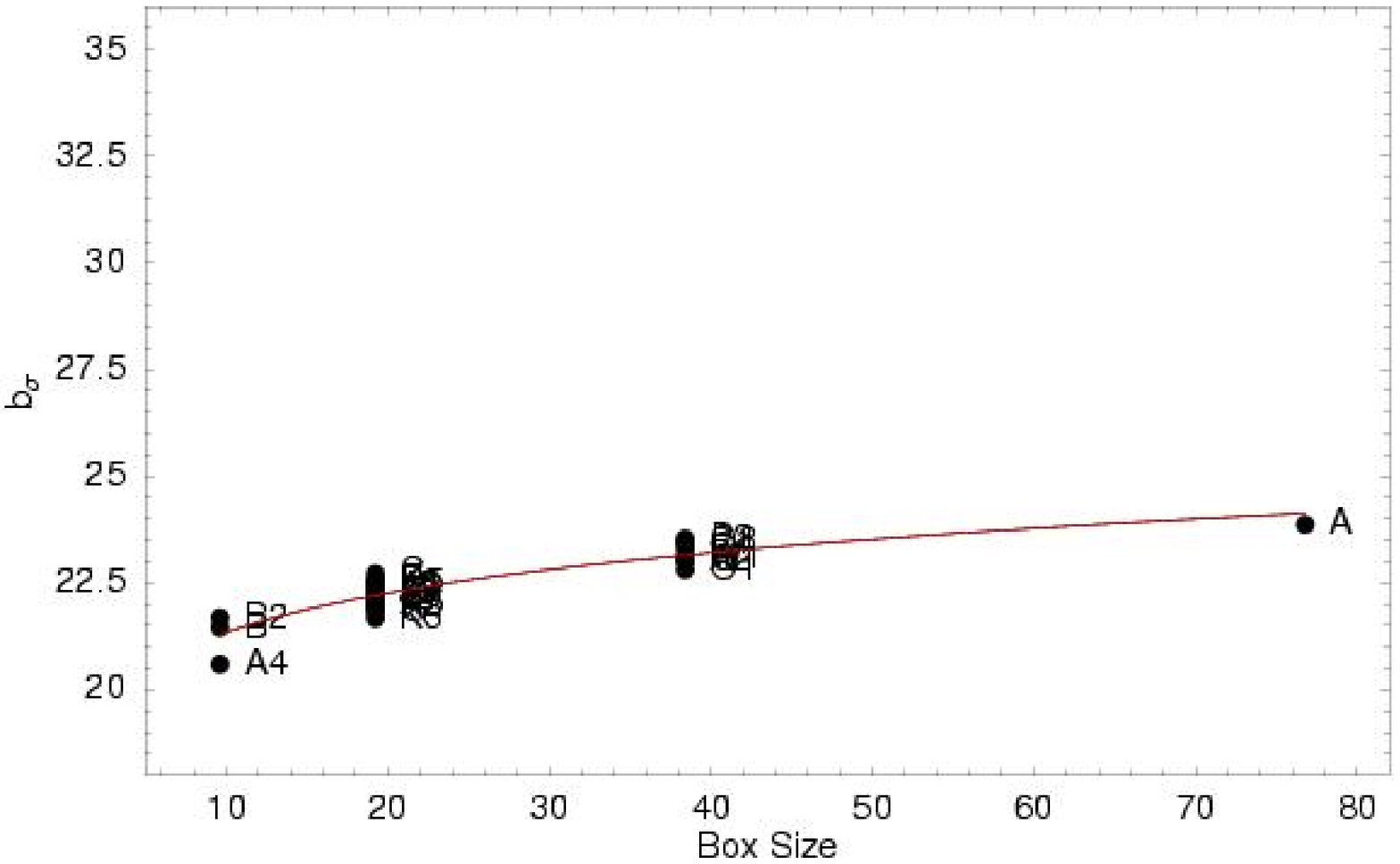}}
\caption{\NOTE{boxbsig.ps}
As Fig. \ref{fig:boxFbar.fig} but for \bsig\ against box size.
}
\label{fig:boxbsig.fig}
\end{figure}

\clearpage
\begin{figure}
{\plotone{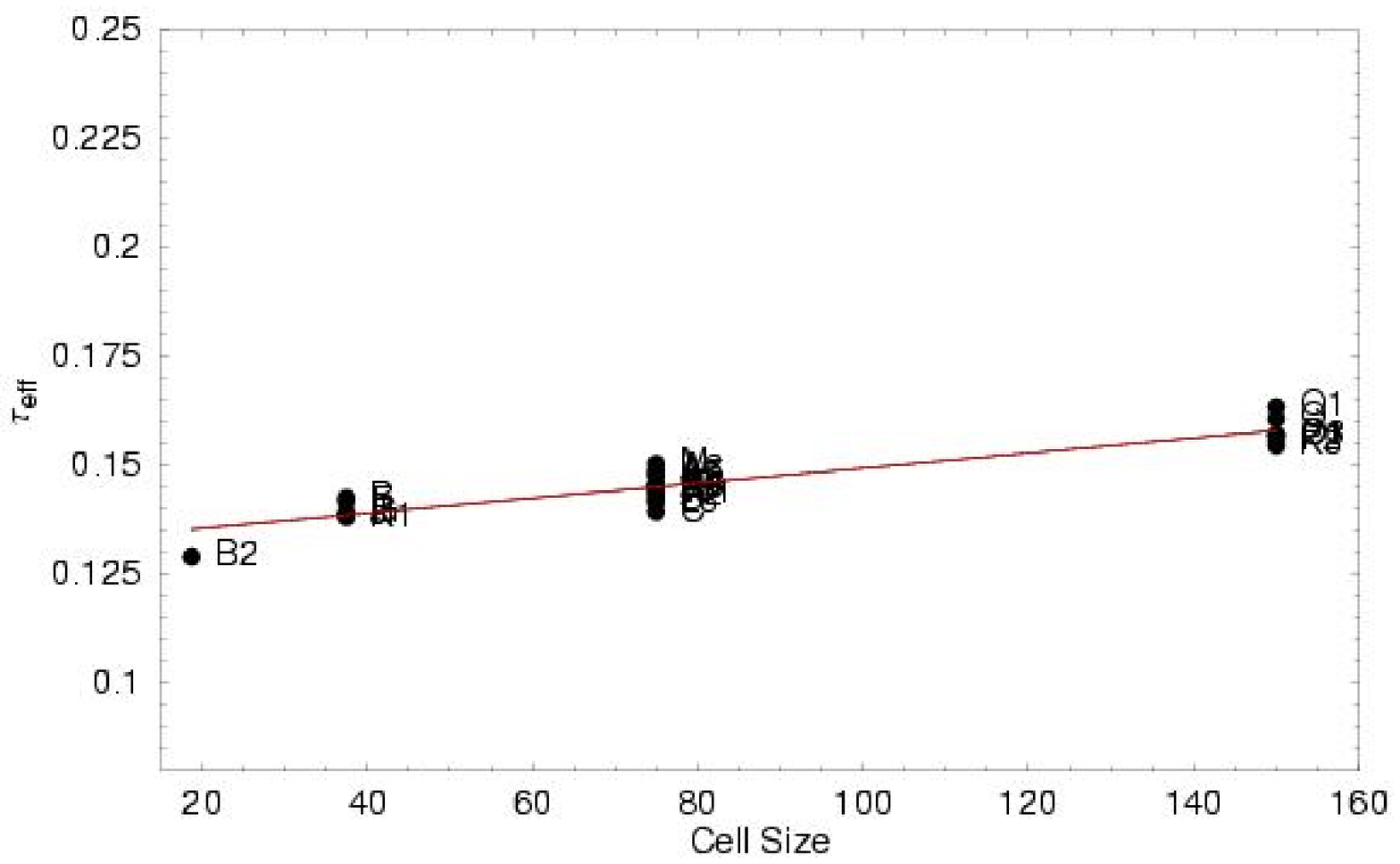}}
\caption{\NOTE{cellFbar.ps}
As Fig. \ref{fig:boxFbar.fig} but for \taueff\ against cell size.
}
\label{fig:cellFbar.fig}
\end{figure}

\clearpage
\begin{figure}
{\plotone{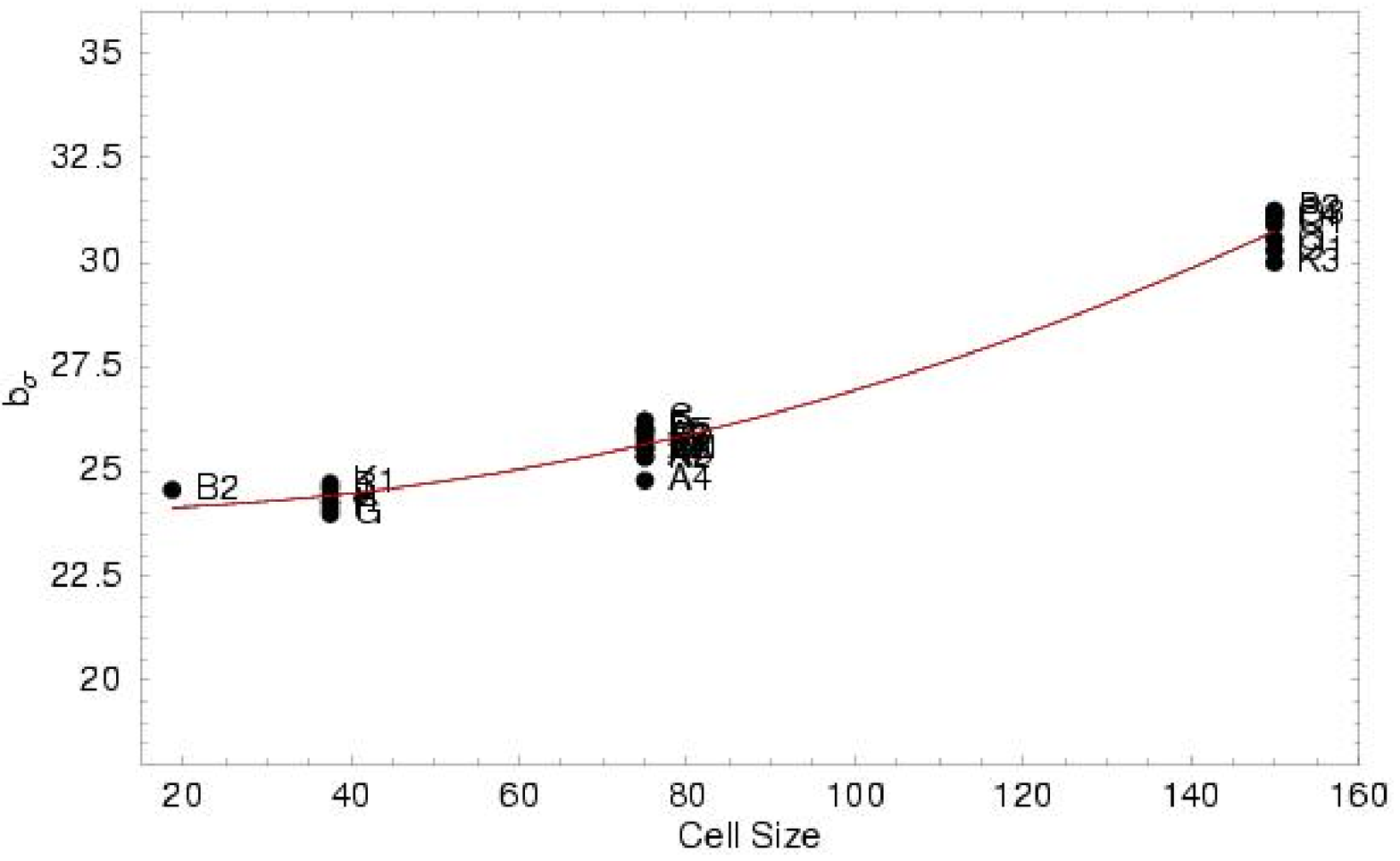}}
\caption{\NOTE{cellbsig.ps}
As Fig. \ref{fig:boxFbar.fig} but for \bsig\ against cell size.
}
\label{fig:cellbsig.fig}
\end{figure}

\clearpage
\begin{figure}
{\plotone{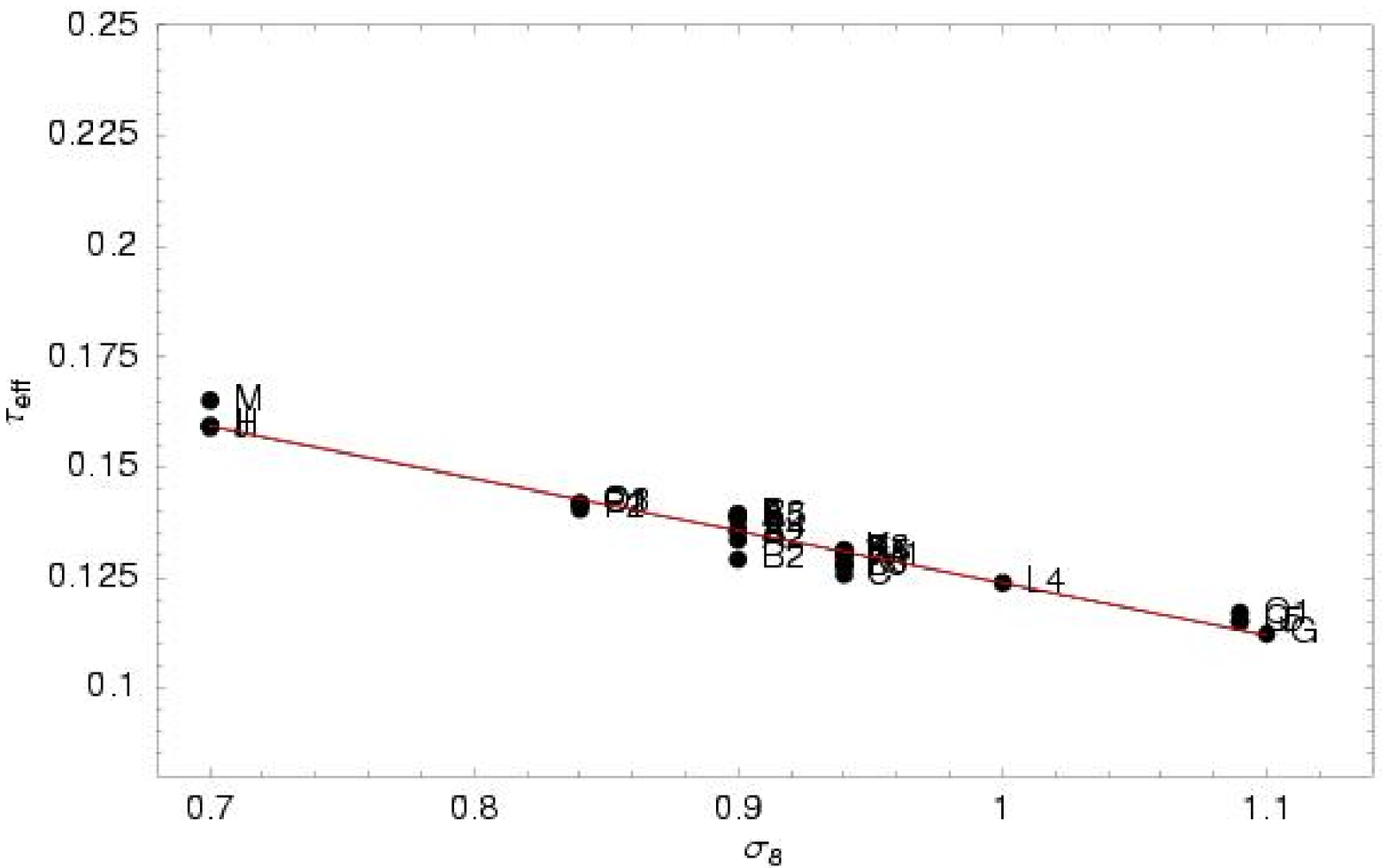}}
\caption{\NOTE{sigma8Fbar.ps}
As Fig. \ref{fig:boxFbar.fig} but for \taueff\ against \sig .
}
\label{fig:sigma8Fbar.fig}
\end{figure}

\clearpage
\begin{figure}
{\plotone{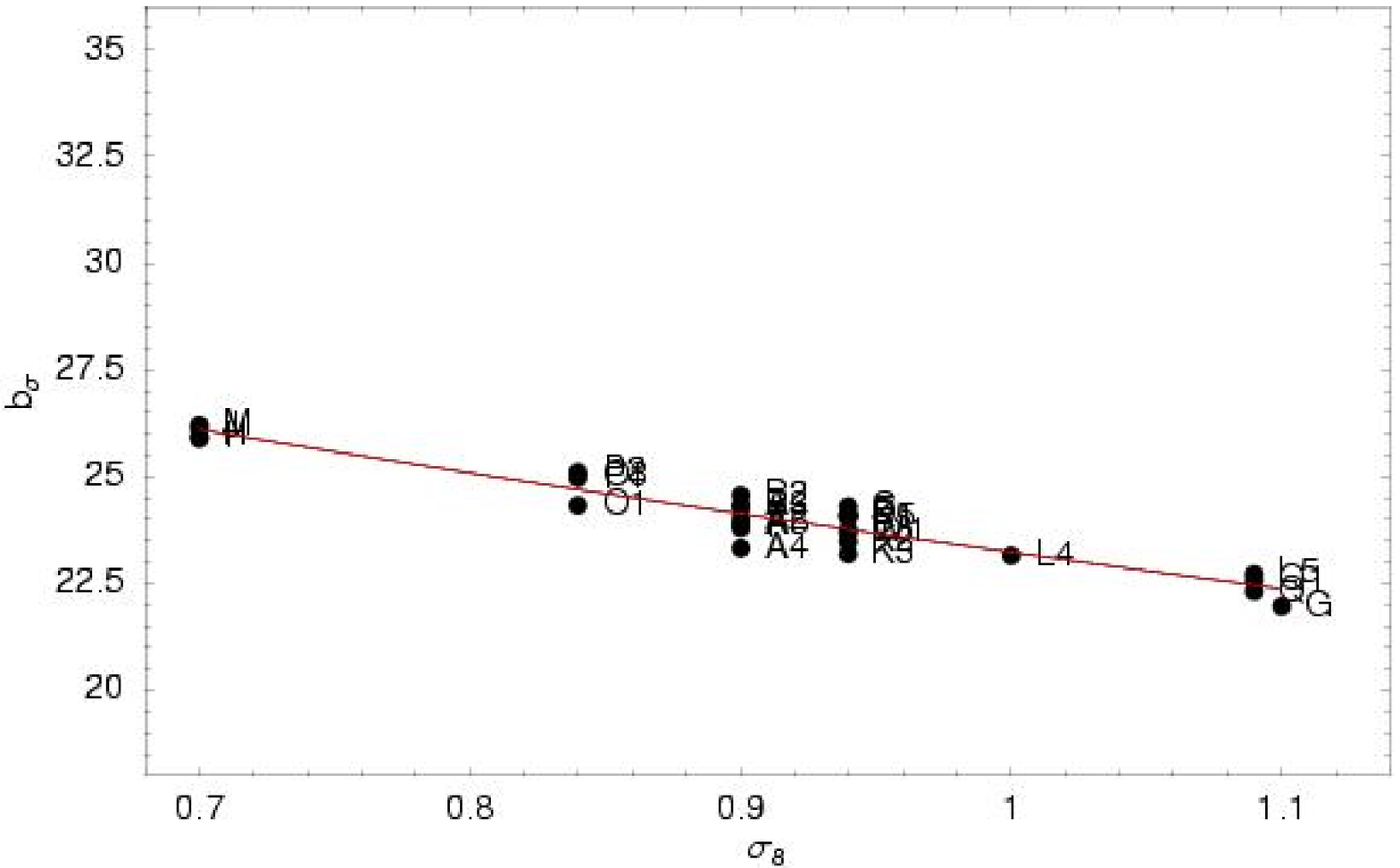}}
\caption{\NOTE{sigma8bsig.ps}
As Fig. \ref{fig:boxFbar.fig} but for \bsig\ against \sig .
}
\label{fig:sigma8bsig.fig}
\end{figure}

\clearpage
\begin{figure}
{\plotone{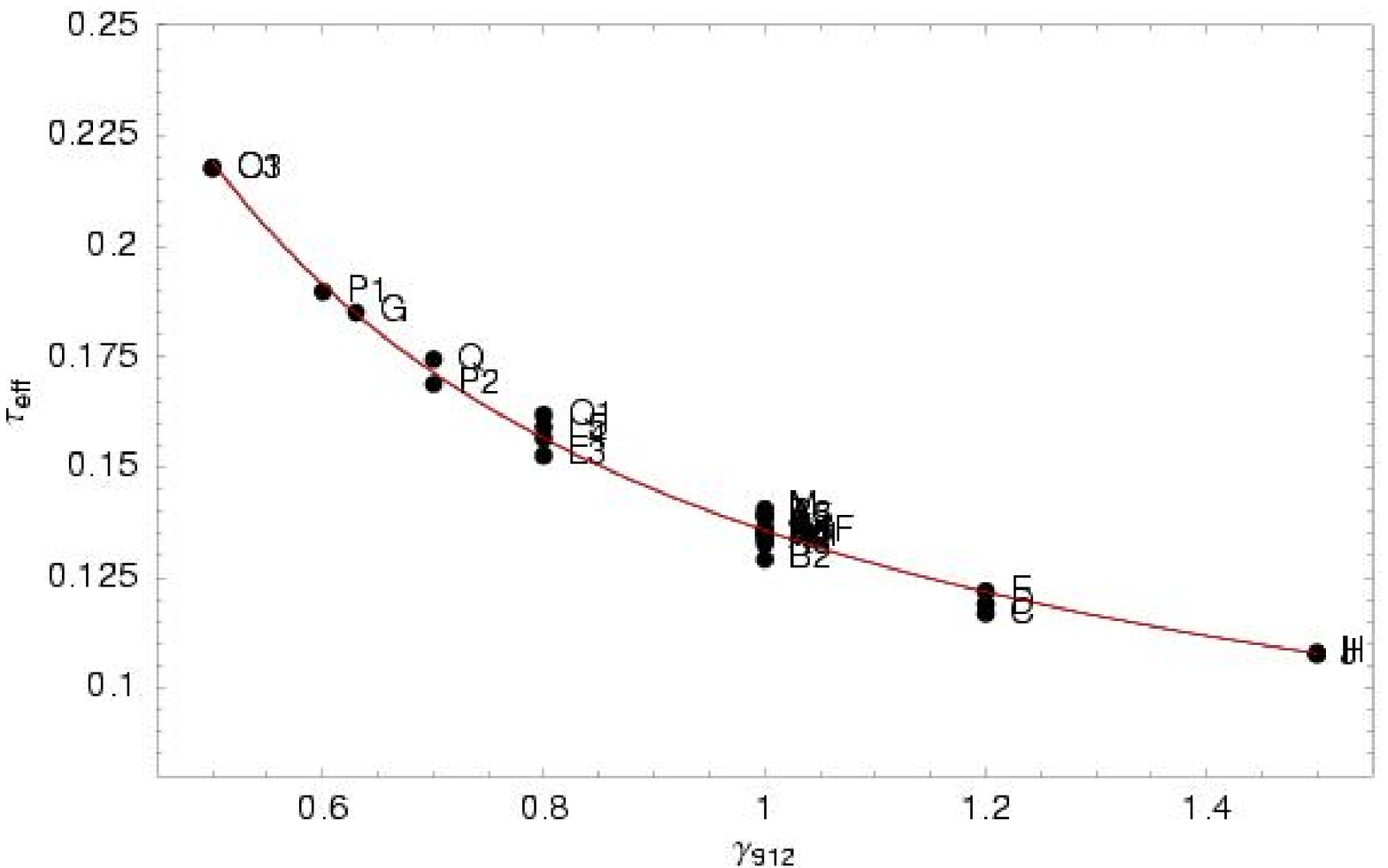}}
\caption{\NOTE{gamma912Fbar.ps}
As Fig. \ref{fig:boxFbar.fig} but for \taueff\ against \gammah .
}
\label{fig:gamma912Fbar.fig}
\end{figure}

\clearpage
\begin{figure}
{\plotone{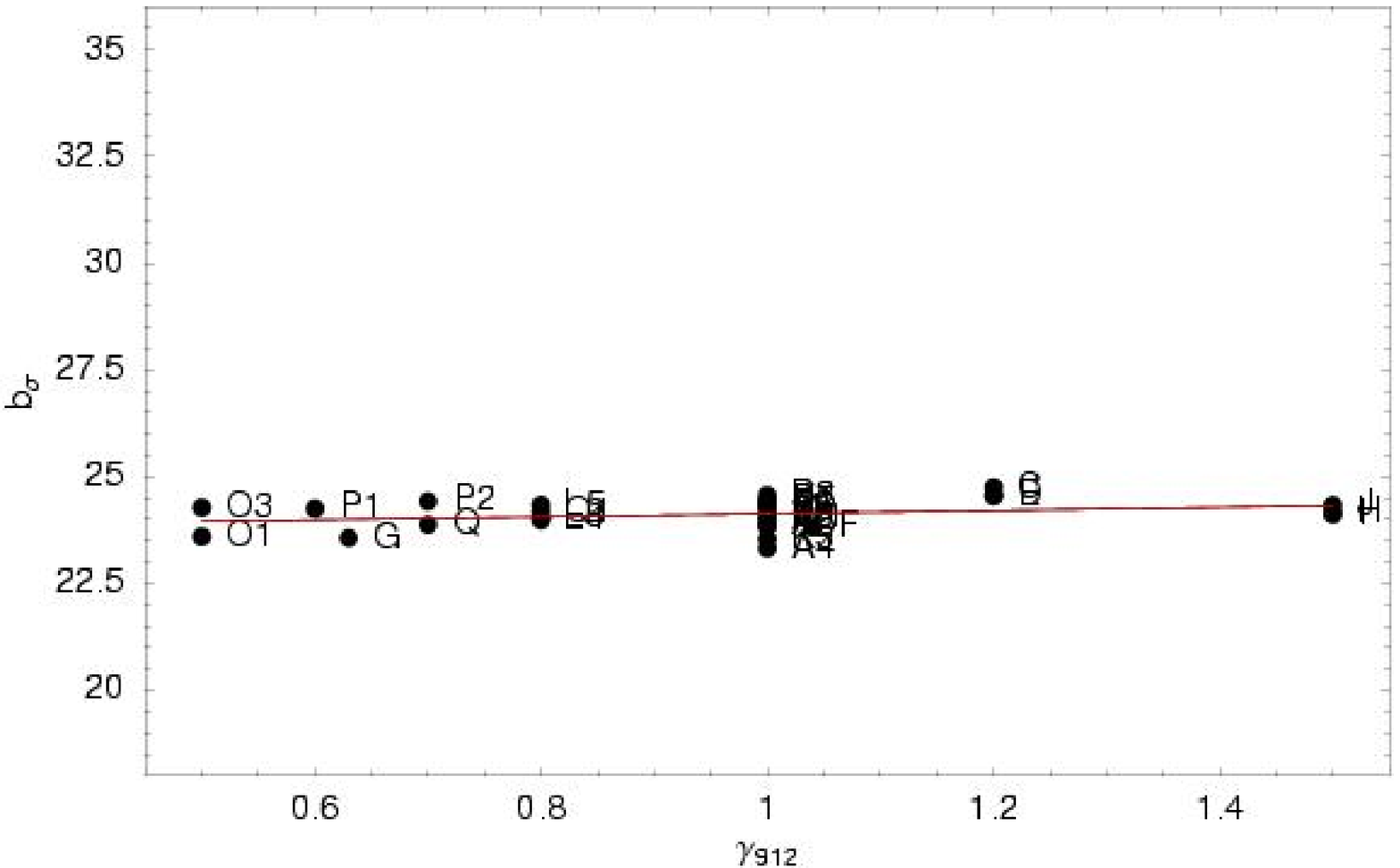}}
\caption{\NOTE{gamma912bsig.ps}
As Fig. \ref{fig:boxFbar.fig} but for \bsig\ against \gammah .
}
\label{fig:gamma912bsig.fig}
\end{figure}

\clearpage
\begin{figure}
{\plotone{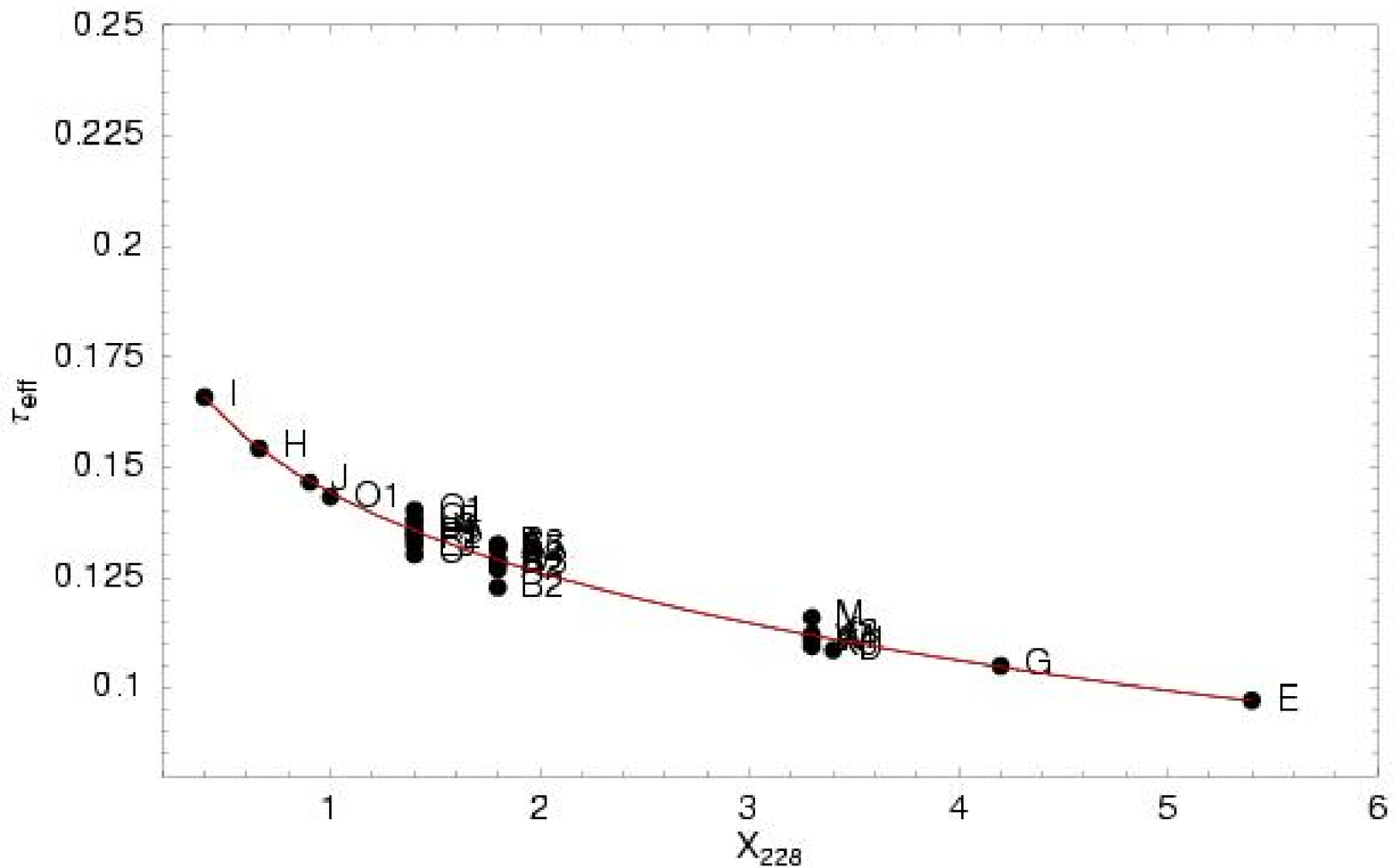}}
\caption{\NOTE{gamma228Fbar.ps}
As Fig. \ref{fig:boxFbar.fig} but for \taueff\ against \gammahe .
}
\label{fig:gamma228Fbar.fig}
\end{figure}

\clearpage
\begin{figure}
{\plotone{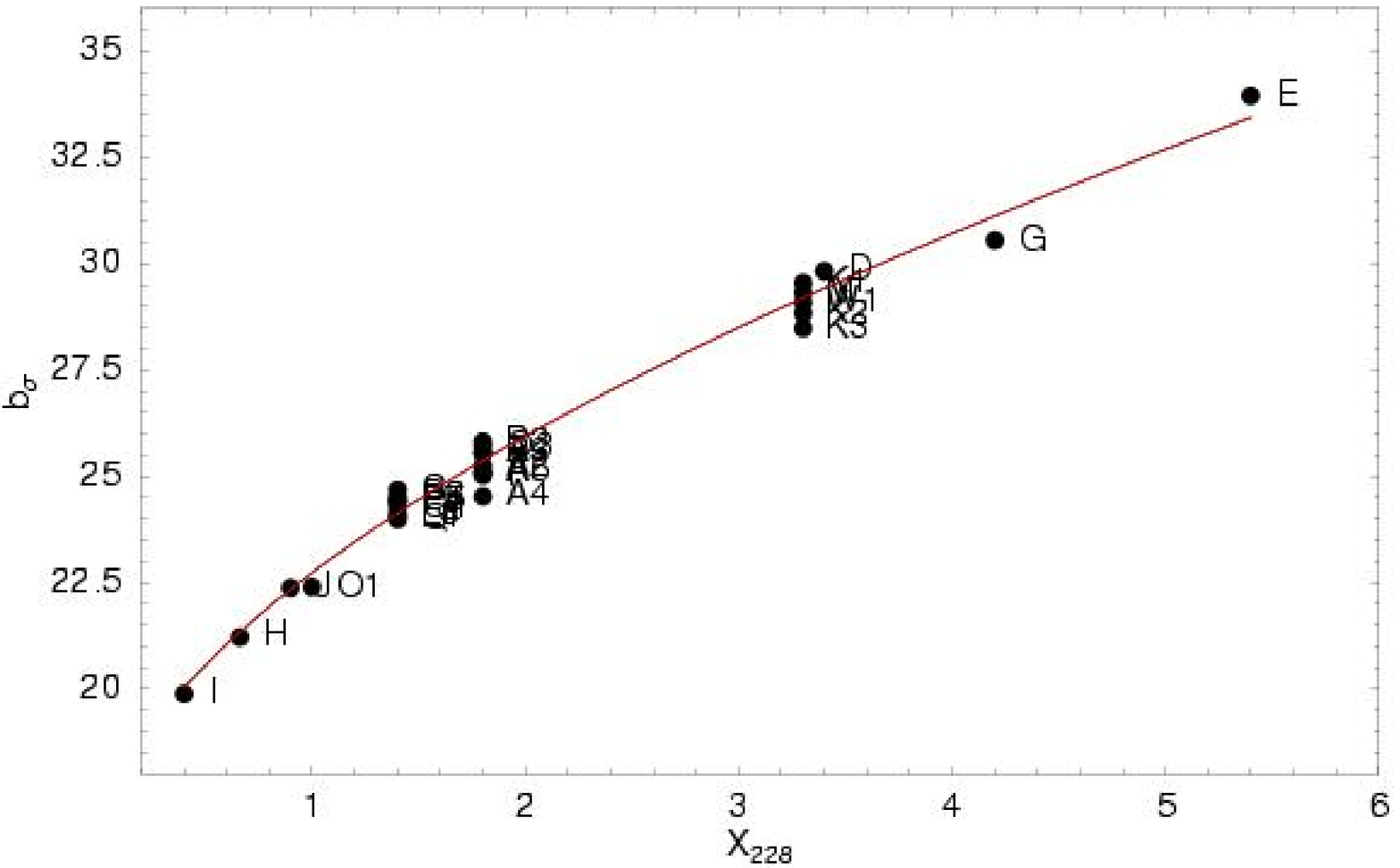}}
\caption{\NOTE{gamma228bsig.ps}
As Fig. \ref{fig:boxFbar.fig} but for \bsig\ against \gammahe .
}
\label{fig:gamma228bsig.fig}
\end{figure}

\clearpage
\begin{figure}
{\plotone{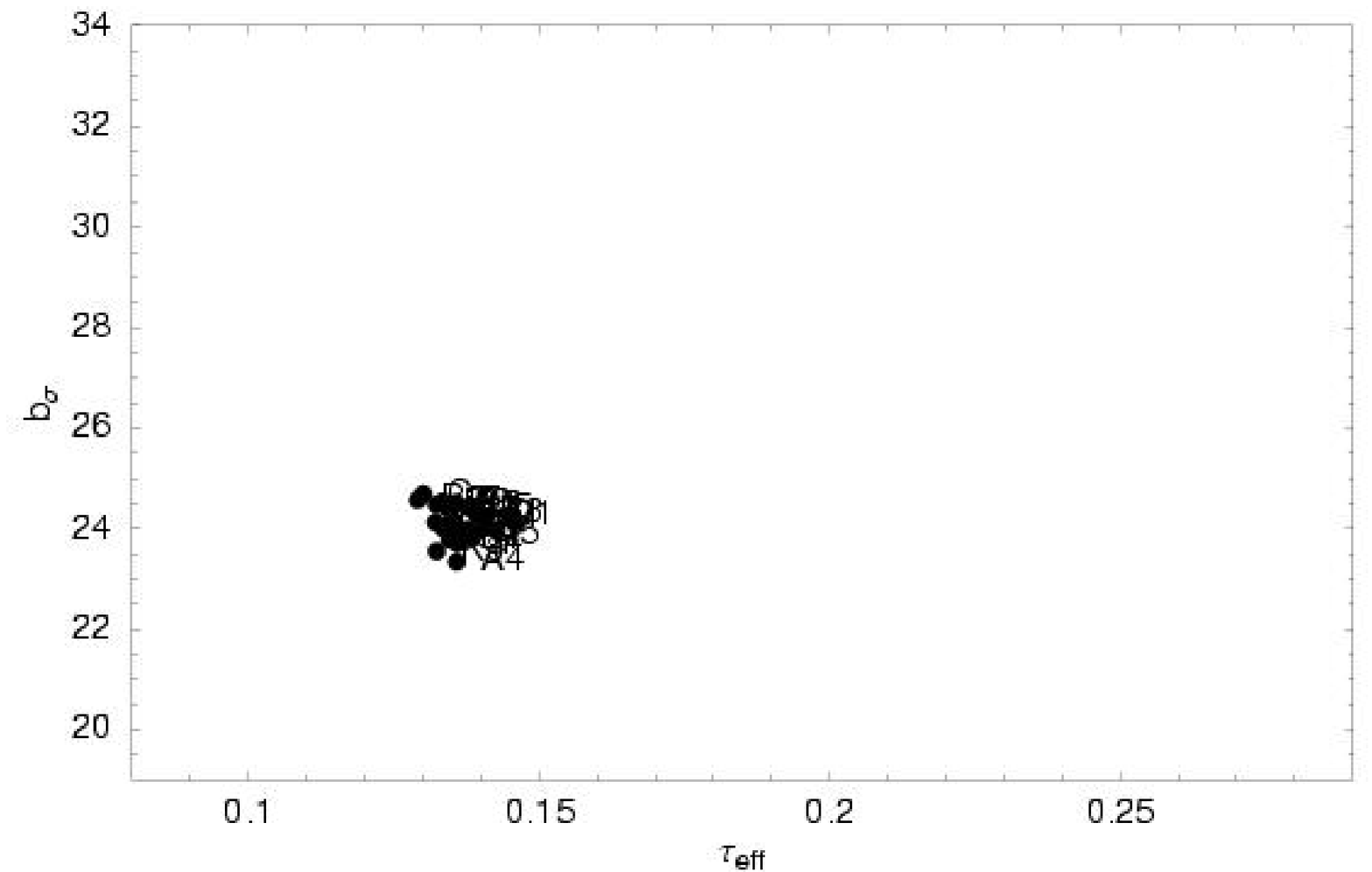}}
\caption{\NOTE{bsigTau2.ps}
Line width parameter \bsig\ against effective optical depth for various
simulations after rescaling \taueff\ and \bsig\ to our standard model.
We show the same simulations as in Fig. \ref{fig:bsigTau1.fig}}
\label{fig:bsigTau2.fig}
\end{figure}
\clearpage

% in-out  section se.tex

\clearpage
\begin{figure}
{\plotone{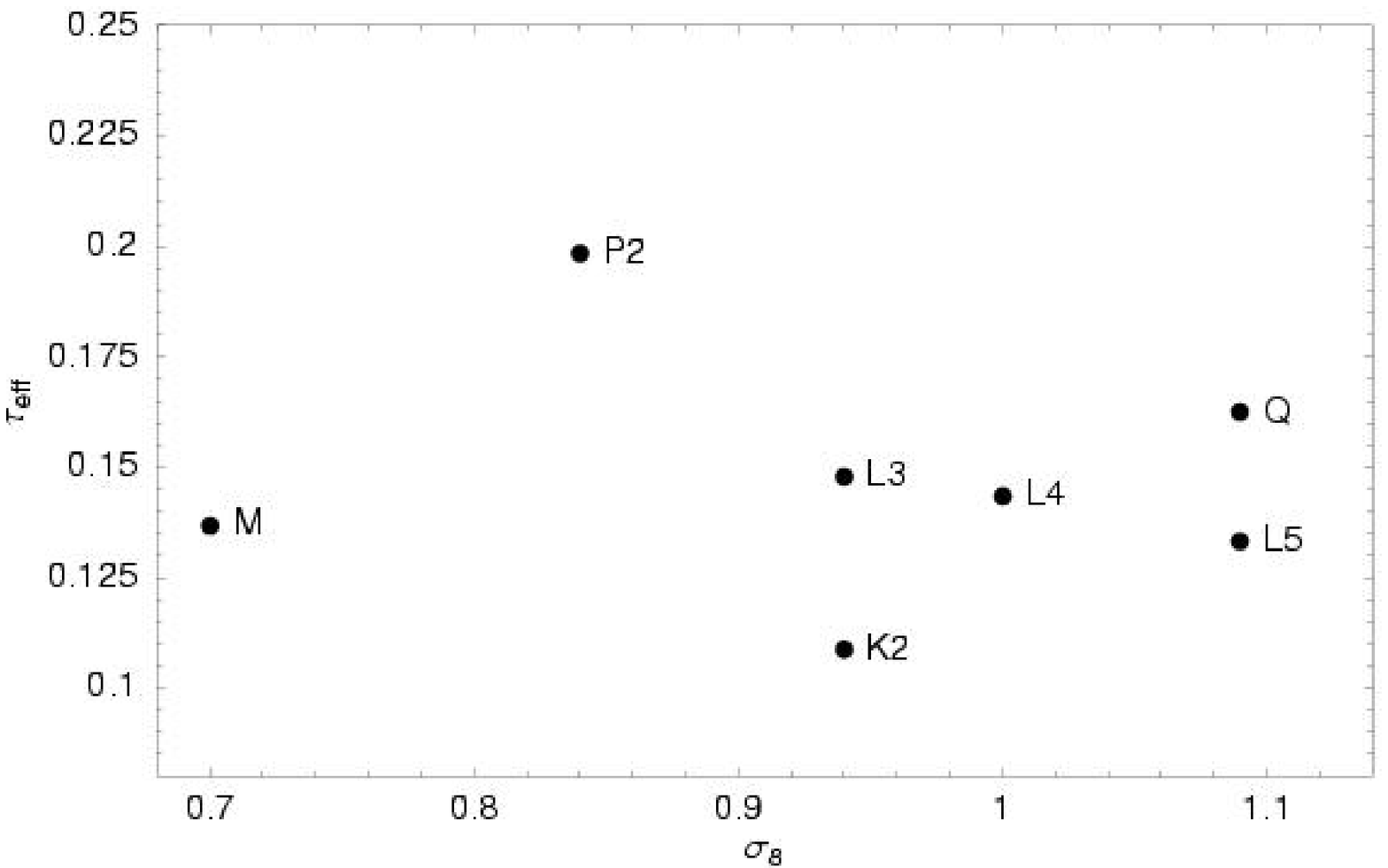}}
\caption{\NOTE{sigma8FbarRaw.ps}
Output parameter \taueff\ as a function of input \sig\ for three
sets of simulations that differ in \sig\ alone.
MK2 have 75 kpc cells and \gammahe $=3.3$.
The L series have 75 kpc cells and \gammahe $=1.4$.
P2Q have 150 kpc cells and \gammahe $=1.4$. 
}
\label{fig:sigma8FbarRaw.fig}
\end{figure}

\clearpage
\begin{figure}
{\plotone{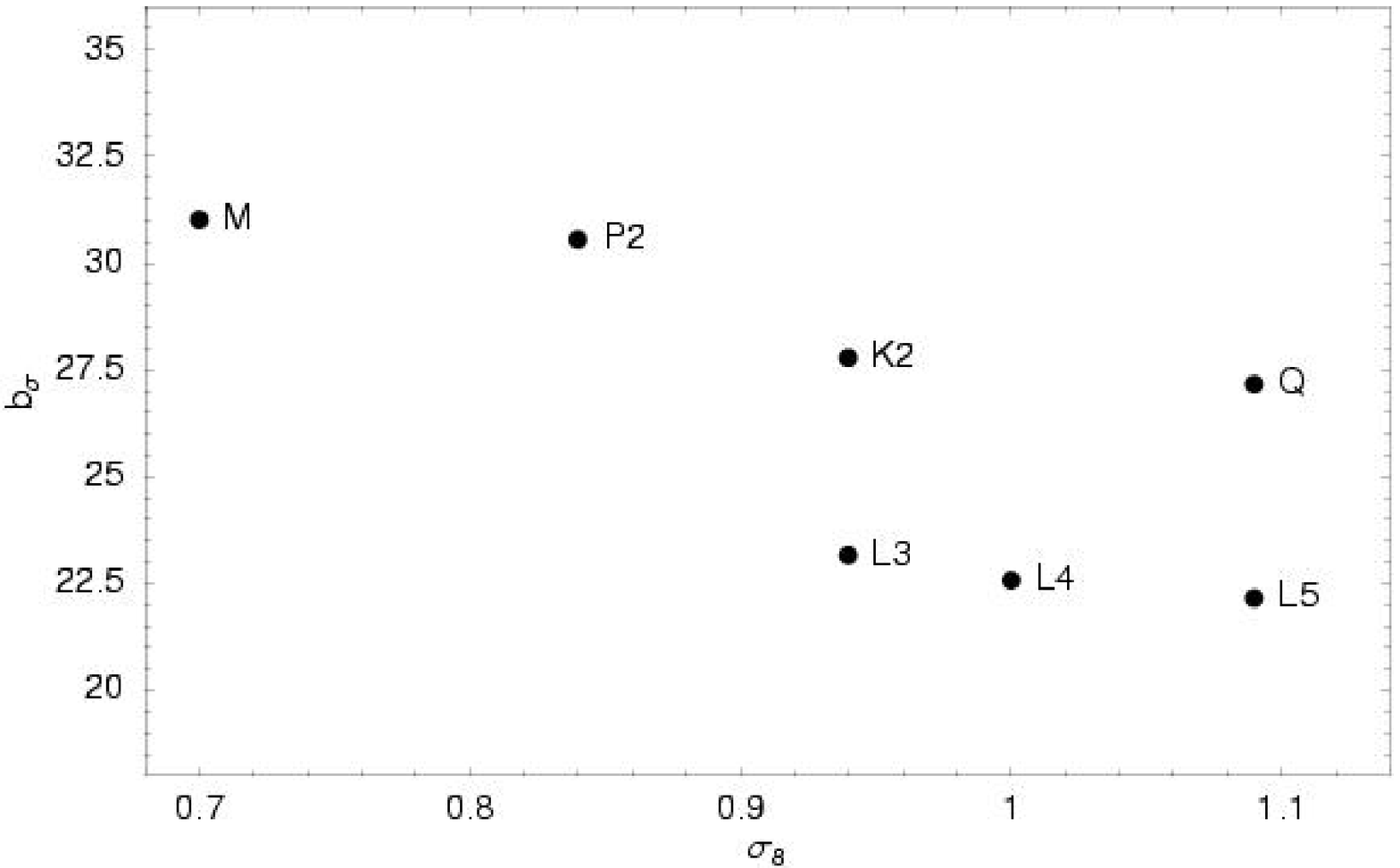}}
\caption{\NOTE{sigma8bsigRaw.ps}
Output parameter \bsig\ as a function of input \sig\ for three
sets of simulations that differ in \sig\ alone.
MK2 have 75 kpc cells and \gammahe $=3.3$.
The L series have 75 kpc cells and \gammahe $=1.4$.
P2Q have 150 kpc cells and \gammahe $=1.4$. 
}
\label{fig:sigma8bsigRaw.fig}
\end{figure}

\clearpage
\begin{figure}
{\plotone{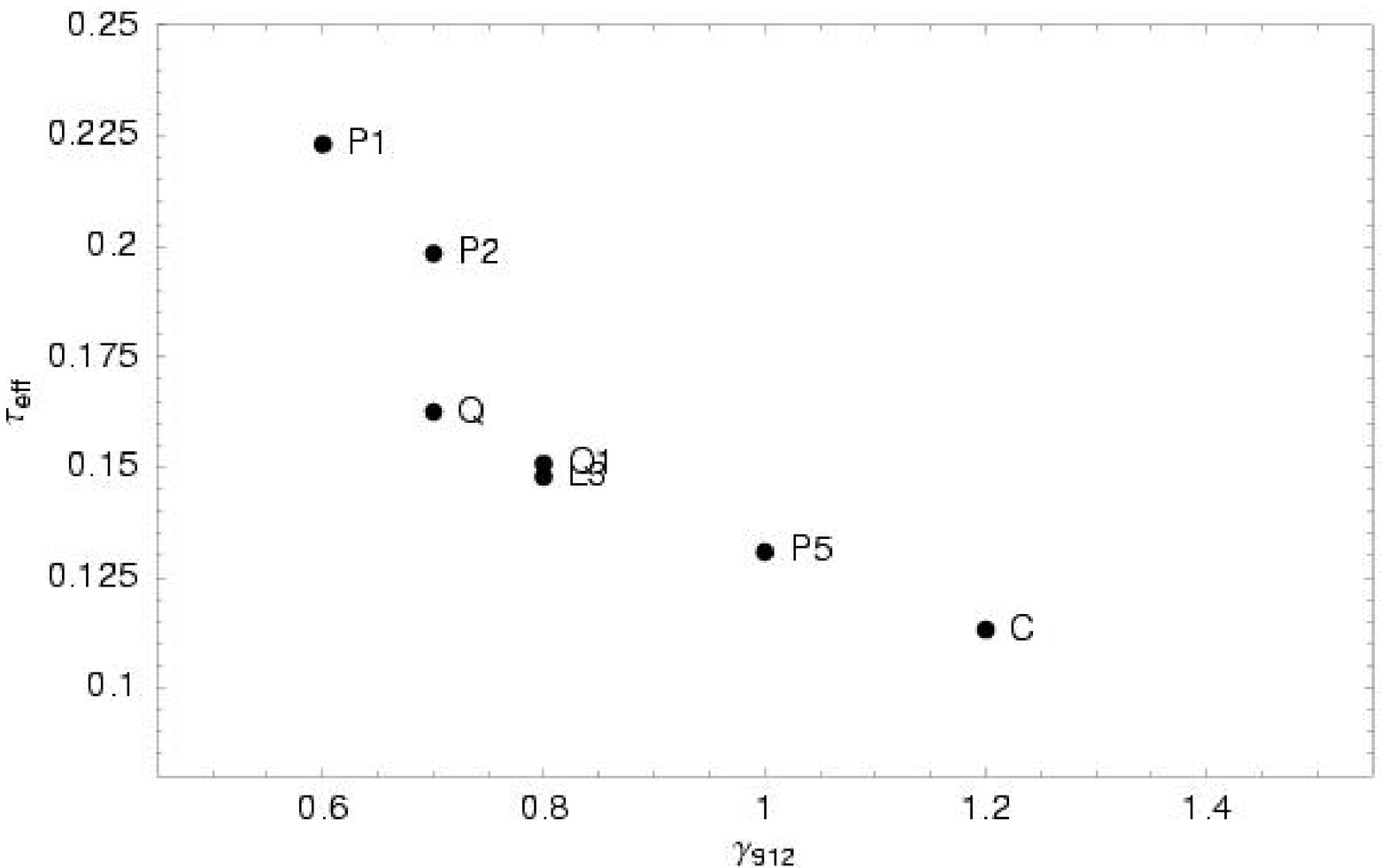}}
\caption{\NOTE{gamma912FbarRaw.ps}
Output parameter \taueff\ as a function of input \gammah\ for three
sets of simulations that differ in \sig\ alone.
L3P5C have 75 kpc cells, \gammahe $1.4$ and \sig $=0.94$.
P1P2 have 150 kpc cells and \gammahe $1.4$ and \sig $=0.84$. 
QQ1 have 150 kpc cells and \gammahe $1.4$ and \sig $=1.09$. 
}
\label{fig:gamma912FbarRaw.fig}
\end{figure}

\clearpage
\begin{figure}
{\plotone{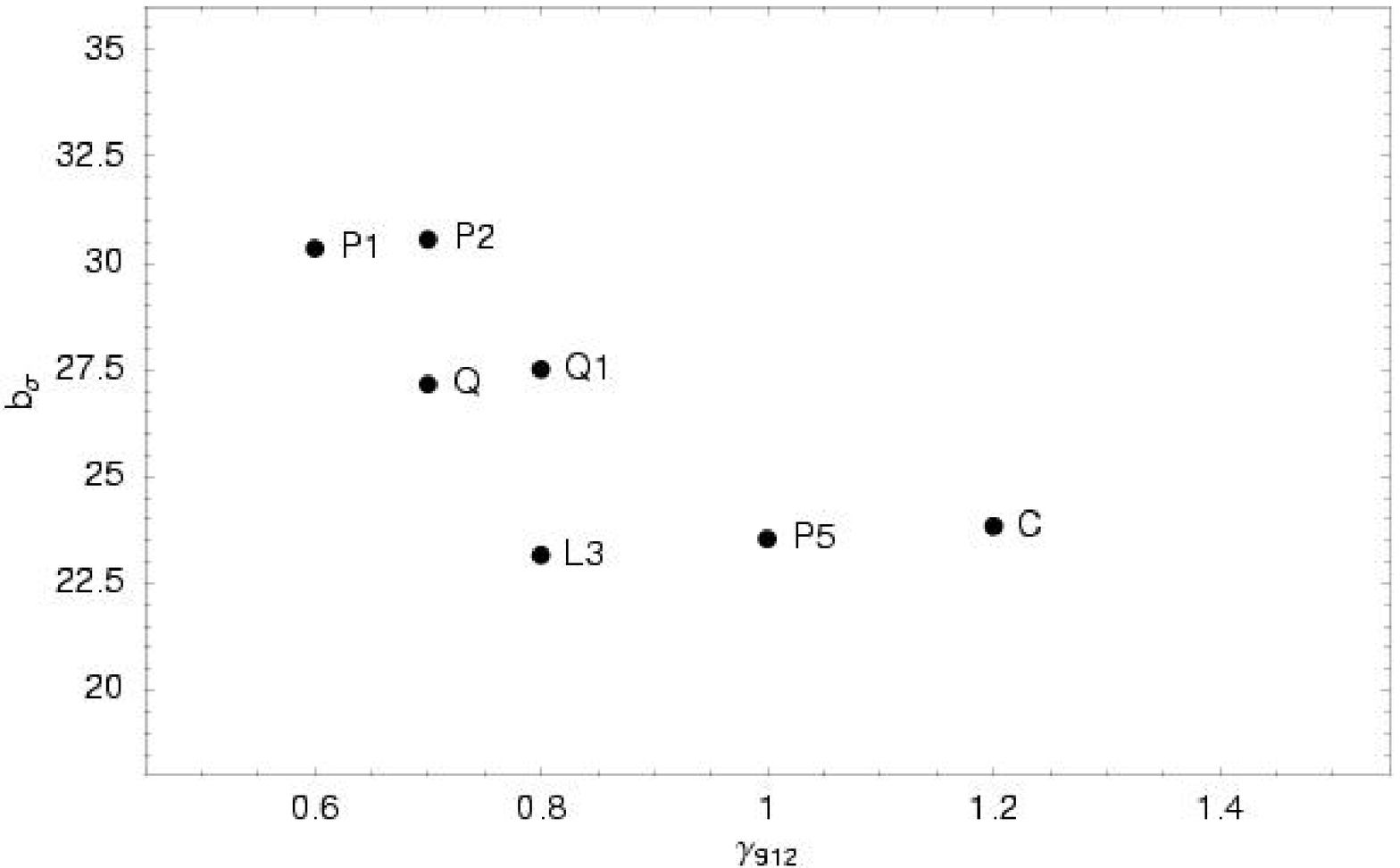}}
\caption{\NOTE{gamma912bsigRaw.ps}
As Fig. \ref{fig:gamma912FbarRaw.fig} but for \bsig .
}
\label{fig:gamma912bsigRaw.fig}
\end{figure}

\clearpage
\begin{figure}
{\plotone{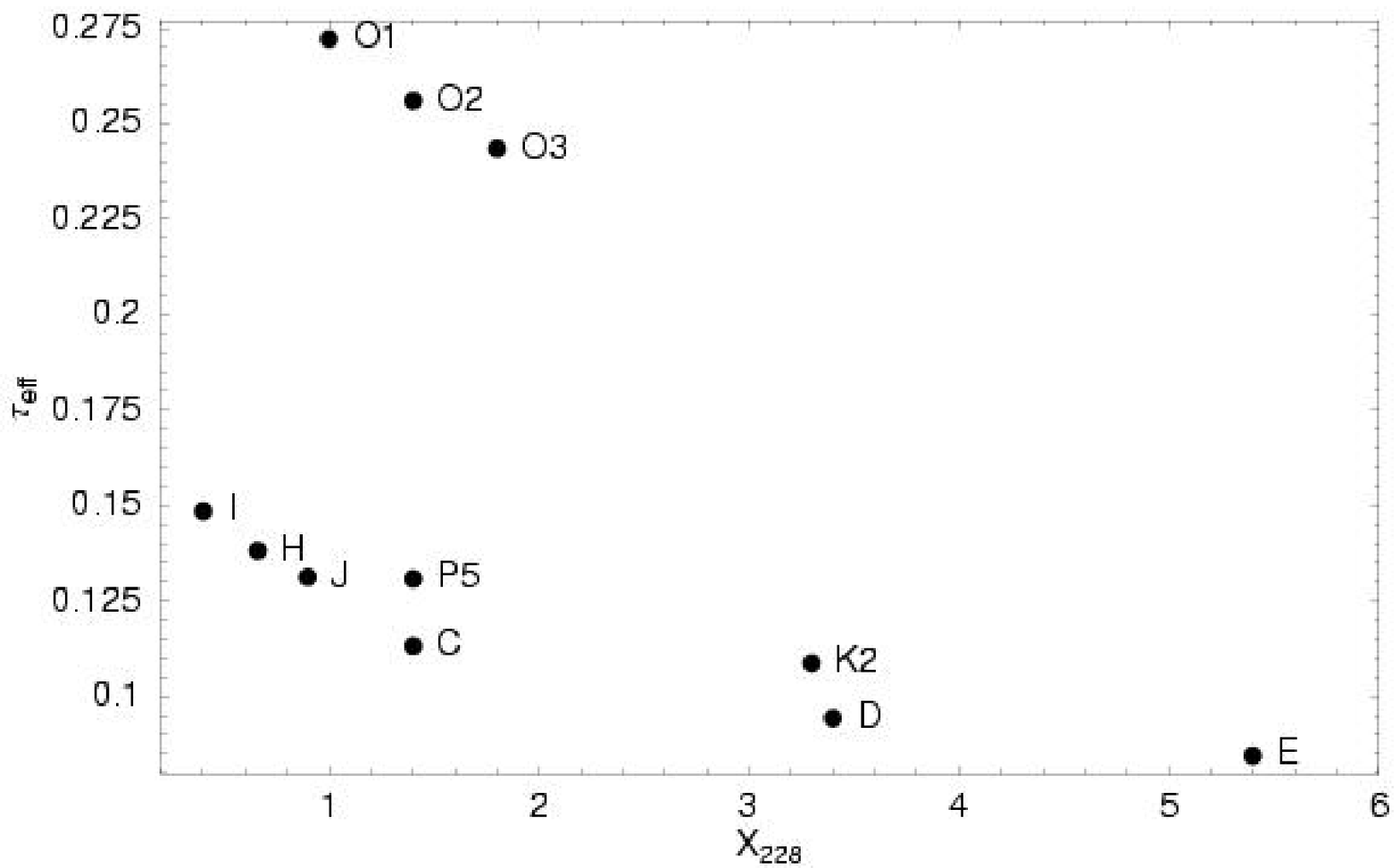}}
\caption{\NOTE{gamma228FbarRaw.ps}
Output parameter \taueff\ as a function of input \gammahe\ for four
sets of simulations that differ in \gammahe\ alone.
IHJ have 37.5 kpc cells, \gammah $=1.5$ and \sig $=0.70$. 
CDE have 37.5 kpc cells, \gammah $=1.2$ and \sig $=0.94$. 
P5K2 have 75 kpc cells, \gammah $=1.0$ and \sig $=0.94$.
O1O2O3 have 150 kpc cells, \gammah $=0.5$ and \sig $=0.84$. 
}
\label{fig:gamma228FbarRaw.fig}
\end{figure}

\clearpage
\begin{figure}
{\plotone{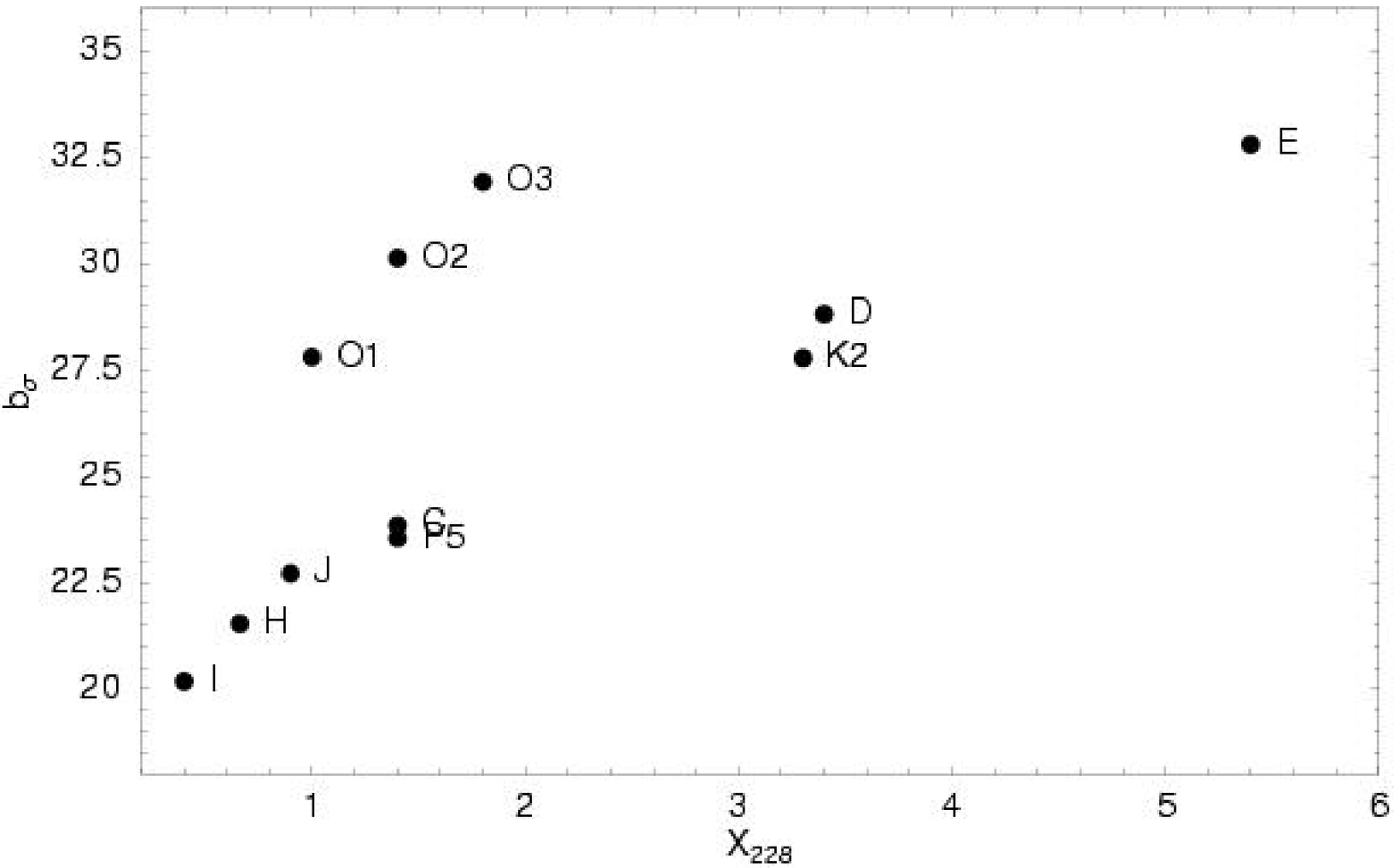}}
\caption{\NOTE{gamma228bsigRaw.ps}
As Fig. \ref{fig:gamma228FbarRaw.fig} but for \bsig .
}
\label{fig:gamma228bsigRaw.fig}
\end{figure}

\clearpage
\begin{figure}
{\plotone{specIJ}}
\caption{\NOTE{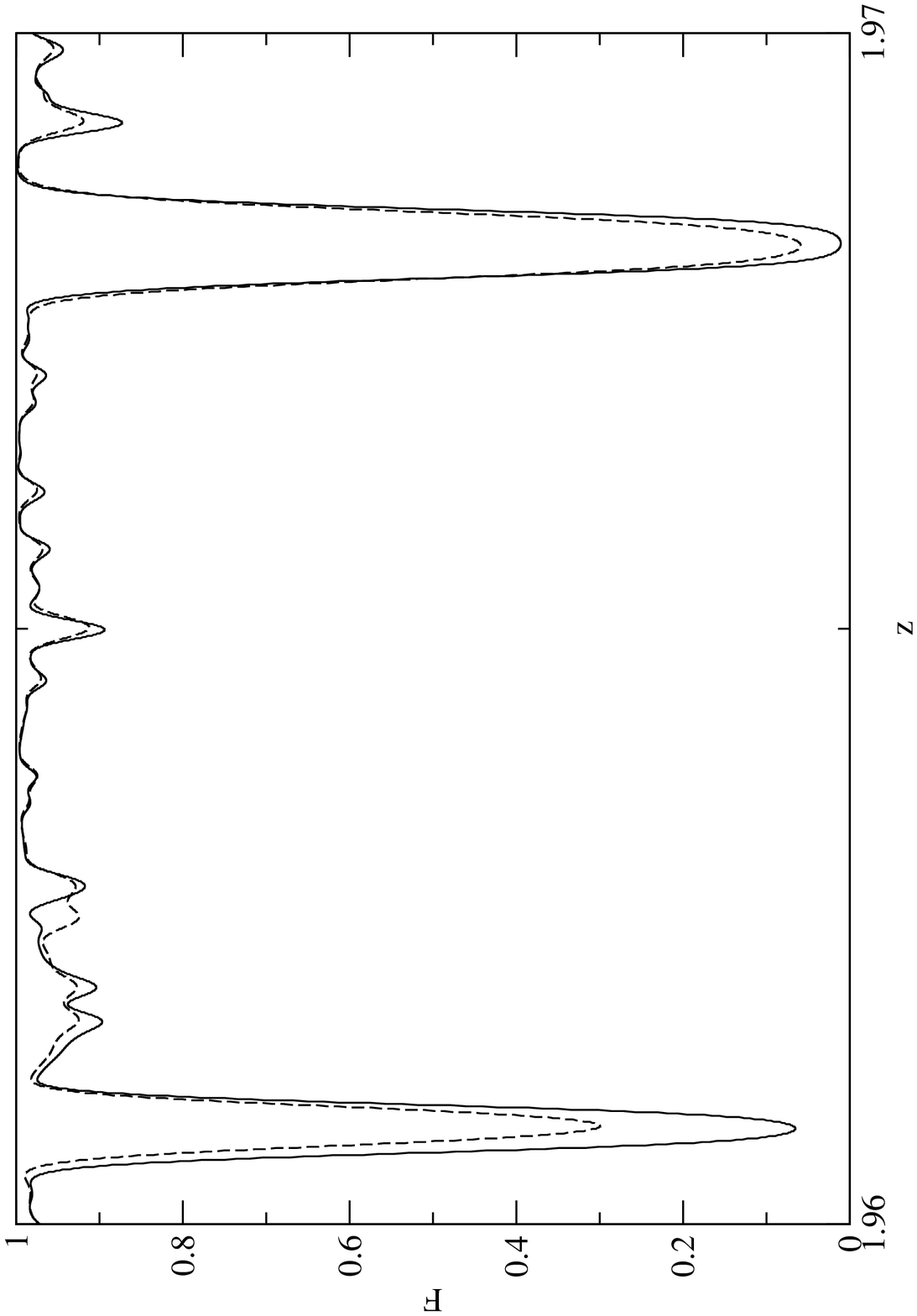}
Sample spectra for two simulations I and J which differ only in
the \gammahe\ parameter. The solid line is simulation I, \gammahe\ 
= 0.4 and the dotted line is simulation J, \gammahe\ = 0.9. 
}
\label{fig:specIJ.fig}
\end{figure}

%---------------------------------------------------
\begin{figure}
{\plotone{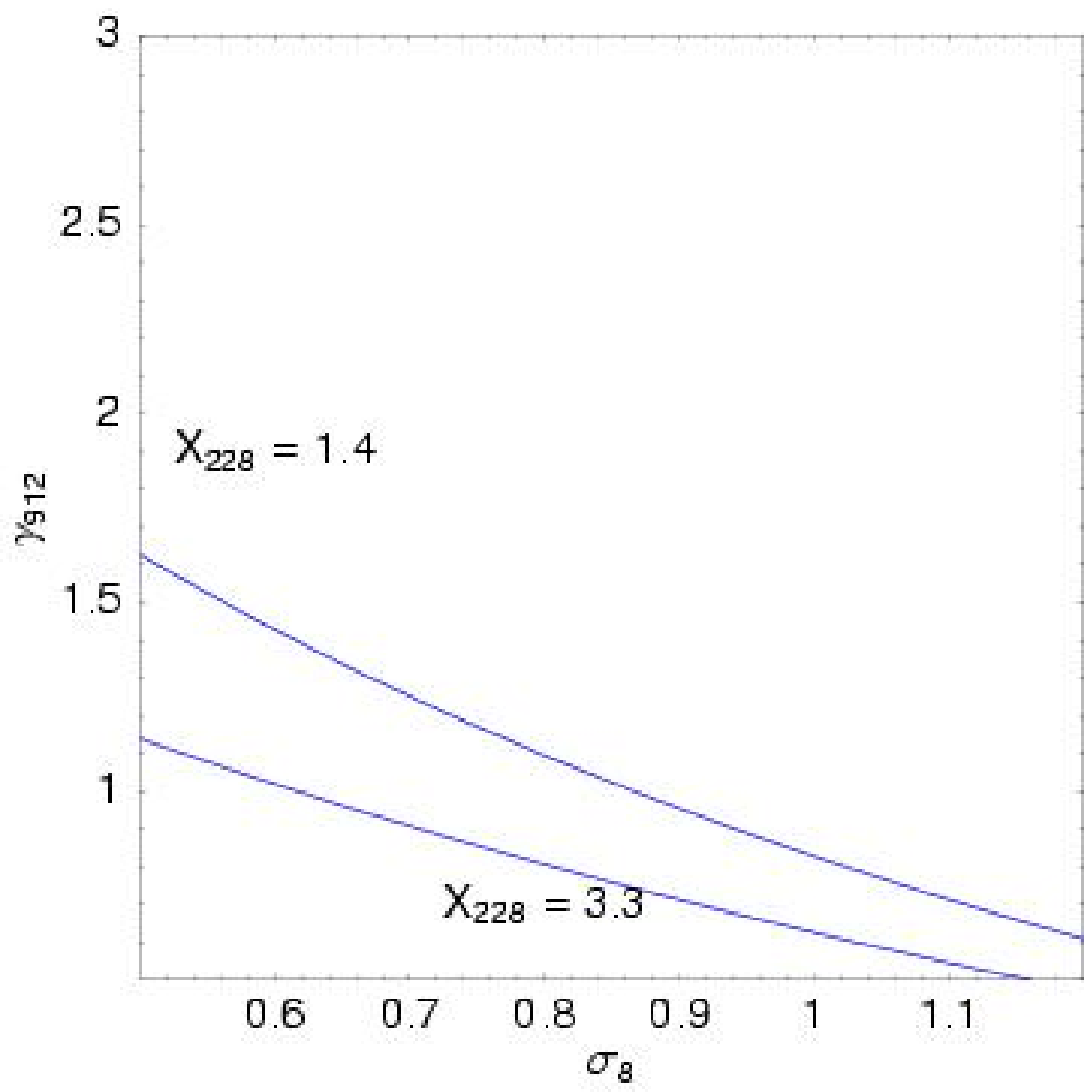}}
\caption{\NOTE{g912s8constg228.ps}
The lines show the value of \gammah\ required to give the observed mean
flux at $z=1.95$ for a range of \sig\ values and our standard model.
All our simulations have \sig\ 
in the range 0.7 -- 1.1, and hence outside this range the scaling relations 
are extrapolated and the errors are unknown and possibly large.
A value of \gammahe $= 1.4$ was used in Fig. 16 of T04b while
\citet[Fig. 3 right]{bolton04b} used 3.3. The models with 
\gammahe $= 3.3$ are all too hot and ruled out by the widths of
absorptions lines, measured with either \bsig\ or \ps . The models with 
\gammahe $=1.4$ are too hot for \sig $< 0.94$ and too cold for \sig $> 0.94$.
Models that are consistent with both the mean flux and the line widths have
\gammahe\ increasing with \sig , as shown in Fig. \ref{fig:g228s8.fig}.}
\label{fig:g912s8constg228.fig}
\end{figure}

\begin{figure}
{\plotone{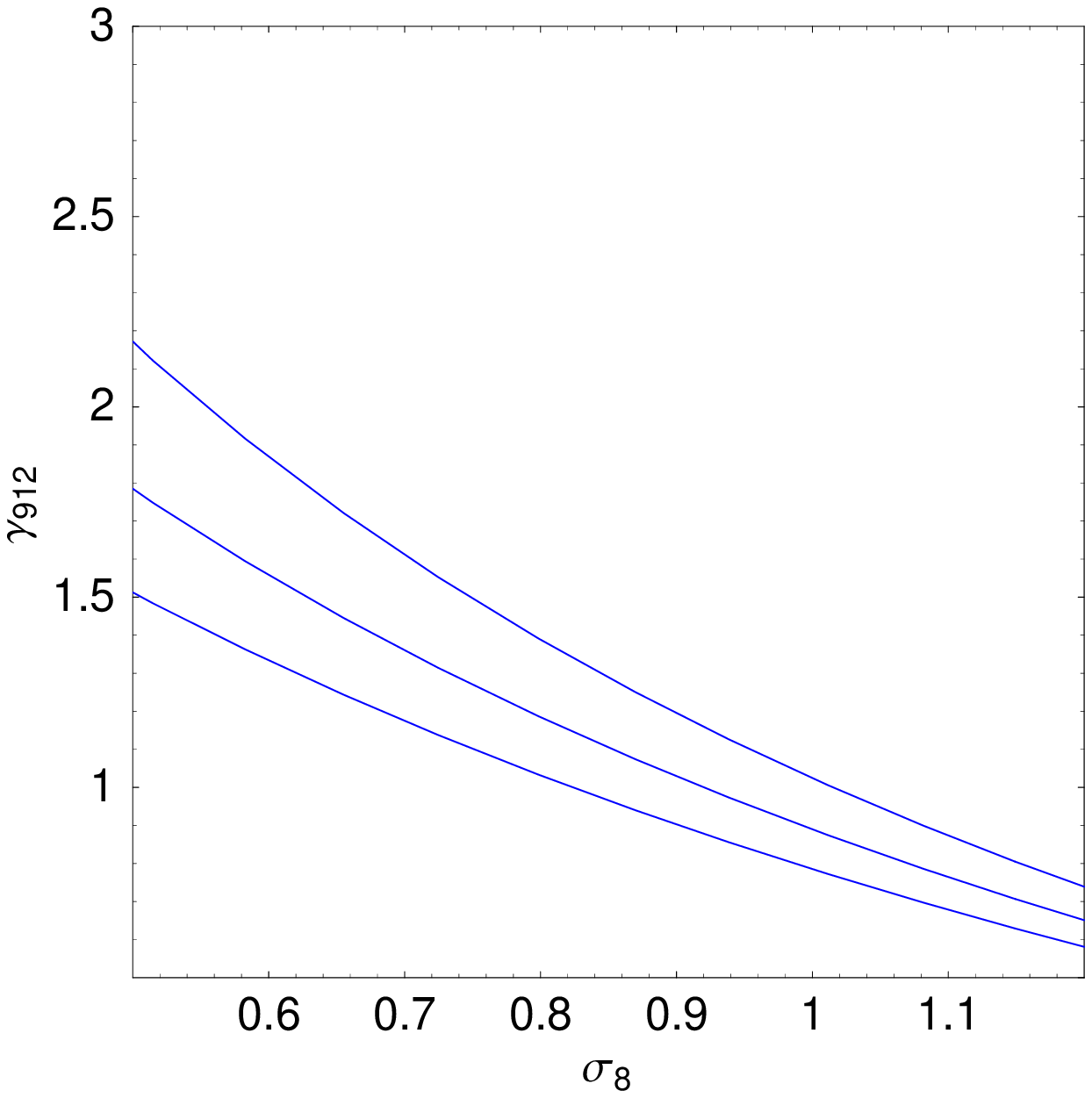}}
\caption{\NOTE{g912s8constg228Error1.ps}
The \gammah\ required to give DA = 0.885, 0.875, and 0.865 for
various \sig\ in our standard model with \gammahe $= 1.4$.
}
\label{fig:g912s8constg228Error1.fig}
\end{figure}

%---------------------------------------------------

\begin{figure}
{\plotone{g912s8}}
\caption{\NOTE{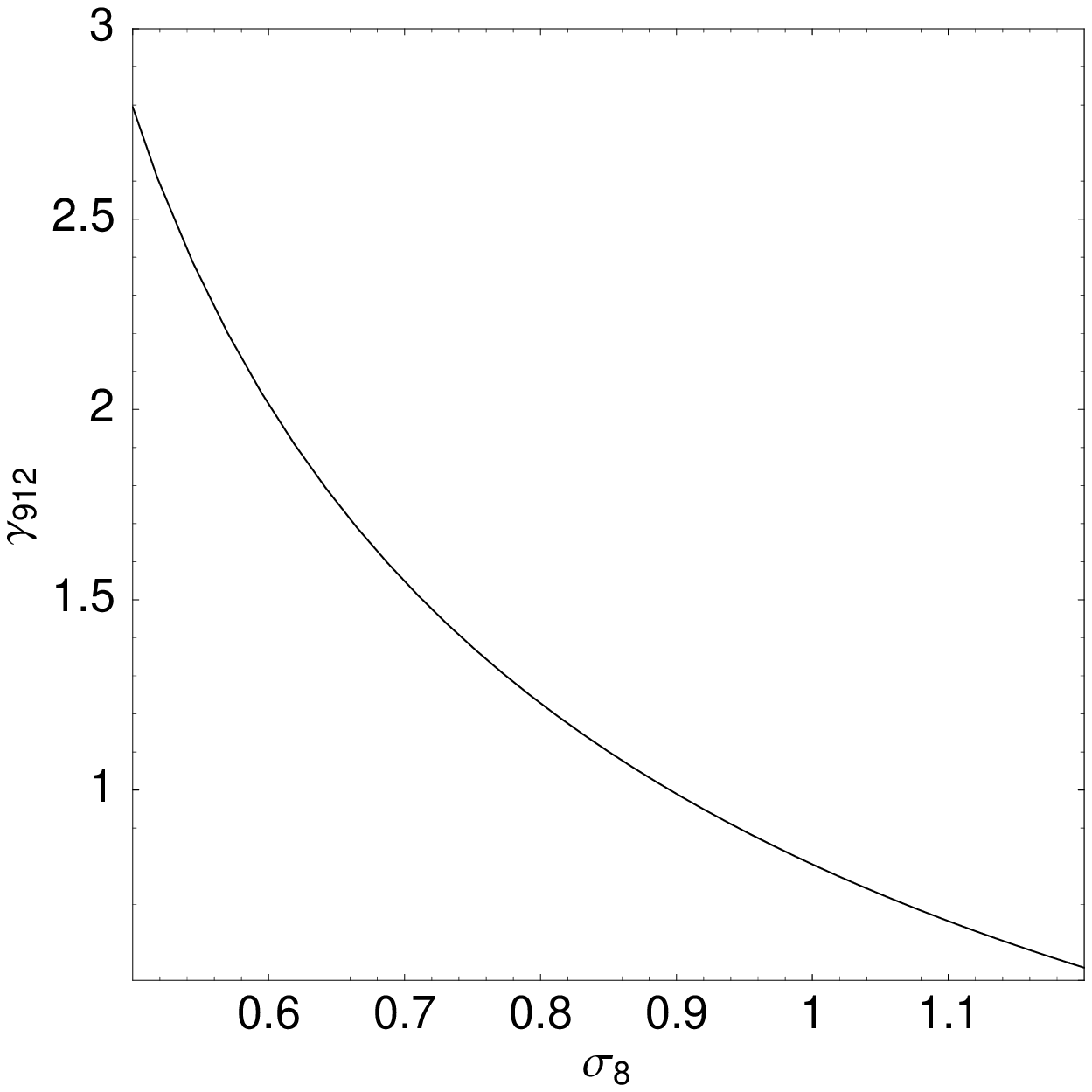} The line shows the values of 
\gammah\ and \sig\ that give simulated spectra that match observed \fbar\ and
\bsig . We increase the \gammahe\ with increasing \sig , 
as shown in Fig. \ref{fig:g228s8.fig}.
}
\label{fig:g912s8.fig}
\end{figure}

\begin{figure}
{\plotone{g228s8}}
\caption{\NOTE{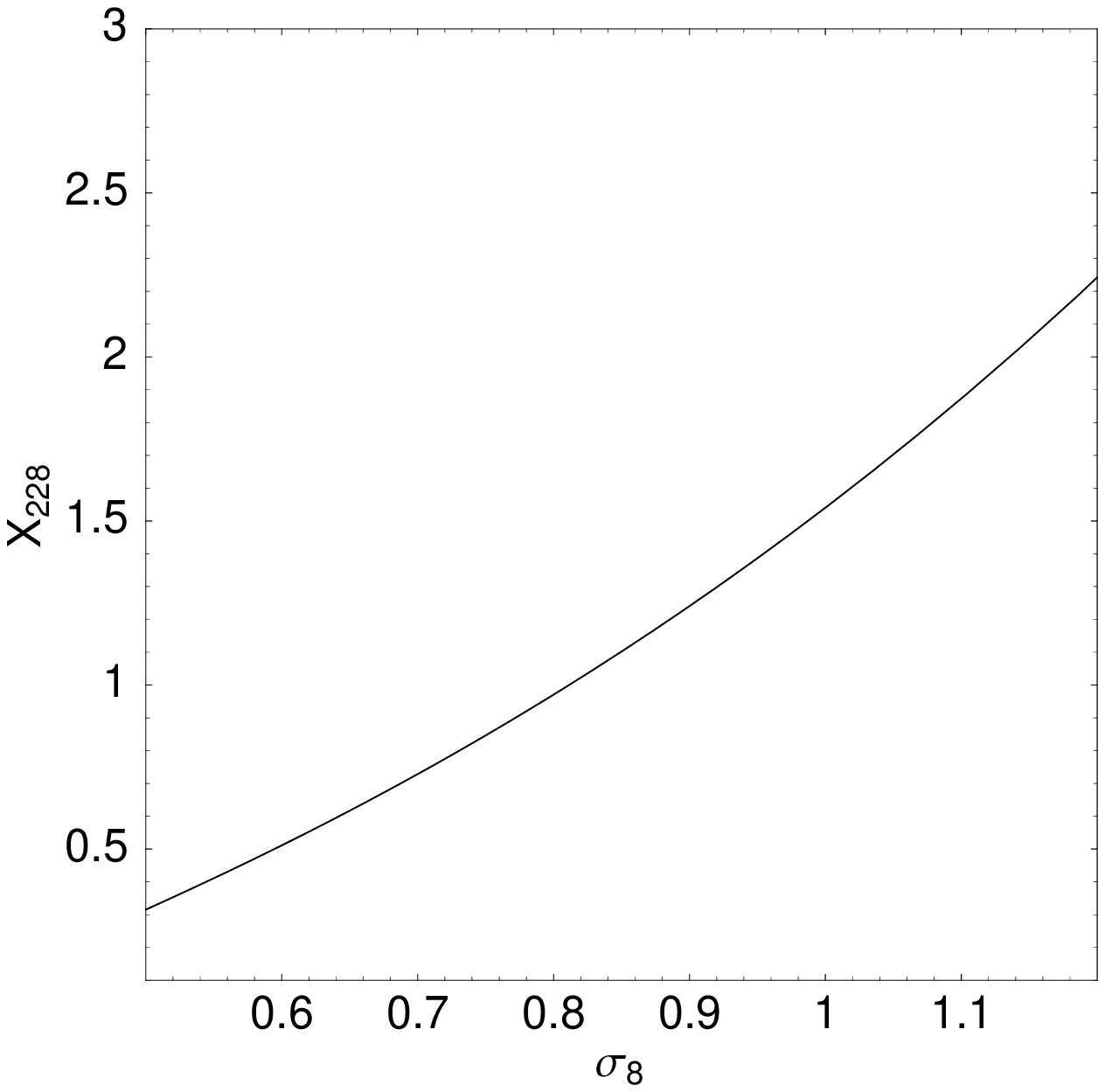} 
As Fig. \ref{fig:g912s8.fig} but for \gammahe\ and \sig .
}
\label{fig:g228s8.fig}
\end{figure}

\begin{figure}
{\plotone{g912g228}}
\caption{\NOTE{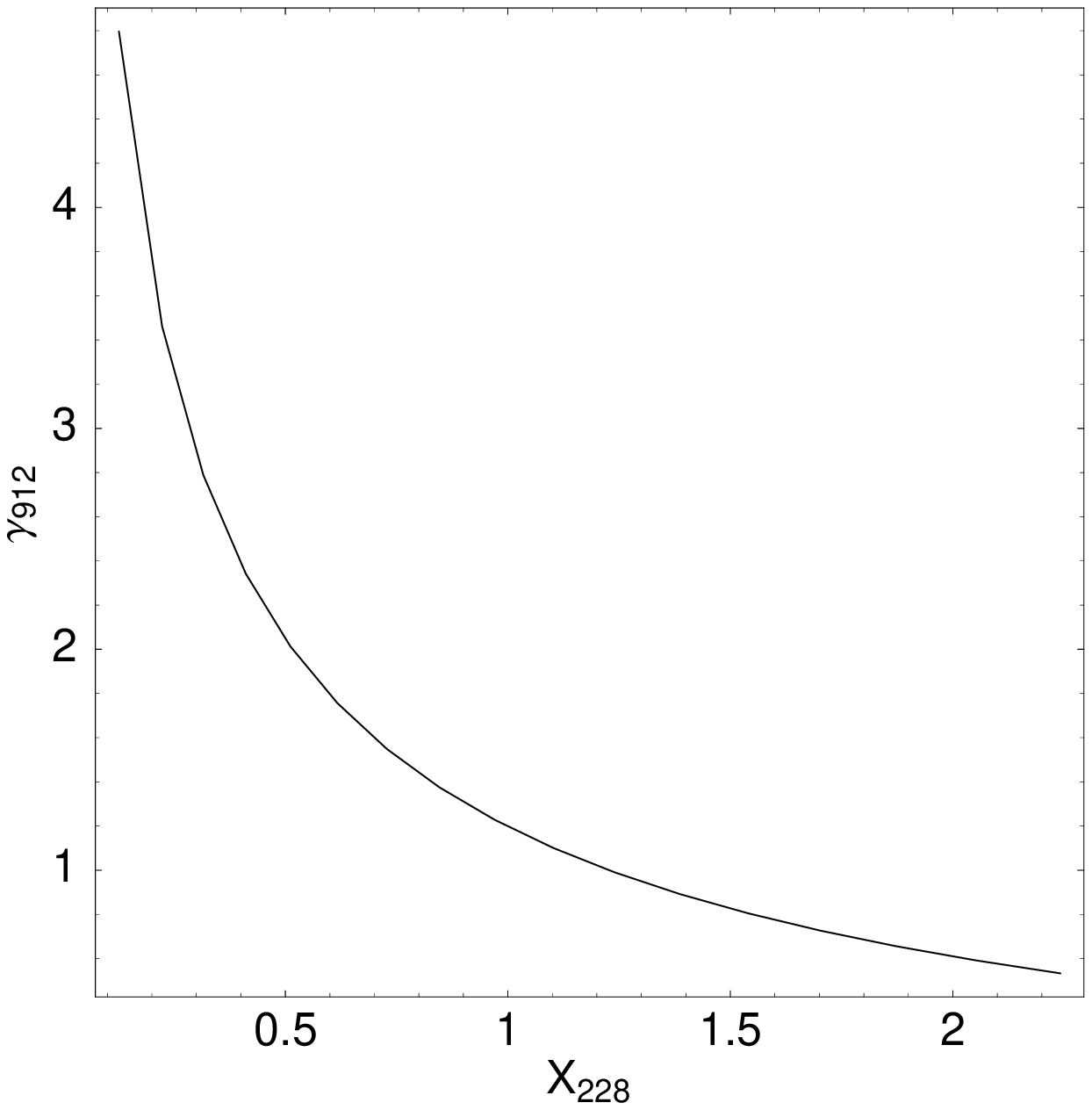} 
As Fig. \ref{fig:g912s8.fig} but for \gammahe\ and \gammah .
}
\label{fig:g912g228.fig}
\end{figure}

%---------------------------------------------------

%\begin{figure}
%{\plotone{fluxstart1}}
%\caption{\NOTE{fluxstart1.ps}}
%\label{fig:fluxstart1.fig}
%\end{figure}

\begin{figure}
{\plotone{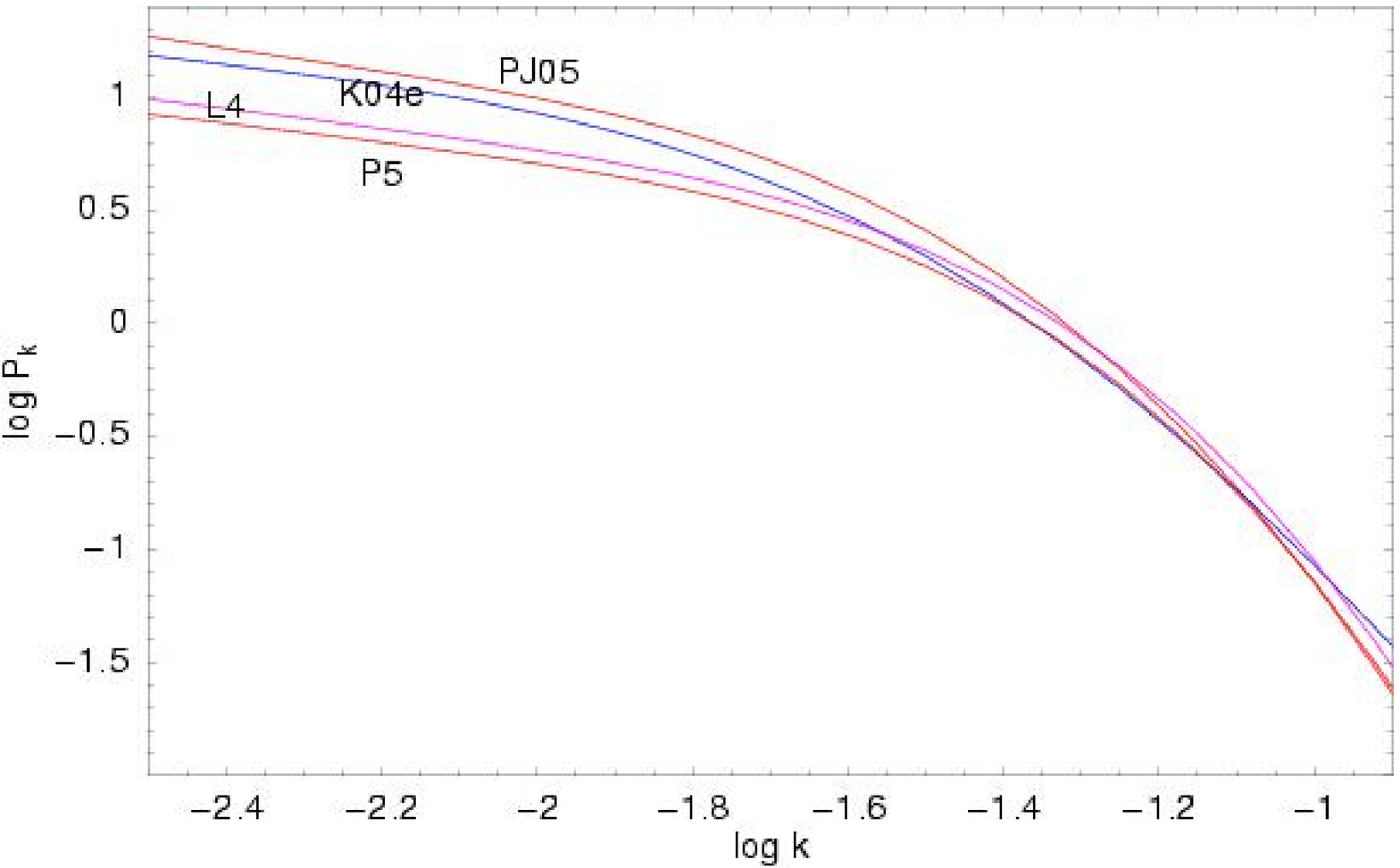}}
\caption{\NOTE{powerdata1.ps}  The flux power spectrum of 
data compared with the flux power spectrum from simulations L4 and P5, all at
$z=1.95$.
The spectrum labelled K04e is PF2I from Table \ref{table:kim.tbl}, and the
spectrum labelled PJ05 is our estimate of the $z=1.95$ power spectrum
from the high resolution data in Table \ref{tbl:obs_table}. 
Both K04e and PJ05 are the power of the low density IGM, with the 
contributions from \lya\ of LLS and metal lines removed.
We gave polynomial fits to these power spectra in Table \ref{table:pfits}.
There is good
agreement between the data and the simulations at $\log k > -1.4$ s/km, but
the simulations have too little power on larger scales or smaller $k$ values. 
The lack of large scale power may be because the boxes are too small.
We show the power at smaller 
$k$ values than is justified by the box size to help illustrate the trends 
and differences.
Both L4 and P5 have 19.2 Mpc boxes that include only modes $n > 2$
for $log k = -2$ s/km (\S \ref{subsec:MeasP}).
}
\label{fig:powerdata1}
\end{figure}

\end{document}